\documentclass[12pt,a4paper]{report}

\usepackage[utf8]{inputenc}
\usepackage[francais]{babel}
\usepackage[T1]{fontenc}
\usepackage{lmodern}

\usepackage{geometry}
\usepackage{graphicx} % allow embedded images
\usepackage{booktabs} % book-quality tables
\usepackage{microtype} % Improves character and word spacing
\usepackage{titling}
\usepackage{enumitem}

\usepackage{varioref}
\usepackage{amsmath}  % extended mathematics
\usepackage[separate-uncertainty = true,multi-part-units=single,group-separator = \text{~}]{siunitx}
\DeclareSIUnit[number-unit-product = {}]\an{an}

\usepackage{fancyvrb} % extended verbatim environments
  \fvset{fontsize=\normalsize}% default font size for fancy-verbatim environments

\usepackage{mdwlist}

\usepackage{pythontex}

\setlist[itemize,1]{label=$\bullet$}
\setlist[itemize]{%
labelsep=8pt,%                           
labelindent=2\parindent,%               
itemindent=0pt,%
leftmargin=*,%                          
listparindent=-\leftmargin% 
}

\newcommand{\RNum}[1]{\uppercase\expandafter{\romannumeral #1\relax}}

\newcommand{\subtitle}[1]{%
  \posttitle{%
    \par\end{center}
    \vspace*{-0.5cm}
    \begin{center}\large#1\end{center}
    \vskip0.5em}%
}

\usepackage{placeins}
\usepackage{float}
\usepackage[caption = false]{subfig}
\usepackage{floatpag}
\title{Incertitudes et Mesures}
\subtitle{Petit guide pédagogique}
\author{R. Legrand}
\renewcommand{\arraystretch}{1.5}

\makeatletter
\renewcommand*\env@cases[1][2.0]{%
  \let\@ifnextchar\new@ifnextchar
  \left\lbrace
  \def\arraystretch{#1}%
  \array{@{}l@{\quad}l@{}}%
}
\makeatother

\begin{document}
%%%%%%%%%%%%%%%%%%%%%%%%%%%%%%%%%%%%%%%%%%%%%%%%%%%%%%%%%%%%%%%%%%%%%%%%%%%%%%%%%%%%%%%%%%%
\maketitle

\tableofcontents
\listoffigures

\chapter*{Introduction}

Ce document est initialement à destination des enseignants de BTS en mesures industrielles et a été rédigé en vu de préparer la réforme du BTS Métiers de la mesure. Il propose une initiation à l'évaluation des incertitudes expérimentales en sciences. La première partie consiste en une revue des notions classiquement abordées en classes de lycée et post-bac. Les autres parties seront centrées sur des études de cas. Les méthodes par inférences sont au centre des techniques et méthodes proposées dans ce cours.

J'ai choisi de ne pas développer l'aspect "mesure" pour me concentrer rapidement sur les implications d'une incertitude statistique relative à cette mesure. Avec grande modestie, je souhaite mettre à plat le fait que l'ensemble des usages associés aux opérations de mesure relève souvent du réflex et ces usages sont "transmis" aux élèves et étudiants avec la même intensité dogmatique que lorsque nous étions nous même étudiants. Nous sommes nous arrêté pour comprendre pourquoi après avoir acquis des données, nous en faisons la moyenne arithmétique ? Et pourquoi ne retenons nous pas plutôt la valeur médiane ou n'importe quel autre estimateur ? Comment et pourquoi éliminer des valeurs aberrantes, et surtout comment les caractériser ?

L'objectif est de réaliser un tour des techniques statistiques employées pour tirer une valeur et une incertitude d'un ensemble de données. Ces notions statistiques sur lesquelles reposent nos calculs d'incertitudes sont loin d'être naïves et ne sont pas toutes abordables en cycle secondaire, ni même en cycle BTS. Ceci étant, les oublier complètement conduit à ne jamais se poser de questions quand à l'utilisation de ces outils.

Ce cours ne contient pas\footnote{Pour le moment\ldots} de notions sur la composition des incertitudes, la distinction entre incertitudes de type A et et de type B, ni même de techniques d'évaluation des incertitudes. Ce sont des sujets importants, et il existe de nombreuses ressources en ligne couvrant ces sujets.

Les chapitres \ref{InferenceBayesienne} et \ref{ApplicationsInferencesBayesiennes} portent exclusivement sur l'approche par analyse bayésienne de l'analyse de données. Il s'agit d'une simple introduction, des ouvrages ou articles bien plus complets existent pour approfondir le sujet. Les ressources utilisées pour la rédaction de ce document sont citées en tête de chapitre.

L'ensemble des méthodes numériques sont réalisées avec Python 3. Ce document est distribuée sous \textbf{licence Creative Commons BY-NC-SA}.

%Mesures et incertitudes
\chapter{Incertitudes et mesures}
\label{MesureEtIncertitudes}

Ce premier chapitre propose une vue d'ensemble des techniques et méthodes utilisées dans les classes de lycée et formations post-bac techniques et scientifiques en sortie de lycée. Souvent présentées comme un ensemble de règles à appliquer, les fondements théoriques sont souvent négligés quand à l'évaluation des incertitudes d'une mesure. Pourtant, il est loin d'être systématique que les conditions d'application des méthodes enseignées soit réunies et cela conduit à une évaluation erronée des incertitudes de mesure.

Enfin, il permettra d'introduire le fait qu'une mesure et son analyse ne peuvent se passer d'éléments statistiques. Ces derniers sont le langage nécessaire et adéquat à la compréhension du processus de mesure et l'élaboration de son incertitude.

Références : 
\begin{itemize}
\item VIM : Vocabulaire international de métrologie, BIPM
\item GUM : guide pour l'expression de l'incertitude de mesure, BIPM
\item Simple demonstration of the central limit theorem using mass measurements, K. K. Gan
\item A simple demonstration of the central limit theorem by dropping balls onto a grid of pins, K. K. Gan
\item Arrondissage des résultats de mesure, Nombre de chiffres significatifs, M.M. Bé, P. Blanchis, C. Dulieu
\item Statistiques, IUT Biothechnologie, 2ème année, Université de La Rochelle, J-C. Breton

\end{itemize}

\section{La mesure en physique et dans l'industrie}

Les sciences physiques sont une science expérimentale. Elles reposent sur la méthode scientifique permettant la validation d'hypothèses à partir d'observations expérimentales. Ces observations expérimentales nécessitent des mesures de grandeurs physiques qui sont alors connues avec un certain degrés d'incertitude. Nous verrons qu'il est nécessaire d'introduire un traitement statistique des données acquises. Ce dernier permet de circonscrire le domaine de connaissances, de certitudes associées aux mesures et au modèle. 

Concernant le domaine industriel, seul l'aspect mesure de grandeurs physiques est mis en jeu dans le cadre de processus de fabrication, de contrôles, d'accord à une norme ou un cahier des charges par exemple. Mais le fond du problème reste identique : Comment quantifier ces incertitudes pour valider ou comparer les mesures réalisées ?

\section{Vocabulaire}
Les notions recouvrants les concepts d'\textit{erreurs} de mesures sont diverses et variés et peuvent être souvent source de confusion. Les organisations de métrologie, regroupées sous l'acronyme \textbf{ISO}\footnote{International Organization for Standardization} ont publié un \textit{Guide pour l'expression de l'incertitude de mesure} ainsi que le \textit{VIM : Vocabulaire international de métrologie} afin de standardiser les usages.

Ces ouvrages utilisent et définissent une centaine de termes et acronymes. 

Voici quelques extraits du VIM :

\subsection{Mesure, mesurage et mesurande}

\begin{quote}
Le  mot  "mesure"  a,  dans  la  langue  française  courante,  plusieurs  significations.  Aussi  n'est-il  pas  employé  seul dans le présent Vocabulaire. C'est également la raison pour laquelle le mot "mesurage" a été introduit pour  qualifier  l'action  de  mesurer.  Le  mot  "mesure"  intervient  cependant  à  de  nombreuses  reprises  pour  former  des  termes  de  ce  Vocabulaire,  suivant  en  cela  l'usage  courant  et  sans  ambiguïté.  On  peut  citer,  par  exemple: instrument de mesure, appareil de mesure, unité de mesure, méthode de mesure. Cela ne signifie pas que l'utilisation du mot "mesurage" au lieu de "mesure" pour ces termes ne soit pas admissible si l'on y trouve quelque avantage. \textit{Extrait du VIM}
\end{quote}

\begin{itemize}
\item \textbf{Mesurage}, m : processus consistant à obtenir expérimentalement une ou plusieurs \textbf{valeurs} que l'on peut raisonnablement attribuer à une \textbf{grandeur}
\item \textbf{Mesurande}, m : \textbf{grandeur} que l'on veut mesurer
\end{itemize}

\subsection{Valeur vraie et Valeur de référence}

\begin{itemize}
\item \textbf{Valeur vraie}, f : \textbf{valeur  d'une  grandeur}  compatible  avec  la  définition de la \textbf{grandeur}
\item \textbf{Valeur de référence}, f : valeur  d'une  grandeur  servant  de  base  de  comparaison  pour  les  valeurs  de  grandeurs  de  même  nature  
\end{itemize}

On notera le caractère prudent et pragmatique de la définition de la \textit{Valeur vraie}. Dans les faits, il est impossible d'avoir accès à la valeur vraie d'une grandeur. En dehors des constantes fondamentales, pour lesquelles on considère qu'il n'existe qu'une seule valeur vraie, il est important de noter que les données acquises lors qu'un mesurage ne permettent que de définir des intervalles dans lesquelles il est raisonnable de penser que la valeur vraie se situe.

Cette notion sera rediscutée dans la partie traitant d'inférence et dans laquelle il ne sera plus question de trouver la valeur vraie, mais plutôt de définir la probabilité qu'une valeur soit vrai.

\subsection{Incertitude et Erreur}

\begin{itemize}
\item \textbf{Incertitude de mesure}, f : paramètre  non  négatif  qui  caractérise  la  dispersion  des \textbf{valeurs}  attribuées  à  un  \textbf{mesurande},  à  partir  des informations utilisées
\item \textbf{Erreur de mesure}, f : différence entre la \textbf{valeur mesurée} d'une \textbf{grandeur} et une \textbf{valeur de référence}
\end{itemize}

Il est à noter que ces définitions ne font pas référence à la \textbf{valeur vraie}. S'il n'existe pas de valeur de référence, il est difficile d'utiliser la notion d'erreur telle que définie dans le \textit{VIM}. Il reste donc cette notion d'incertitude de mesure qui est une grandeur déterminée par analyse des données et permettant de définir un degré de confiance pour la valeur mesurée.

\section{Conséquences des incertitudes dans la mesure}

Les sources des incertitudes de mesures sont répertoriées dans les documents ISO consacrés aux incertitudes de mesure : conditions environnementales, biais humain, mesurande intrinsèquement aléatoire, etc.

L'ensemble de ces sources sont globalement à regrouper en deux catégories. Celles qui ont un effet systématique sur la mesure, celles qui ont un effet aléatoire.

\begin{itemize}
\item Les effets systématiques conduisent à un décalage constant de la valeur mesurée par rapport à la valeur théorique. Une fois identifiés, ces effets systématiques peuvent être simplement retranchés à la valeur mesurée. Toute la difficulté réside dans l'identification et l'évaluation de ces effets systématiques.
\item Les effets aléatoires conduisent à une erreur aléatoire dispersée autour de la valeur vraie. Ces effets peuvent être liés à l'objet d'étude\footnote{Par exemple, la taille d'individus dans une population.} ou au processus de mesure\footnote{Mesure de l'accélération de la pesanteur en un lieu précis.}. Dans les deux cas, il sera possible d'étudier les mesures avec une méthode statistique pour en tirer des informations utiles sur le mesurande.
\end{itemize}

\textbf{Remarque didactique}

\begin{figure}[h]
\begin{center}
\includegraphics[width = 0.5\textwidth]{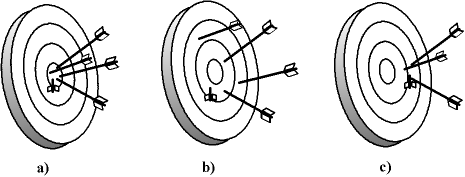}
\end{center}
\caption[Représentations des incertitudes de mesures illustrée par flèches et cible]{Représentations des incertitudes de mesures classiquement présentées aux étudiants et illustrée par flèches et cible. Cette représentation a comme défaut majeur de mettre de côté le fait qu'en générale, l'expérimentateur n'a aucune idée de la position précise du centre de la cible.}
\label{Incertitude_illustree_par_fleches_et_cible}
\end{figure}

Il est courant de voir les notions d'incertitude aléatoire et systématique représentées comme des flèches atteignant une cible donc le centre serait la \textit{valeur vraie} (figure \ref{Incertitude_illustree_par_fleches_et_cible}).

\begin{figure}[h]
\begin{center}
\includegraphics[width = 0.4\textwidth, angle = 90]{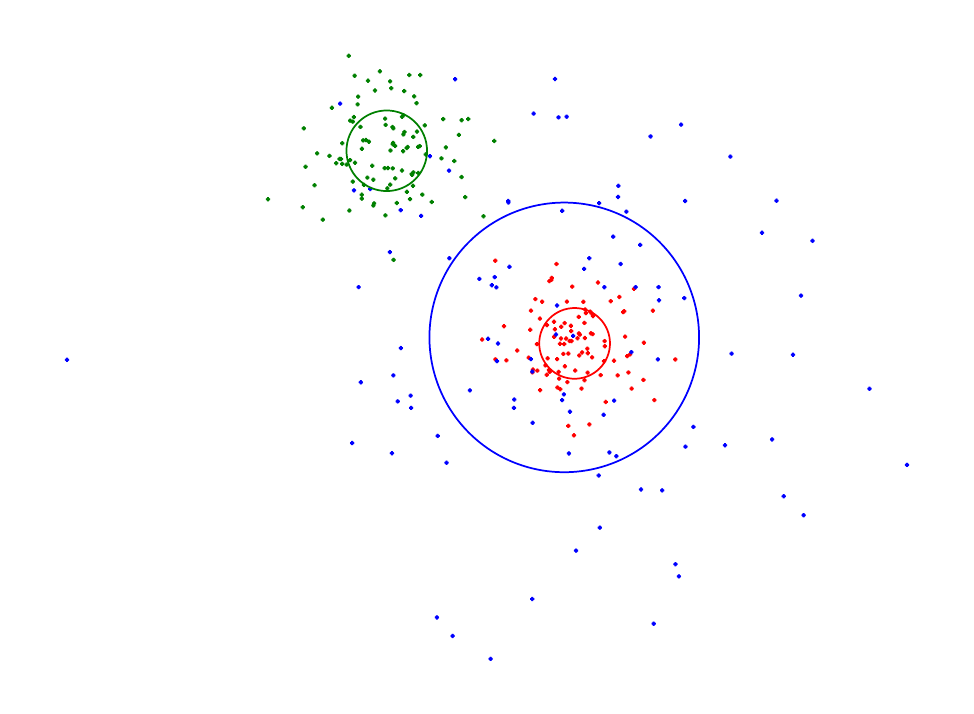}
\end{center}
\caption[Représentation des incertitudes de mesure avec plusieurs mesurages]{Trois jeux de données constituant le mesurage d'un même mesurande sont représentés en rouge, bleu et vert. A chaque jeu de donnée, on associe un cercle dont le centre est la moyenne des points et le rayon représente la dispersion. Les données vertes et rouge ont sensiblement la dispersion, mais ne sont pas centrée sur la même valeur : il existe un effet systématique sur au moins l'un des deux mesurages. Le mesurage représenté en bleu est bien plus dispersée que les deux autre mesurages, mais est compatible avec le mesurage rouge.}
\label{MesureDisper}
\end{figure}

La difficulté de cette représentation est qu'elle sous-entend une connaissance à priori de la valeur vraie et de ce fait, la \textit{qualité} de la mesure est caractérisée à partir de cette connaissance. Dans les faits, il ne faut pas oublier que la valeur vraie n'est jamais connue et que cette valeur ne peut qu'être estimée à partir des données. La figure \ref{MesureDisper} représente, à mon sens, tout aussi schématiquement mais plus correctement la conception qu'il faut se faire d'une analyse de données. Sur cette figure, les données sont des points représentés par des couleurs différentes : trois jeux de données sont respectivement représentés en rouge, vert et bleu. La position moyenne de chaque jeu de donnée est utilisée comme centre d'un cercle dont le rayon est l'écart type.
\begin{itemize}
\item Les jeux de mesures vert et rouge présente la même dispersion, donc les mêmes incertitudes de mesure, mais ne sont pas centrées sur la même moyenne : mise en évidence d'une erreur systématique.
\item Les jeux de mesures rouge et bleu sont "inclus" l'un dans dans l'autre, ces deux mesures sont compatibles mais présente une dispersion, donc une incertitude aléatoire différente
\item A partir de ces trois jeu de données, il est raisonnable de penser que seuls les mesures vertes présentent une erreur systématique par rapport à la valeur vraie.
\end{itemize}

\subsection{Caractériser l'aléatoire}

\subsubsection{Distribution statistique et mesures}

La dispersion des valeurs issues du mesurage forme une distribution statistique. Cette distribution est souvent modélisable par une loi de probabilité décrivant le caractère aléatoire d'une expérience. 

La description de cette loi de probabilité se fait avec un nombre restreint de paramètres. En particulier, dans le cadre d'une mesure, valeur moyenne et de l'écart type de la loi de probabilité sont les deux paramètres permettant de caractériser principalement le résultat de la mesure. Ces deux grandeurs sont formellement définies par les relations suivantes :

Pour un nombre N d'éléments contenus dans la distribution : 

\begin{equation}
\text{Espérance : }\mu =  \dfrac{1}{N} \sum_i^N x_i
\end{equation}

\begin{equation}
\text{Écart type : }\sigma = \sqrt{\dfrac{1}{N} \sum_i^N (\mu - x_i)^2}
\end{equation}

Le nombre d'éléments contenu dans cette distribution statistique peut être infini si la grandeur physique mesurée est continue. Ainsi, en raisonnant sur l'ensemble des éléments accessibles au mesurage, il est possible de définir la moyenne et l'écart type d'une loi de probabilité $f(x)$ par : 

\begin{equation}
\text{Espérance : }\mu =  \int f(x) dx
\end{equation}

\begin{equation}
\text{Écart type : }  \sigma = \int f(x) (\mu - x)^2 dx
\end{equation}

\textbf{Remarque :} La variance est définie comme le carré de l'écart type $V = \sigma^2$. Les deux termes recouvrent des concepts similaires.

Ces distributions sont caractéristique du mesurage et du mesurande.

\subsubsection{Loi Normale}

Les lois normales jouent un rôle central. Elles sont parmi les loi de probabilité les plus adaptées pour modéliser les phénomènes naturels issus d'événements aléatoires. Elles prennent une place particulière car elles sont la limite de suites de tirages aléatoires indépendants et dont le comportement individuel ne suit pas forcément une loi normale (voir théorème de la limite centrale).

%\begin{pycode}
%import numpy as np
%from scipy import stats
%import matplotlib.pyplot as plt
%from matplotlib import rc
%
%plt.clf()
%
%moy = 50
%EcType = 10
%
%# plot a best-fit Gaussian
%F_fit = np.linspace(0, 100, 100)
%pdf = stats.norm(moy, EcType).pdf(F_fit)
%
%
%#fig, ax = plt.subplots()
%
%plt.rc('text', usetex=True)
%plt.rc('font', family='serif')
%
%plt.plot(F_fit, pdf, '-k')
%plt.xlabel("x"); plt.ylabel(r'$\mathcal{N}(x)$')
%plt.vlines([moy], 0, 1*max(pdf), linewidth=2, alpha=0.4, color = 'b')
%plt.text(moy+2,1*max(pdf), r'Moyenne $\mu$ de la loi')
%
%plt.vlines([moy+EcType], 0, 0.6*max(pdf), linewidth=2, alpha=0.4, color = 'r')
%plt.vlines([moy-EcType], 0, 0.6*max(pdf), linewidth=2, alpha=0.4, color = 'r')
%arr_width = .009
%#ax.arrow(moy-EcType, 0.005,2*EcType, 0,width =0.0001, head_width=0.001, head_length=1, fc='k', ec='k',length_includes_head=True)
%plt.annotate('', xy=(moy-EcType,0.005), xytext=(moy+EcType,0.005),
%            arrowprops={'arrowstyle': '<->'}, va='center')
%plt.text(moy-8,0.006, r'Écart type $\sigma$')
%plt.suptitle('Loi normale $\mathcal{N}(x)$')
%
%plt.savefig("C:/Users/Romain/Desktop/Mesures/Images/LoiNormale.pdf", bbox_inches="tight")
%plt.clf()
%
%
%
%\end{pycode}

\begin{figure}[h!]
\begin{center}
\includegraphics[width = 0.6\textwidth]{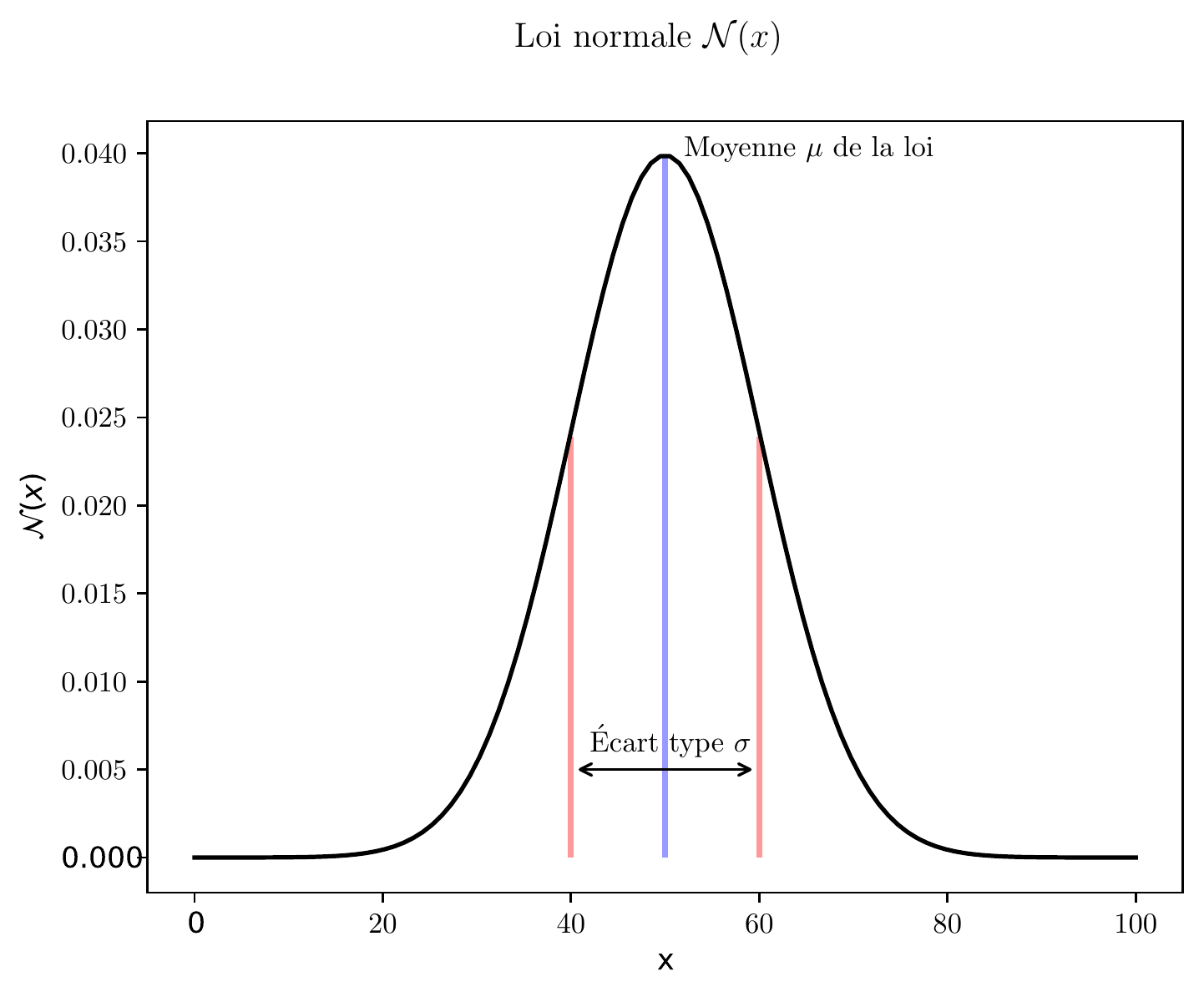}
\end{center}
\caption[Exemple de loi normale]{Exemple de loi normale centrée sur $\mu = 50$ et dont l'écart type $\sigma = 10$.}
\label{LoiNormale}
\end{figure}

La loi normale représentée en figure \ref{LoiNormale} est définie par la distribution suivante : 
\begin{equation}
\mathcal{N}(x) = \dfrac{1}{\sigma \sqrt{2\pi}} e^{-\dfrac{1}{2} \left(\dfrac{x - \mu}{\sigma}\right)^2}
\end{equation}

Il s'agit d'une courbe symétrique dont les valeurs les plus probables sont situées à quelques écarts types de la valeur moyenne. Pour la loi normale, on peut citer quelques intervalles de confiance : 
\begin{itemize}
\item $P(\mu - \sigma \leq x \leq \mu + \sigma) = \num{0.68}$
\item $P(\mu - 2\sigma \leq x \leq \mu + 2\sigma) = \num{0.95}$
\item $P(\mu - 3\sigma \leq x \leq \mu + 3\sigma) = \num{0.997}$
\end{itemize}

Un mesurage réalisant cette distribution aura comme caractéristiques : \nolinebreak
\begin{itemize}
\item résultat du mesurage : espérance $\mu$
\item incertitude de mesure caractérisée par l'écart type $\sigma$
\end{itemize}

Cela signifie qu'un mesurage aura 68\% de chance de fournir une valeur située à une distance $\sigma$ de l'espérance $\mu$.

\subsubsection{Construction d'estimateurs à partir de données}

En réalité, il est impossible d'avoir accès à la loi de probabilité dans sont ensemble mais uniquement à une sous partie, à un échantillon réalisant la loi de probabilité : c'est l'objectif du mesurage, qui permet d'obtenir des données représentative de la distribution à laquelle obéit le mesurage en question.

A partir de ces données, le rôle des estimateurs est de fournir une \textit{estimation}\footnote{Grandeur permettant d'évaluer un paramètre statistique inconnu à partir d'un échantillon. Divers paramètres caractérisent les estimateurs : convergence, biais, efficacité et robustesse. L'objectif de ce document n'est pas de refaire un cours démontré de statistique mais de donner des éléments permettant un approfondissement du sujet.} de l'espérance et de l'écart-type. Ces deux grandeurs sont représentatives de la valeur vraie d'une part et de l'incertitude associée au mesurage d'autre part.

Pour une loi statistique d'espérance $\mu$ et d'écart type $\sigma$ et à partir d'un ensemble de données ${x_i}$ de cardinal $n$, on définit les estimateurs suivants : 

L'estimateur $\overline{x}$ sans biais de l'espérance est la moyenne arithmétique  : 
\begin{equation}
\overline{x} = \dfrac{1}{N} \sum_i^N x_i
\end{equation}

L'estimateur $\sigma_{n-1}$ sans biais de l'écart-type lorsque l'espérance est inconnue :
\begin{equation}
\sigma_{n-1}^2 = \dfrac{1}{n - 1} \sum_i^n (x_i - \overline{x})^2
\end{equation}

\textbf{Remarque :} cet estimateur est aussi appelé "écart type expérimental", pour éviter la confusion il est conseillé d'utiliser le terme d'estimateur de l'écart type.

Dans le cas où l'espérance $\mu$ est connue, l'estimateur $S$ de l'écart type est alors :
\begin{equation}
S^2 = \dfrac{1}{n} \sum_i^n (x_i - \mu)^2
\end{equation}

Cet estimateur est rarement utilisé dans le cadre d'une mesure physique puisque le but d'une mesure est de déterminer cette valeur moyenne qui sera ensuite assimilée à l'espérance qui est la valeur vraie du mesurage et de caractériser le tout par un écart type.

\subsubsection{Intervalle de confiance}

Si un mesurage ${x_i}$ contenant $n$ valeurs suit une loi statistique d'espérance $\mu$ et d'écart type $\sigma$ alors, l'application du théorème de la limite centrale indique que la moyenne arithmétique $\overline{x}$ suit une loi normale d'espérance $\mu$ et d'écart type $\dfrac{\sigma}{\sqrt{n}}$.

Il est classiquement conclu le résultat suivant concernant $\mu$ la valeur vraie de la mesure

\begin{equation}
\overline{x} = \mu  \pm \dfrac{\sigma}{\sqrt{n}}
\end{equation}

Ce qui, en terme d'intervalle, correspond à la probabilité suivante pour une loi normale : 

\begin{equation}
P(\mu - \dfrac{\sigma}{\sqrt{n}} \leq \overline{x} \leq \mu + \dfrac{\sigma}{\sqrt{n}}) = 68 \%
\end{equation}

Cela signifie que statistiquement, la valeur moyenne des données à 68\% de chance d'être située à $\dfrac{\sigma}{\sqrt{n}}$ de l'espérance de la distribution, qui peut être assimilée à la valeur vraie recherchée.

Classiquement, un retournement de cette relation est réalisé pour déterminer l'intervalle de confiance. En assimilant $\overline{x}$ à $\mu$, on obtient l'intervalle de confiance suivant : 

\begin{equation}
P( \overline{x}- \dfrac{\sigma}{\sqrt{n}} \leq \mu \leq \overline{x} + \dfrac{\sigma}{\sqrt{n}}) = 68 \%
\end{equation}

Cet intervalle correspond à l'incertitude type $u$ de la mesure : 
\begin{equation}
u(x) = \dfrac{\sigma}{\sqrt{n}}
\end{equation}

\textbf{Remarque : } Ces résultats ne sont utilisables que lorsque la variance de la valeur du mesurage est connue. Cette variance peut être éventuellement déterminée au travers d'un processus de composition des incertitudes.

\subsubsection{Incertitude élargies $U$}

L'incertitude élargie correspond à des intervalles de confiance défini pour une probabilité donnée que la valeur vraie du mesurande soit dans l'intervalle. Généralement, on définit un facteur d'élargissement $k$ tel que l'incertitude élargie  $U = k u$, avec $u$ l'incertitude type.

Pour un mesurage suivant une loi normale, il est possible de définir les incertitudes élargies suivantes :

\begin{center}
\begin{tabular}{c|c}
facteur d'élargissement $k$ & niveau de confiance en \% \\ 
\hline 
\hline 
1 & \num{68.3} \\ 

2 & \num{95.5} \\ 

3 & \num{99.7} \\ 

\end{tabular} 
\end{center}

Ces facteurs d'élargissement ne sont valables que dans le cas d'une loi normale. Usuellement en physique, la valeur moyenne est issue de processus de mesure dont il est possible de connaitre les caractéristiques, et donc d'après le théorème de la limite centrale, la valeur moyenne tend vers une statistique suivant une loi normale. Cependant, cela n'est vrai que si le nombre de mesures est grand et si les conditions d'application du théorème de la limite centrale sont vérifiées.

%\subsubsection{Inégalité de Bienaymé-Tchebychev}
%Dans le cas où la distribution statistique est inconnue, il est toujours possible de minorer le niveau de confiance au moyen de l'inégalité de Bienaymé-Tchebychev. 
%\begin{center}
%
%\begin{tabular}{c|c}
%\hline 
%facteur d'élargissement $k$ & niveau de confiance en \% \\ 
%\hline 
%$k$ & $1 - \dfrac{1}{k^2}$ \\ 
%
%2 & \num{75} \\ 
%
%3 & \num{89} \\ 
%\hline 
%\end{tabular} 
%\end{center}
%
%Dans les faits, cette inégalité est rarement exploitable dans le domaine de la mesure physique puisqu'elle suppose de connaitre l'espérance et la variance de la loi.

\section{Théorème de la limite centrale}

\subsection{Énoncé et lien avec la mesure physique}

Ce théorème permet de connaitre la statistique limite de la valeur moyenne de variables indépendantes possédant une espérance et un écart type borné. La valeur moyenne tend vers une loi normale dont l'écart type est de la forme $\dfrac{\sigma}{\sqrt{n}}$.

Ce résultat est d'une importance fondamentale dans le domaine de la mesure. En effet, il permet de comprendre l'intérêt d'évaluer la valeur moyenne d'une série de données : cette valeur moyenne tend vers une valeur limite \textbf{et} l'incertitude associée à cette valeur moyenne décroit avec le nombre de mesures en $\dfrac{1}{\sqrt{n}}$.

\subsection{Exemple}
Imaginons qu'un étudiant reçoive une note comprise en 0 et 10. Bien que ce ne soit pas flatteur pour l'enseignant, cette probabilité est constante\footnote{Il s'agit d'une distribution uniforme et continue sur l'intervalle [0,10]} sur l'ensemble des notes accessibles à l'étudiant. L'étudiant reçoit $N$ notes.

\begin{center}
\begin{center}
 Quelle loi statistique suit la valeur moyenne ?
\end{center}
\end{center}

Pour répondre à ce problème, nous allons exploiter à notre avantage la possibilité d'utiliser des programmes informatiques capables "d'explorer" pour nous ce type de distribution. L'idée est de simuler l'existence de centaines de milliers d'étudiants recevant $N$ notes dont nous allons calculer la valeur moyenne.

Le choix d'une distribution uniforme et continue de notes peut sembler éloigner de la réalité, mais les résultats sont identiques avec une distribution discrète et uniforme. 

Pour $N=3$, nous allons tirer au sort 3 notes, puis en calculer la moyenne pour la stocker dans un tableau. Ensuite le processus recommence avec un nouveau jeu de note. Il est ainsi possible de s'intéresser à la statistique de la moyenne.

\begin{figure}[h!]
\begin{center}
\includegraphics[width = 0.6\textwidth]{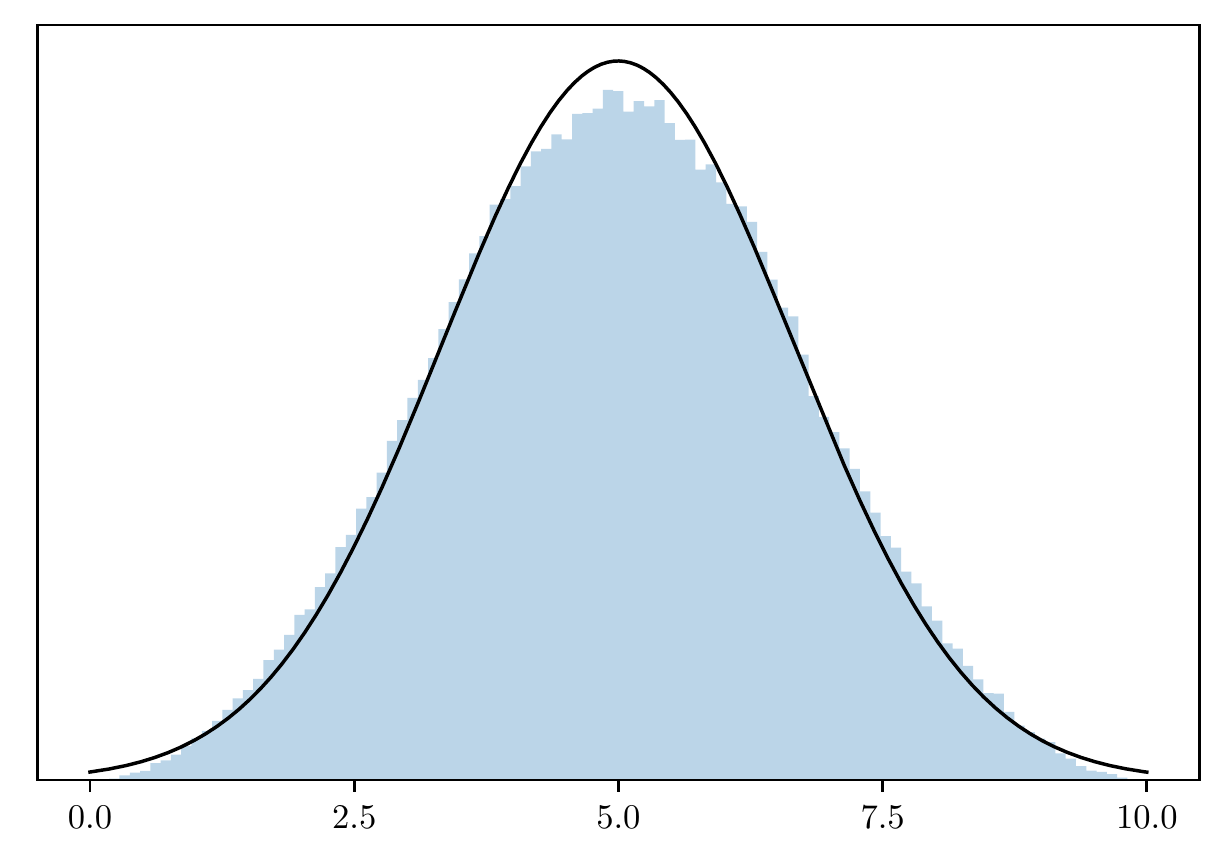}
\end{center}
\caption[Distribution d'une moyenne de trois notes]{Distribution d'une moyenne de trois notes comprises entre 0 et 10. Le tirage aléatoire est réalisé \num{300000} fois, ce qui permet d'explorer la loi statistique issue du calcul de la moyenne.}
\label{TheoLimiteCentrale2}
\end{figure}

Pour $N=3$, le résultat obtenu est celui présenté en figure \ref{TheoLimiteCentrale2}.

La distribution possède une forme de courbe en cloche, la valeur la plus probable est 5, sans surprise. La courbe représentée en noir est une loi normale de même valeur moyenne et de même écart type. La courbe obtenue, bien que proche, ne suit pas une loi normale. Outre quelques écarts, la différence majeur est que la loi normale autorise des valeurs allant à l'infini. Ici, c'est impossible dans la mesure où les notes et la valeur moyenne sont comprises entre 0 et 10.

Réaliser ce même travail pour différentes valeurs de N (figure \vref{TheoLimiteCentrale1}). Pour $N=1$, nous retrouvons la distribution initiale uniforme sur l'intervalle [0, 10], aux fluctuations statistiques prés qui s'atténuent en augmentant le nombre de tirages.

Pour $N=2$, la distribution est triangulaire, avec un maximum de probabilité à $5$. Pour $N\geq 3$, la forme générale s'arrondit et la courbe s'affine de plus en plus : son écart type se réduit.
Pour chaque valeur de $N$, une courbe de loi normale de même moyenne et de même écart type est tracée. Très rapidement, la forme générale tend vers celle d'une loi normale. Nous assistons à la convergence vers une loi normale comme indiquée par le théorème de la limite centrale.

Cette distribution est la distribution de Bates\footnote{$\sum^n_{k=0} -1^k C^n_k \left( \dfrac{x-a}{b-a} - k /n \right)^{n-1} \text{sgn}\left(\dfrac{x-a}{b-a} -k/n\right) $}, qui est obtenue en réalisant la moyenne d'une distribution uniforme.

\subsection{Intervalle de confiance}

Le théorème de la limite centrale stipule que l'écart type de la distribution statistique de la valeur moyenne tend vers $\tfrac{\sigma}{\sqrt{N}}$, où $\sigma$ est l'écart type de la distribution initiale\footnote{Ici, pour une distribution uniforme : $\sigma = \tfrac{\text{Valeur Maximum} - \text{Valeur minimum}}{\sqrt{12}}$, soit ici $\tfrac{10}{\sqrt{12}} \approx \num{3.46}$}. 

\begin{figure}[h!]
\begin{center}
\includegraphics[width = 1\textwidth]{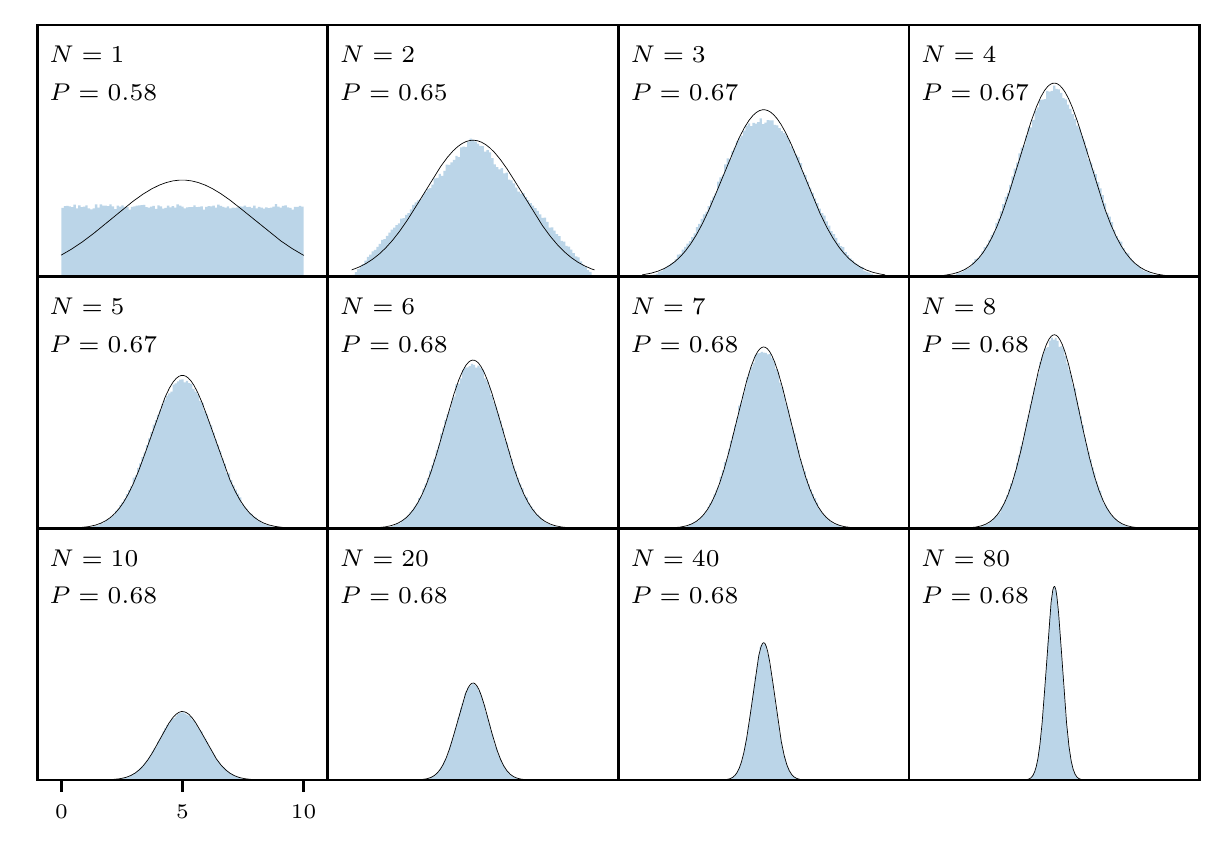}
\end{center}
\caption[Évolution de la distribution d'une moyenne]{Évolution de la distribution d'une moyenne de $N$ notes comprises entre 0 et 10. Le tirage aléatoire est réalisé \num{300000} fois, ce qui permet d'explorer la loi statistique issue de ce calcul de moyenne. $P$ représente le ratio du nombre de moyennes comprise dans l'intervalle  $\mu \pm \tfrac{\sigma}{\sqrt{N}}$. On constate rapidement que $P$ tend vers \num{0.68} qui est l'intervalle de confiance d'une loi normale.}
\label{TheoLimiteCentrale1}
\end{figure}

Pour vérifier ce résultat, il suffit de réaliser un comptage pour connaitre le ratio de moyennes qui sont situées à moins d'un écart type de la valeur centrale. Pour une loi normale, cela correspond à l'intervalle de confiance à $68\%$. Ces résultats sont donnés par la valeur $P$ précisée sur la figure \vref{TheoLimiteCentrale1}. Très Rapidement, ce ratio et donc l'intervalle de confiance s'approche de la valeur théorique d'une loi normale.

\subsection{Conditions d'application du théorème de la limite centrale}

Les conditions d'application de ce théorème sont faibles. Dans sa formulation classique, il est demandé aux données issues du mesurage d'être non corrélées et de suivre une même statistique possédant un écart type constant et fini. Ces conditions sont très largement répandues et c'est ce qui explique la prédominance des lois normales pour décrire la plupart des phénomènes physiques ou naturelles.

D'autres formulations\footnote{Condition de Liapounov ou condition de Lindeberg} ont des hypothèses encore plus faibles. Le théorème de la limite centrale continue d'être valable si les variables sont indépendantes et suivent une statistique d'écart type fini : il n'est pas obligatoire que les statistiques ou même que les écart type soient identiques.

Enfin, dans le cas de variables faiblement corrélés, il est démontré\footnote{Influence of global correlations on central limite théoréms ans entropic extensivity, Marsh, Fuentes, Moyano, Tsallis} que la valeur moyenne continue de tendre vers des lois statistiques centrées et dont l'écart type décroit avec le nombre de données. Dans ce cas, l'incertitude sera d'autant plus réduite que le nombre de mesures sera important, mais les intervalles de confiance sont beaucoup plus difficiles à évaluer.

\begin{figure}[h!]
\begin{center}
\includegraphics[width = 0.5\textwidth]{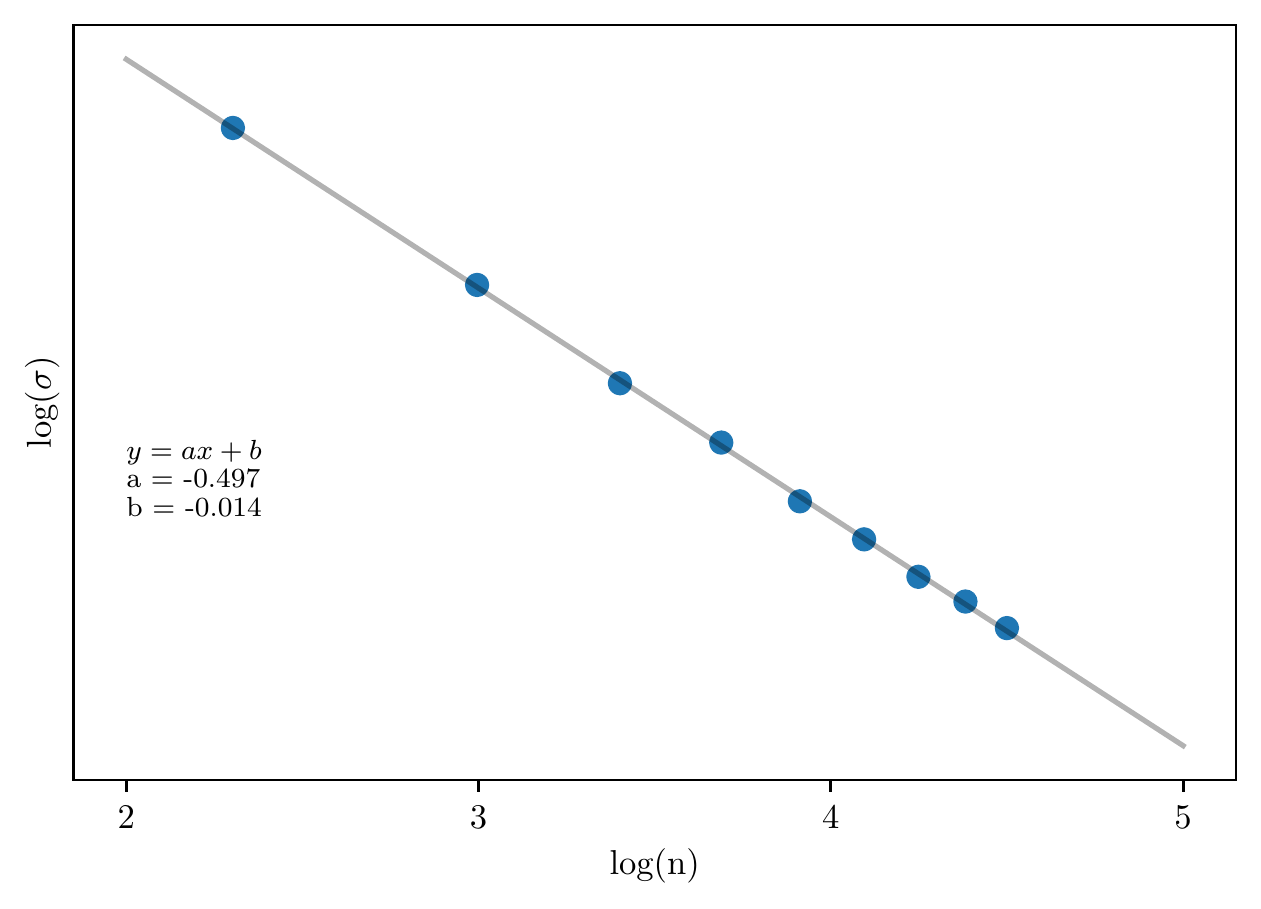}
\end{center}
\caption[Évolution de l'écart type d'une moyenne de $n$ tirages issus de processus aléatoire]{Évolution de l'écart type d'une moyenne de $n$ tirages issus de processus aléatoire suivant une loi normale d'écart type constant. Le graphique est en échelle log/log afin de mettre en évidence une droite de pente $-\tfrac{1}{2}$.}
\label{TheoLimiteCentrale3}
\end{figure}

Le graphique \vref{TheoLimiteCentrale3} représente l'écart type de la valeur moyenne d'un tirage aléatoire suivant une loi normale d'écart type $\sigma = 1$ en fonction du nombre de variables. Les variables sont indépendantes et l'écart type constant. L'écart type suit bien une évolution en $\dfrac{1}{\sqrt{n}}$ caractérisée par une droite de pente de $-\tfrac{1}{2}$.

\subsection{Cas de variables non indépendantes} 

Lorsque l'opération de mesurage à un effet sur les mesurages successifs, les données acquises ne sont plus indépendantes les unes des autres. Par exemple, lors d'une mesure électrique, l'auto-échauffement peut contribuer à modifier le résultat de mesures successives.

Dans ce cas, il est difficile de conclure quoi que ce soit. Dans la plupart des cas, la corrélation entre les mesures sera faible et une forme, même faible, du théorème de la limite centrale continuera de s'appliquer et il pourra être observé  une réduction des incertitudes. Ceci étant, il n'est pas possible d'affirmer que les intervalles de confiance sont ceux définis par une loi normale, ni même de s'assurer une convergence en $\tfrac{1}{\sqrt{n}}$.

\begin{figure}[h!]
\begin{center}
\includegraphics[width = 0.5\textwidth]{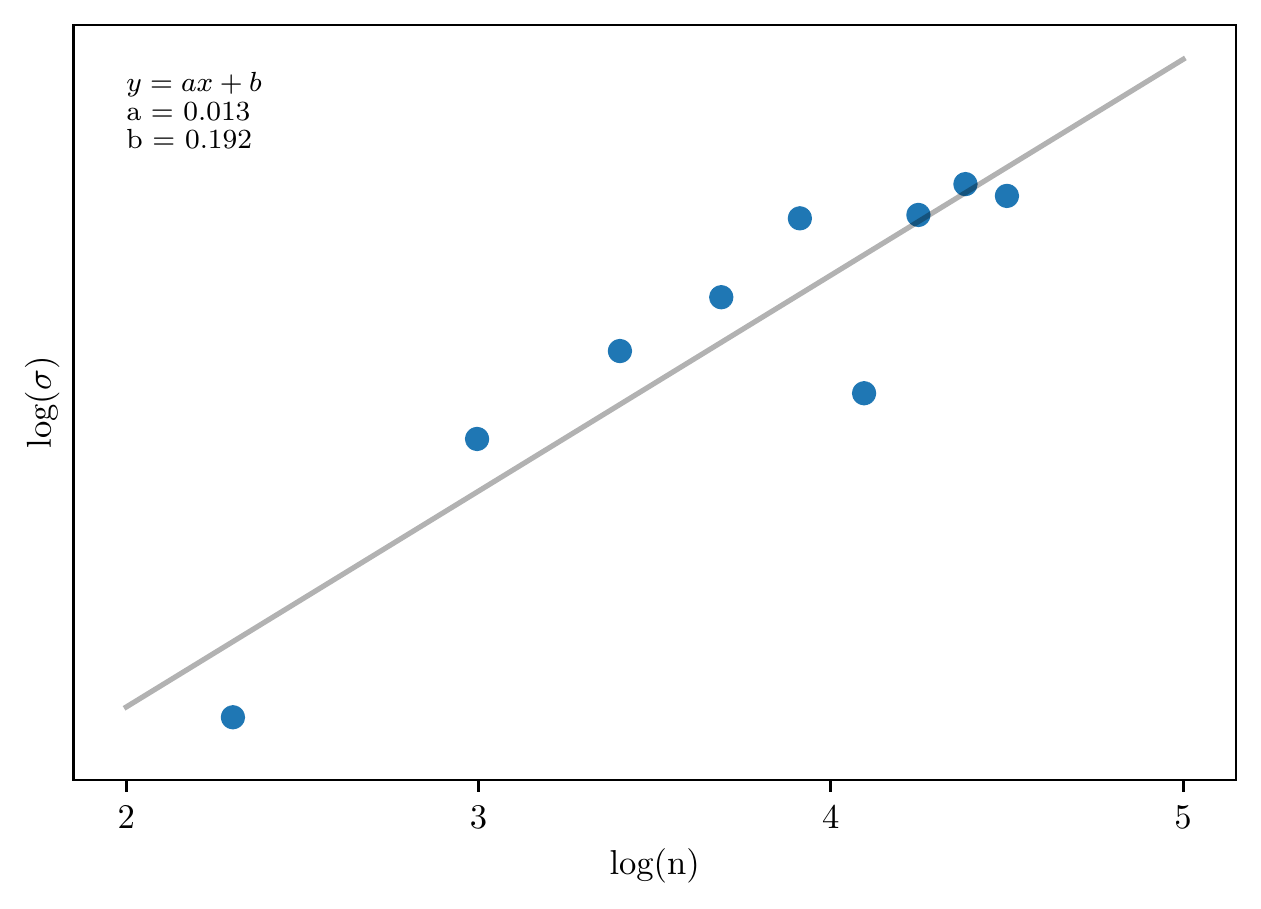}
\end{center}
\caption[Évolution de l'écart type d'une moyenne de $n$ tirages corrélés]{Évolution de l'écart type d'une moyenne de $n$ tirages corrélés issus de processus aléatoire suivant une loi normale d'écart type constant. Le graphique est en échelle logarithmique. L'écart type ne converge pas pour ce type de tirage aléatoire.}
\label{TheoLimiteCentraleCorrelated}
\end{figure}

Le graphique \vref{TheoLimiteCentraleCorrelated} représente un teste similaire à la figure \vref{TheoLimiteCentrale3}, à ceci prés que les variables sont corrélés entre elle par la relation suivante : 
\begin{equation}
x_{i+1} = \mathcal{N}\left(\overline{x}_{0\rightarrow i}, \sigma = 1\right)
\end{equation}

La variable aléatoire $i$ suit une loi normale d'écart type constant $\sigma = 1$ et d'espérance $\overline{x}_{0\rightarrow i}$ la valeur moyenne de l'ensemble des tirages aléatoires précédents.

\subsection{Loi aléatoire sans écart type}

Il n'est pas possible de défini un écart type pour certaines distribution. En particulier, les lois suivant une statistique Lorentzienne encore appelé lois de Cauchy ne possède pas d'écart type et sont adaptés pour modéliser les raies d'émission en spectroscopie : 
\begin{equation}
f(x, x_c, a) = \dfrac{1}{\pi}\cdot \dfrac{a^2}{\left(x-x_c\right)^2 + a^2}
\end{equation}

\begin{figure}[h!]
\begin{center}
\includegraphics[width = 0.5\textwidth]{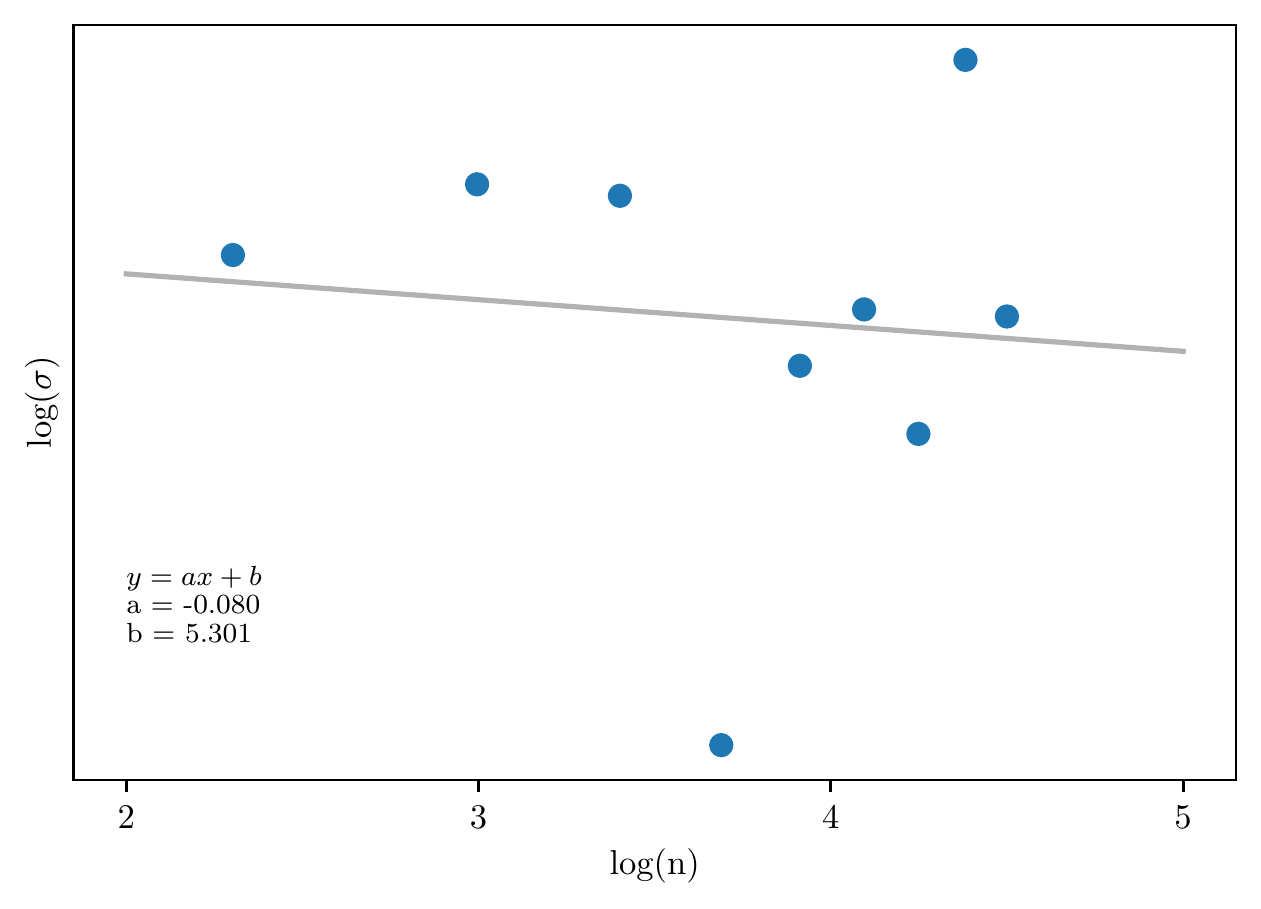}
\end{center}
\caption[Écart type de la moyenne de $n$ tirages non corrélés issus de processus aléatoire suivant une loi de Cauchy]{Écart type de la moyenne de $n$ tirages non corrélés issus de processus aléatoire suivant une loi de Cauchy de valeur centrale $x_c = 0$ et de facteur d'échelle $a=1$. Le graphique est en échelle logarithmique. L'écart type ne converge pas pour ce type de tirage aléatoire.}
\label{TheoLimiteCentraleCauchy}
\end{figure}

Le théorème de la limite centrale ne peut pas s'appliquer si ce type de bruit est présent lors du mesurage. Voir figure \vref{TheoLimiteCentraleCauchy}

\FloatBarrier

\subsection{Mise en évidence expérimentale du théorème de la limite centrale.}

D'un point de vu didactique, il est intéressant de réaliser une démonstration expérimentale de ce théorème. La démonstration formelle nécessite une maitrise des notions d'intégration, de convergence de série par majorant. Elle n'est pas à portée d'étudiants en début de cycle et ne contribue pas à comprendre son mécanisme de fonctionnement.

L'objectif n'est pas de commenter en détails les dispositifs pédagogiques, mais de simplement lister quelques idées fonctionnant en phase de travaux pratiques avec les étudiants. Il est à garder en mémoire que la mise en évidence de ce théorème nécessite forcement un grand nombre de répétitions. Il est donc important de trouver des systèmes simples et rapides à mettre en place.

Pour assurer une convergence relativement rapide, il est important de respecter les conditions suivantes : 
\begin{itemize}
\item les mesurages doivent être indépendants;
\item les écart types doivent être relativement homogène.
\end{itemize}

\subsubsection{Réalisation expérimentale}

\begin{itemize}
\item Mesurer la masse de \SI{100}{\milli\L} d'eau dans une éprouvette graduée de \SI{400}{\milli\L}. L'expérience est réalisée par deux étudiants, le premier remplissant l'éprouvette pendant que le second réalise la mesure de la masse. Un total de 30 mesures doit être réalisé;
\item Mesure de la masse d'un plateau percé de 9 trous de rayons différents remplis de petites billes d'aciers.
\item Planche de Galton\footnote{Permet rapidement de mettre en évidence le théorème de la limite centrale}
\item Mesures de la valeur d'une résistance\footnote{Automatisable avec mesures réalisées par un multimètre ineffaçable avec ordinateur}
\end{itemize}

\begin{figure}[h!]
\begin{center}
\includegraphics[width = 0.49\textwidth]{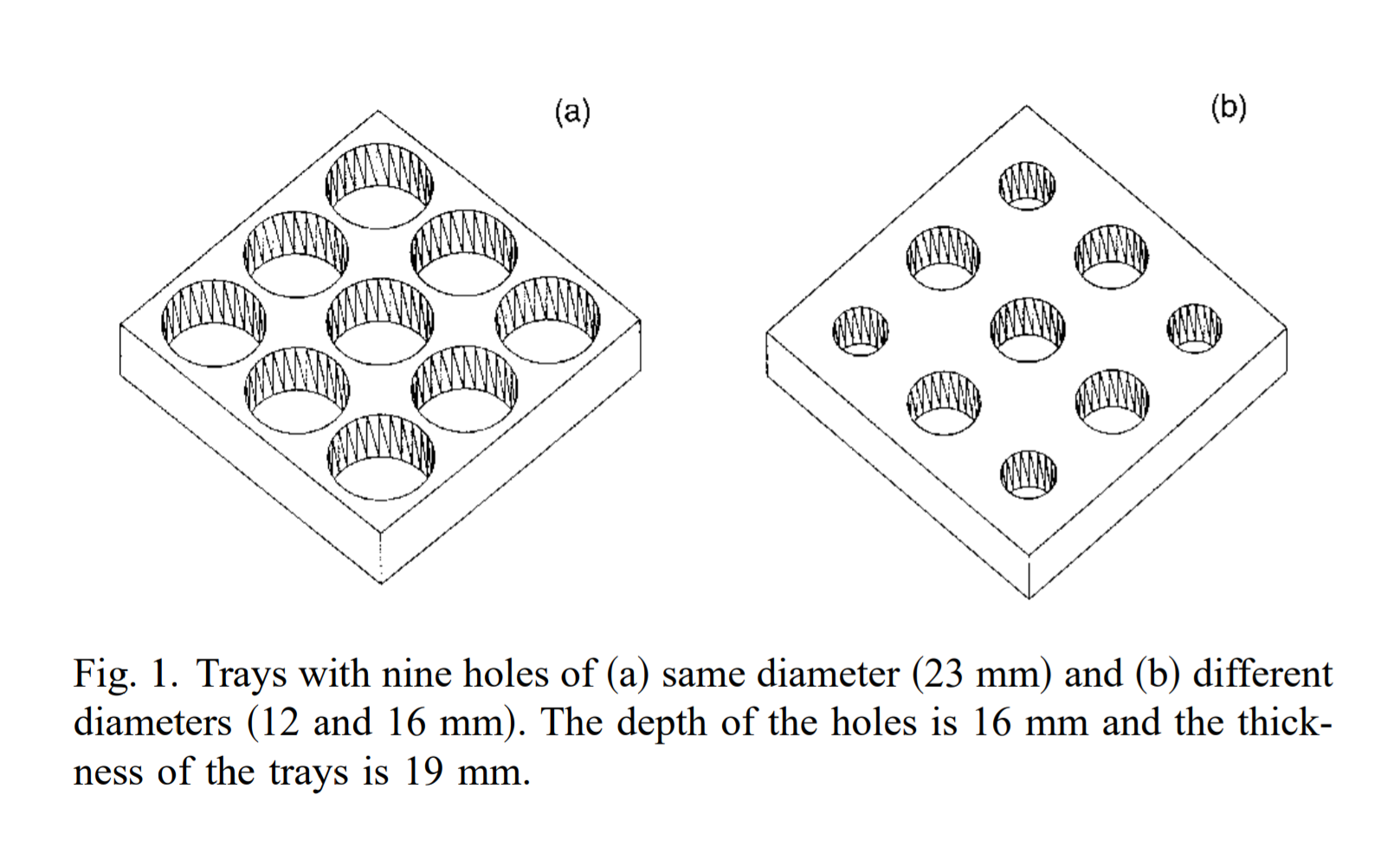}
\includegraphics[width = 0.49\textwidth]{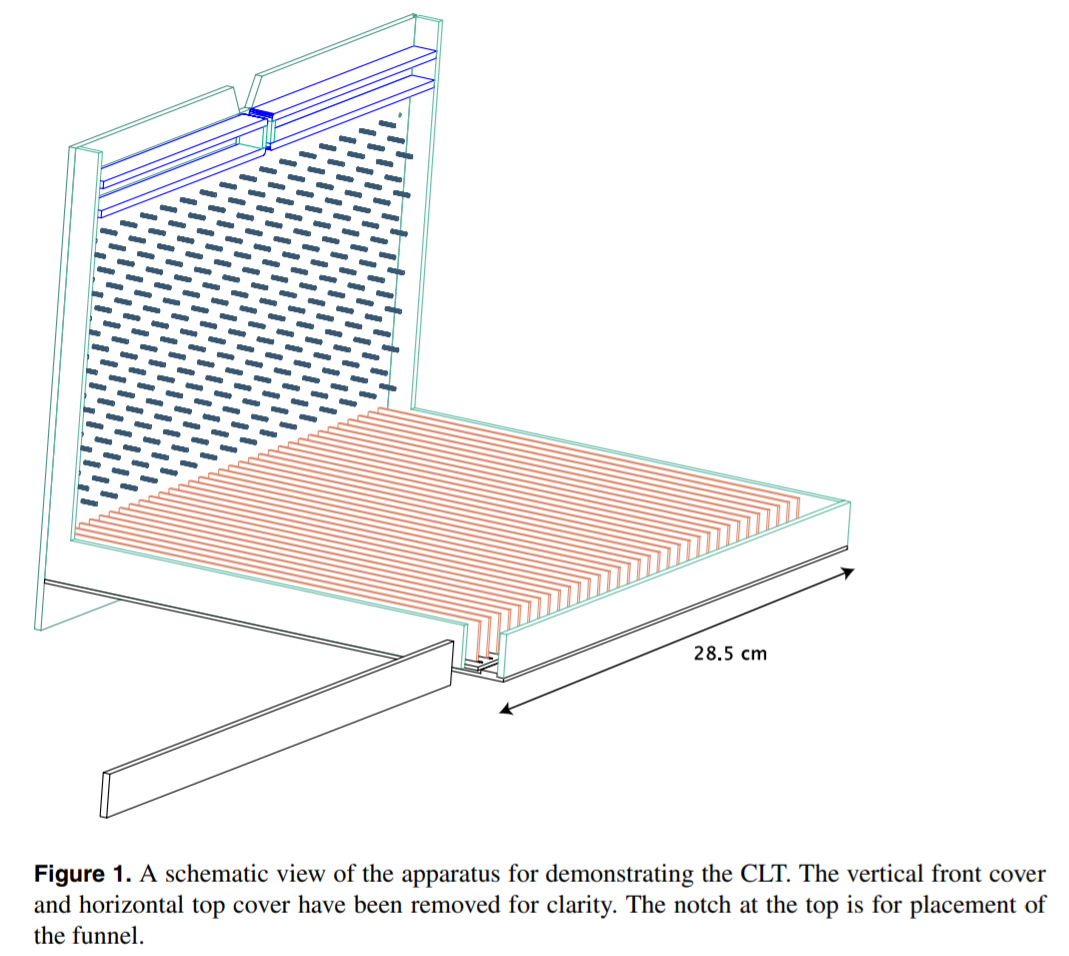}
\end{center}
\caption[Dispositifs expérimentales pour mise en évidence du théorème de la limite centrale]{Dispositifs expérimentales pour mise en évidence du théorème de la limite centrale. Images réalisées par K. K. Gan.}
\label{CLT_exp}
\end{figure}

Voir \textit{Simple demonstration of the central limit theorem using mass measurements and A simple demonstration of the central limit theorem by dropping balls onto a grid of pins, K. K. Gan}.

\subsubsection{Simulations numériques}

\textbf{Tirage discret :}

Avec un tableur, il est relativement simple de réaliser la moyenne d'un tirage discret aléatoire et d'en répéter l'expérience un grand nombre de fois pour voir apparaître une loi normale.

\textbf{Tirage continue : }

Il est possible avec n'importe quel langage de programmation de réaliser une mise en évidence de ce théorème. Les tirages pseudo-aléatoires sont suffisamment pour reproduire l'aspect aléatoire d'une mesure. Le principe de l'algorithme est le suivant : 
\begin{enumerate}
\item Définir un nombre n qui sera le nombre de mesure réalisés. n peut être compris entre 10 et 100, la forme gaussienne de la loi normale est atteinte pour $n=10$. 
\item Définir un nombre p qui représente le nombre de répétitions des n tirages aléatoires. p doit être grand, au minimum 1000.
\item dans une boucle allant de 1 à p : 
\begin{itemize}
\item Réaliser n tirages aléatoires d'une variable x : $x_i$, i allant de 1 à n. La distribution utilisée pour ce tirage peut être quelconque tant qu'elle possède une valeur moyenne et un écart type défini. Une simple loi uniforme comprise entre 0 et 10 suffit.
\item Calculer la valeur moyenne des n tirages précédents : $\overline{x}_j$, j allant de 1 à p et stocker cette valeur dans un tableau à p entrées.
\end{itemize}
\item Réaliser un histogramme des valeurs de $\overline{x}_p$
\end{enumerate}

La figure \vref{TheoLimiteCentrale1} a été réalisée pour différentes valeurs de $n$ à partir d'une loi uniforme comprise entre 0 et 10.

Le code ci-dessous est une version python de l'algorithme suggéré.

\begin{quote}
\begin{pyverbatim}
"""
Convergence vers la loi normale de la moyenne d'un tirage aléatoire
continue et uniforme
"""

#Modules
import numpy as np
from scipy import stats
import matplotlib.pyplot as plt

#Parametres 
np.random.seed(8)  #Pour la répétabilité du tirage aléatoire
P = 3000 # La moyenne est réalisée P fois
N =  5 # Nombres de mesure


#Partie principale du programme

#Réalisation de P moyennes de N tirages
x0 = (10*stats.uniform.rvs(0,1,N*P)) 
#Calcul de la moyenne suivant les N tirages
#X est un tableau contenant les P moyennes
X = np.mean(np.reshape(x0, [P,N]),1) 

#Ajustement par une courbe gaussienne
G_fit = np.linspace(0, 10, 1000)
pdf = stats.norm(np.mean(X), np.std(X)).pdf(G_fit)

#Tracé graphique
plt.xticks([0,2.5,5,7.5,10])
plt.hist(X, bins=101, histtype="stepfilled", alpha=0.3, density=True)
plt.xlim([-0.5,10.5])
plt.plot(G_fit, pdf, '-k', lw = 1)
plt.text(0,np.max(pdf)/1.1,'Moyenne : {0:.2f}\nEcart type : {1:.2f}'
.format(np.mean(X),np.std(X)))
\end{pyverbatim}
\end{quote}

\subsubsection{Comprendre le théorème de la limite centrale}
C'est un résultat surprenant de voir surgir une distribution de valeurs de plus en plus piquées sur une valeur centrale alors que le tirage réalisé est uniforme. Il ne faut pas oublier que cette distribution n'est plus la distribution initiale. La variable étudiée est la valeur moyenne $\overline{x}$ et non les ${x_i}$. 

Ce résultat apparait intuitivement en raisonnant en terme de combinaisons. Prenons un tirage aléatoire et uniforme $x_i$ d'une note entre 0 et 10 :
\begin{itemize}
\item La meilleur moyenne possible $\overline{x}$ est la valeur 10. Cette moyenne est réalisée si tous les $x_i$ obtenus valent 10. Il n'y a qu'une seule combinaison possible.
\item La plus mauvaise moyenne possibles $\overline{x}$ est la valeur 0. Cette moyenne est réalisée si tous les $x_i$ obtenus valent 0. Il n'y a qu'une seule combinaison possible.
\item La moyenne médiane de 5, quand à elle, est obtenue par un plus grand nombre de combinaison, rien qu'en prenant 2 notes : $\{0,10\}$, $\{1,9\}$, $\{2,8\}$, $\{3,7\}$\ldots Cette moyenne peut être obtenues par un grand nombre de tirages différents.
\end{itemize}

Les deux premiers cas constituent les extrémités de la distribution, ils sont obtenus pour une combinaison unique de ${x_i}$. Intuitivement, il est facile de comprendre que cette réalisation est rare, beaucoup plus rare qu'une moyenne de 5 qui est réalisée par un grand nombre de combinaisons.

Le théorème de la limite centrale est donc un résultat combinatoire. Il est hautement plus probable d'avoir une moyenne proche du centre de la distribution car un grand nombre de tirages permette de la réaliser plutôt qu'une moyenne éloignée du centre de la distribution.

\section{Loi de Student}

Le théorème de la limite centrale permet de calculer l'incertitude associée à la moyenne d'un mesurage lorsque l'incertitude ou l'écart type du mesurage est connue. Ce n'est pas toujours le cas, et la loi de Student permet de déterminer un intervalle de confiance lorsque l'écart type est inconnu. Ce dernier est estimé par : 

\begin{equation}
\sigma_{n-1}^2 = \dfrac{1}{n - 1} \sum_i^n (x_i - \overline{x})^2
\end{equation}

Les intervalles de confiance sont alors données par la relation : 
\begin{equation}
\left[\overline{X} - t \dfrac{\sigma_{n-1}}{\sqrt{n}} ; \overline{X} + t \dfrac{\sigma_{n-1}}{\sqrt{n}} \right]
\end{equation}

$n$ est le nombre d'éléments que contient l'échantillon, et $t$ est le coefficient de Student. Ce coefficient dépend à la fois de $n$ et du niveau de confiance souhaité.

\subsubsection{Exemple}
Pour $n=10$ avec un niveau de confiance de $95\%$ :

\begin{equation}
\left[\overline{X} - \num{2.26} \dfrac{\sigma_{n-1}}{\sqrt{n}} ; \overline{X} + \num{2.26} \dfrac{\sigma_{n-1}}{\sqrt{n}} \right]
\end{equation}

\begin{figure}
\begin{center}
\begin{tabular}{cccc}
 
$1-\alpha$ & 75\% & 95\% & 99\% \\ 
 
k &  &  &  \\ 
\hline 
\hline 
1 & \num{1.000} & \num{6.314} & \num{31.82} \\ 
 
2 & \num{0.816} & \num{2.920} & \num{6.965} \\ 

3 & \num{0.765} & \num{2.353} & \num{4.541} \\ 
 
4 & \num{0.741} & \num{2.132} & \num{3.747} \\ 

5 & \num{0.727} & \num{2.015} & \num{3.365} \\ 
 
10 & \num{0.700} & \num{1.812} & \num{2.764} \\ 

20 & \num{0.687} & \num{1.725} & \num{2.528} \\ 

50 & \num{0.679} & \num{1.676} & \num{2.403} \\ 

100 & \num{0.677} & \num{1.660} & \num{2.364} \\ 
\hline 
$\infty$ & \num{0.674} & \num{1.645} & \num{2.326} \\ 

\end{tabular} 
\end{center}
\caption{Coefficients de Student donnés pour différents intervalles de confiance $1 - \alpha$ et un nombre de degré de liberté $k$. Pour un calcul de valeur moyenne $k=N-1$, avec $N$ le nombre de points à disposition. La dernière ligne correspond à un nombre infini de points de mesure et donc au cas d'une loi normale continue.}
\label{CoeffStudent}
\end{figure}

Cette relation est proche de celle donnée par une loi normale. Les facteurs d'élargissement sont simplement remplacés par les coefficients $t$ donnés qui dépendent du nombre de mesures.

%\section{Deux types d'incertitude : A et B}
%
%[A faire; def : VIM]

%\section{Composition des incertitudes}
%[A faire]

\section{Incertitudes et maximisation de l'entropie}

Il existe un lien fort entre le principe de maximisation de l'entropie au sens de Shanon en théorie de l'information, les distributions rencontrées jusqu'à présent et les mesures qu'elles sont censées décrire. Ce lien a été réalisé par Jaynes en 1957, et il stipule qu'il n'y a pas de différences entre l'entropie de Shanon et l'entropie définie en mécanique statistique\footnote{Voir le principe d'entropie maximale défini par E.T. Jaynes dans \textit{Information Theory and Statistical Mechanics}, 1963}.

Un ensemble de données issues d'un mesurage définit un certain nombre de contraintes sur la distribution statistique sous-jacente. Par exemple, moyenne et écart type sont deux contraintes qui fixent certaines caractéristiques de la distribution. Il est possible d'en déduire tout un ensemble de distributions satisfaisant ces contraintes : loi uniforme, triangulaire, loi de Cauchy, loi normale etc.

Le principe d'entropie maximale stipule qu'il existe une distribution particulière maximisant l'entropie de Shanon, et donc minimisant les informations \textit{a priori} conduisant à cette distribution. Il s'agit donc de la distribution ajoutant le moins d'informations et d'hypothèses à la mesure tout en tenant compte des contraintes fixées par cette dernière.

\subsection{Loi normale}

Connaissant l'espérance (valeur moyenne) et l'écart type d'une distribution, la loi normale est la distribution non bornée d'entropie maximale. Ce qui explique la prédominance de cette loi quand il s'agit de décrire une mesure physique.

\subsubsection{Loi de Student}

La loi de Student est une distribution définie à partir de la loi normale. En effet, si les variables aléatoires suivent une loi normale, alors connaissant l'espérance (valeur moyenne) et le nombre de degrés de liberté (lié au nombre de mesures), la loi de Student est la distribution d'entropie maximale. Ce qui explique son intérêt dans le domaine de la mesure où ce sont généralement deux informations connues et facile à déduire d'un ensemble de données expérimentales.

\subsection{Distribution exponentielle}
Sur le domaine $\left[0, \infty\right]$, pour une valeur moyenne fixé, la loi de décroissance exponentielle $\lambda e^{-\lambda t}$ est celle maximisant l'entropie. Ce qui explique qu'elle est adaptée pour la description de phénomène aléatoire comme la décroissance radioactive ou pour décrire des défaillances de système électronique ou mécanique. 

En effet, ces derniers fixent les contraintes suivantes : 
\begin{itemize}
\item l'événement se produit entre l'instant initiale et l'infini
\item les temps mis pour que les événements se produisent ont une certaine valeur moyenne
\end{itemize}

De ce fait, la loi de décroissance radioactive peut être vue comme le principe de maximisation de l'entropie pour des événements respectant les contraintes précédentes.

\section{Quelques remarques}

Nous venons de voir que les outils statistiques reposent largement sur l'exploitation de la loi normale. Cette distribution a comme particularité d'être assez peu informative\footnote{C'est même la moins informative possible, étant donné certaines contraintes.}, elle nécessite peu de connaissance quand à la forme de la distribution statistique des incertitudes de mesure.

Il est à souligner qu'il s'agit d'un choix, raisonnable, souvent non explicite, mais il s'agit avant tout d'un choix relevant d'un \textit{a priori} sur la mesure : processus de mesurage donnant des valeurs non corrélées, existence d'un écart type et d'une moyenne. 

D'une manière générale, cette analyse est mise de côté au nom d'une forme d'impartialité scientifique : il faudrait être capable d'aborder l'analyse de données sans aucun avis ou opinion sur ces dernières. Dans les faits, ce n'est jamais le cas, et dans une démarche classique, ces \textit{a priori} existent et sont souvent occultés. Il suffit de se rappeler de toutes ces données éliminées car \textit{jugées} aberrantes pour constater que cette conception de la science impartiale et aveugle quant aux données n'est pas complètement vrai et le propos mérite d'être nuancé.

Ce point sera développé dans le chapitre \vref{InferenceBayesienne} sur l'inférence bayésienne qui présente l'avantage de rendre ces choix explicites.

%Modéle et regression
\chapter{Modèles et régressions}
\label{ModelesEtRegressions}

La régression est un ensemble de techniques consistant à déterminer les paramètres modélisant un phénomène à partir de données mesurées. La plus utilisée est la régression linéaire consistant à déterminer les coefficients d'un polynôme à partir de données présentant des incertitudes au moyen de relation d'algèbre linéaire.

Ce chapitre est consacré à l'ajustement d'un modèle polynôme de degrés 1 aux mesures dans le cadre de la régression linéaire. L'objectif est principalement de comprendre les hypothèses sous-jacentes à cette opération et donc les conditions d'applications et de validités.

Les notions développées dans ce chapitre sont issues des ressources suivantes : 
\begin{itemize}
\item BUP 796 : Régression linéaire et incertitudes expérimentales, D. Beaufils, 1997
\item arXiv 1008.4686, astrophysics : Data analysis recipes : fitting a model to data, D. W. Hoog, J. Bovy, D. Lang, 2010
\item Économétrie Giraud et Chaix
\end{itemize}

L'objectif de ce chapitre est de prendre conscience des conditions extrêmement étroites d'application des techniques de régressions linéaires. Il est difficile de se rendre compte que l'utilisation de ces méthodes est souvent fausse puisque les résultats de mesures sont estimés avec une incertitude. Ainsi, il est facile de se convaincre que le résultat est "pertinent" puisqu'il n'est qu'à "un ou deux écart type" de la valeur théorique, et une analyse plus poussée est de ce fait souvent écartée.

\newpage
\section{Régression linéaire : méthode des moindres carrés}
\subsection{Présentation}
L'objectif de cette opération est de déterminer les paramètres $a$ coefficient directeur et $b$ ordonnée à l'origine d'une droite qui soit la \textit{meilleur possible} pour un ensemble de points $(x_i, y_i)$ donnés présentant un bruit aléatoire.

\begin{equation}
\begin{cases}
f(x) &= a x + b \\
y_i &= a x_i + b + \epsilon_i
\end{cases}
\label{residu}
\end{equation}

\begin{figure}[h]
\begin{center}
\includegraphics[width = 0.65\textwidth]{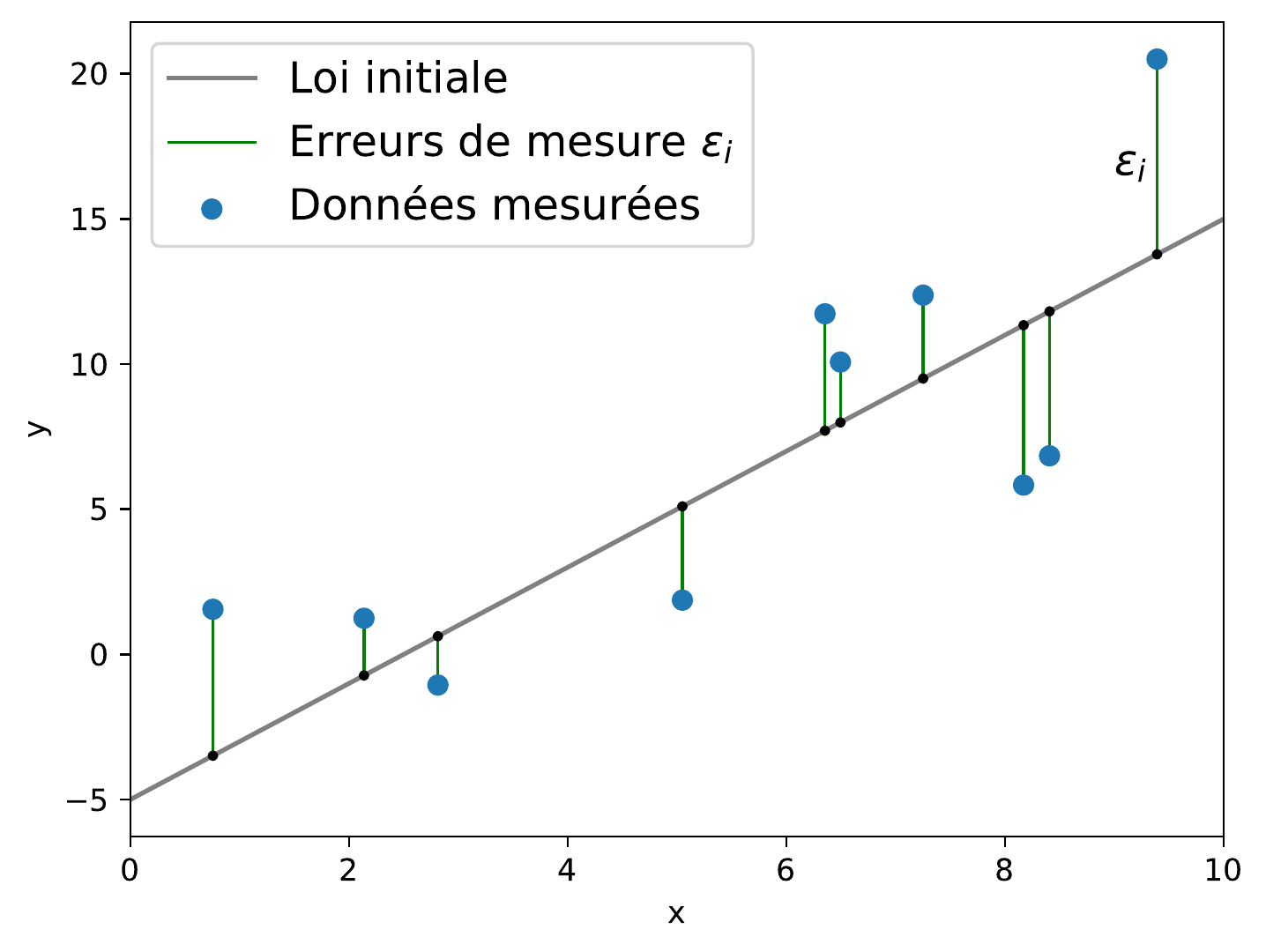}
\end{center}
\caption[Exemple d'ajustement linéaire]{Exemple d'ajustement linéaire d'un ensemble de points à une droite. Les traits verts représentent les résidus ou erreurs $\epsilon_i$ associés à chaque point.}
\label{ResidusMesureLineaire}
\end{figure}

Avec $\epsilon_i$, le résidu du point, une erreur inconnue associée à chaque point $i$. La figure \ref{ResidusMesureLineaire} représente graphiquement cette erreur pour chaque point de mesure par rapport à une loi linéaire initiale.

Cette procédure n'est pas arbitraire. \textit{"Meilleur possible"} consiste à minimiser les écarts quadratiques verticaux entre les données et la droite moyenne :
\begin{equation}
\chi^2 = \sum \limits_{i=1}^{N} \left[y_i - f(x_i)\right]^2
\end{equation}

Cette minimisation conduit à résoudre le système suivant : 
\begin{equation}
\begin{cases}
\dfrac{\partial \chi^2}{\partial a} & =0 \\
\dfrac{\partial \chi^2}{\partial b} & =0
\end{cases}
\end{equation}

Ce système possède les solutions suivantes : 
\begin{equation}
\begin{cases}
a & = \dfrac{\sum\left(y_i - \overline{y}\right)\left( x_i - \overline{x}\right) }{\sum \left(x_i -\overline{x}\right)^2} \\
b & = \overline{y} - a \overline{x}
\end{cases}
\end{equation}

Réaliser un ajustement linéaire consiste, pour le logiciel de traitement de données, à calculer les coefficients précédents.

\FloatBarrier
\subsection{Incertitude constante suivant $y$}

\begin{figure}[h]
\begin{center}
\includegraphics[width = 1\textwidth]{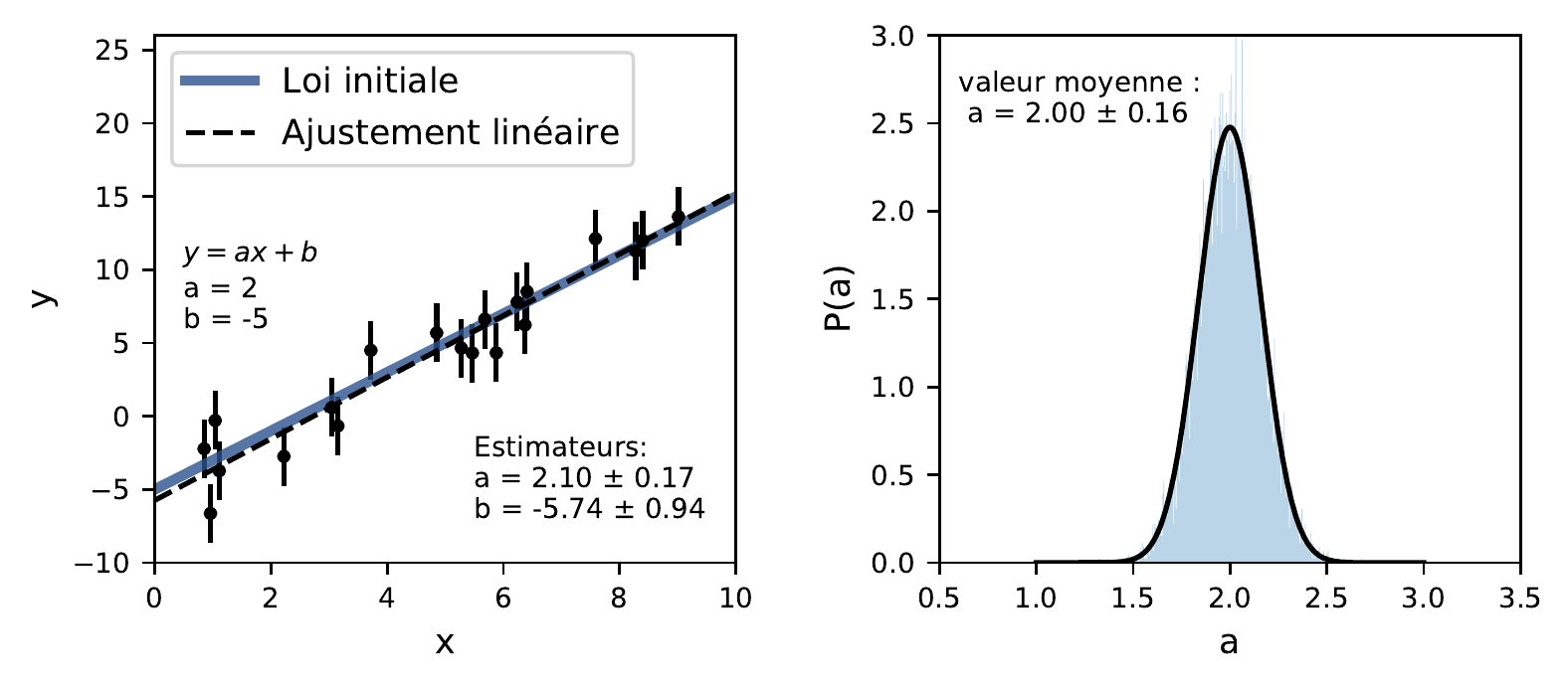}
\end{center}
\caption[Distribution statistique d'un coefficient déterminé par régression linéaire : incertitudes constantes]{Données générées aléatoirement avec un écart type constant. A gauche : exemple de régression linéaire et estimateurs associés. A droite : distribution complète du coefficient $a$.}
\label{LinearFit1}
\end{figure}

Dans l'exemple suivant est présenté en figure \ref{LinearFit1}. Nous générons un ensemble de données aléatoirement à partir d'une loi connue de type $y = ax + b$ dite \textit{loi initiale}. Une erreur suivant une loi normale d'écart type fixe est ensuite ajoutée à l'ensemble des données $y$ pour reproduire un comportement aléatoire. Cette situation vérifie donc toutes les conditions d'application de la méthode des moindres carrés

La procédure numérique permet de calculer l'incertitude\footnote{Numériquement, les relations indiqués précédemment sont généralisées à n'importe quelle loi polynomiale et réécrite dans un formalisme d'algèbre linéaire. Calculer les coefficients $a$ et $b$ consiste donc à des calculs matriciels, facilement réalisés par un système informatique. L'incertitude apparait naturellement comme les coefficients de la matrice de covariance du système.} sur les coefficients $a$ et $b$.

\subsection{Incertitudes suivant $y$}

Les relations présentées dans la partie précédente suppose une incertitude constante suivant l'axe $y$. Dans les faits, cette hypothèse est rarement réalisée. Il est possible de reprendre l'analyse précédente en minimisant la somme des écarts quadratiques normalisés par l'écart type de chaque point.

\begin{equation}
\chi^2 = \sum \limits_{i=1}^{N} \dfrac{\left[y_i - f(x_i)\right]^2}{\sigma_{y_i}^2}
\label{Chi2AvecSigma}
\end{equation}

De même que précédemment, cette minimisation conduit à la détermination de l'ordonnée à l'origine $b$ et la pente $a$ de la régression linéaire.

\subsubsection{Exemple}

\begin{figure}[h]
\begin{center}
\includegraphics[width = 1\textwidth]{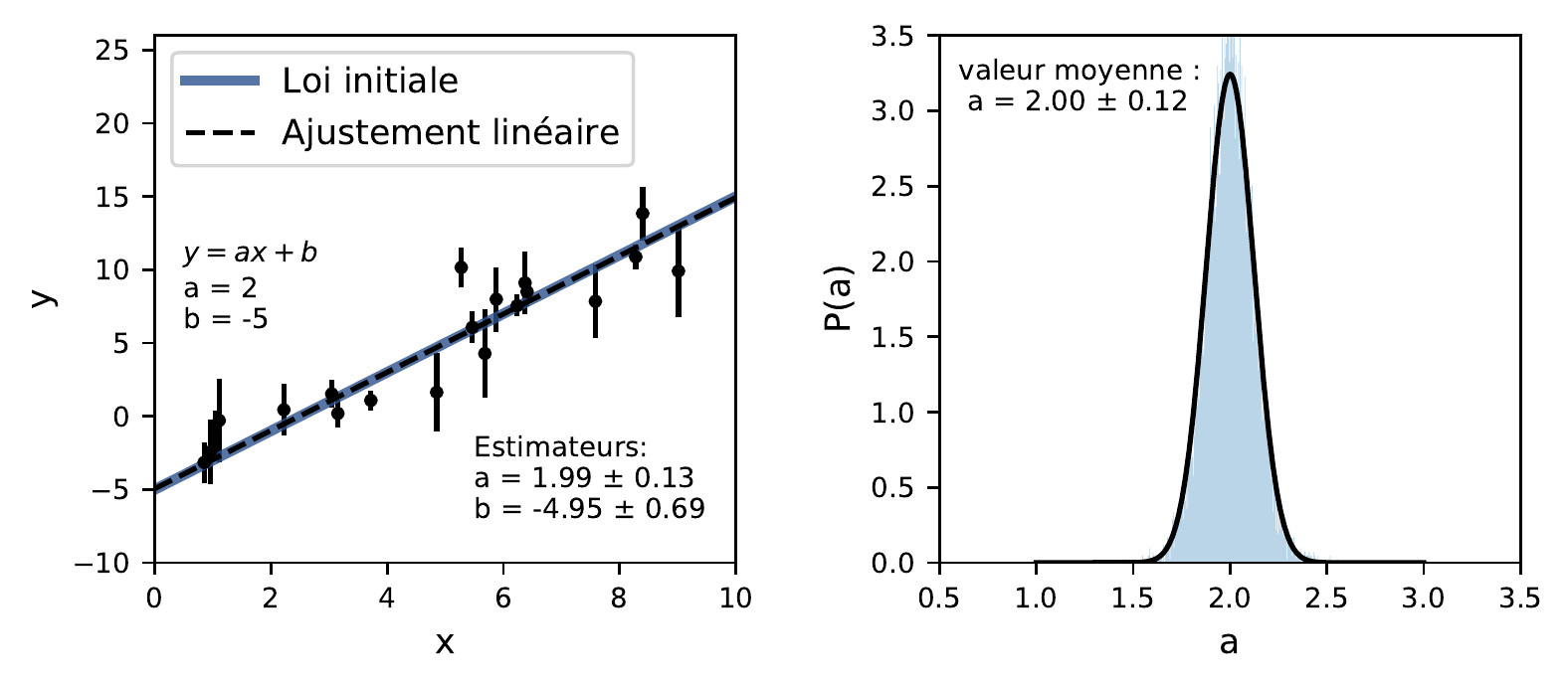}
\end{center}
\caption[Distribution statistique d'un coefficient déterminé par régression linéaire : incertitudes non constantes]{Données générées aléatoirement avec un écart type aléatoire suivant une loi normale. A gauche : exemple de régression linéaire et estimateurs associés. A droite : distribution complète du coefficient $a$.}
\label{LinearFit2}
\end{figure}

La figure \ref{LinearFit2} est un ajustement linéaire à partir de données construite de manière similaire au cas présenté en figure \ref{LinearFit1}, à la différence de l'incertitude sur les données $y$ qui ici n'est pas constante. La procédure des moindres carrés permet de prendre en compte des incertitudes différentes suivant l'axe $y$.

\section{Paramètres influençant les intervalles de confiance des coefficients $a$ et $b$}

Le modèle de la régression linéaire permet d'obtenir des incertitudes sur les coefficients $a$ et $b$. Ces incertitudes sont calculées analytiquement dans le cadre du modèle et doivent s'interpréter comme caractéristiques de la loi normale que suivent les coefficients $a$ et $b$.

Généralement, ces coefficients sont négligés au profit des seuls valeurs $a$ et $b$ alors que l'évaluation des incertitudes occupent une place importante dans le raisonnement scientifique.

\subsection{Estimateur de $a$}

L'écart type de l'estimateur de $a$ est : 

\begin{equation}
\sigma_a = \dfrac{\sigma}{\sqrt{\sum \left( x_i - \overline{x} \right)^2}}
\label{sigmaA}
\end{equation}

où $\sigma$ est l'écart type associé au mesure $y_i$.

\begin{itemize}
\item L'écart type $\sigma$ des mesures influence directement les écart type $\sigma_a$. Réduire $\sigma$ permet de réduire les incertitudes sur la pente.
\item $\sum \left( x_i - \overline{x} \right)^2$ est un terme chiffrant l'écart total entre les points $x_i$ et leur valeur moyenne. L'augmentation de ce dernier conduit à une diminution des incertitudes sur $a$. Ce denier peut être augmenter de deux manières : 
\begin{itemize}
\item en augmentant le nombre de points $N$;
\item en augmentant l'intervalle de prise de mesure $x_i$. Plus les points de mesure seront dispersés et éloignés les uns des autres, plus l'écart type $\sigma_a$ sera faible.
\end{itemize}
\end{itemize}

\subsection{Estimateur de $b$}
L'écart type de l'estimateur de $b$ est : 

\begin{equation}
\sigma_b = \sigma \sqrt{\dfrac{1}{N} + \dfrac{\overline{x}^2}{\sum \left( x_i - \overline{x} \right)^2}} 
\label{sigmaB}
\end{equation}

où $\sigma$ est l'écart type associé au mesure $y_i$ et $N$ le nombre de points.

\begin{itemize}
\item L'écart type $\sigma$ des mesures influence directement les écart type $\sigma_b$. Réduire $\sigma$ permet de réduire les incertitudes sur l'ordonnée à l'origine;
\item $\dfrac{\overline{x}^2}{\sum \left( x_i - \overline{x} \right)^2}$ est un terme chiffrant la dispersion des valeurs de $x_i$ et l'éloignement à l'ordonnée à l'origine. Pour réduire $\sigma_b$, il conviendra de :
\begin{itemize}
\item prendre des points $x_i$ proche de l'ordonnée à l'origine pour réduire le terme $x_i$;
\item augmenter l'intervalle de mesure des $x_i$ ainsi que le nombre de points $n$ qui agit sur le terme en $\dfrac{1}{n}$
\end{itemize}
\end{itemize}

%\FloatBarrier

\section{Valeurs aberrantes}
\subsection{Sensibilité aux données}

La faiblesse de la méthode des moindres carrés réside dans le fait qu'elle donne trop de poids à des données aberrantes par rapport à l'ensemble des données. Cela signifie qu'un faible nombre de valeurs aberrantes peut conduire à une variation significative du résultat final.

\begin{figure}[h]
\begin{center}
\includegraphics[width = 1\textwidth]{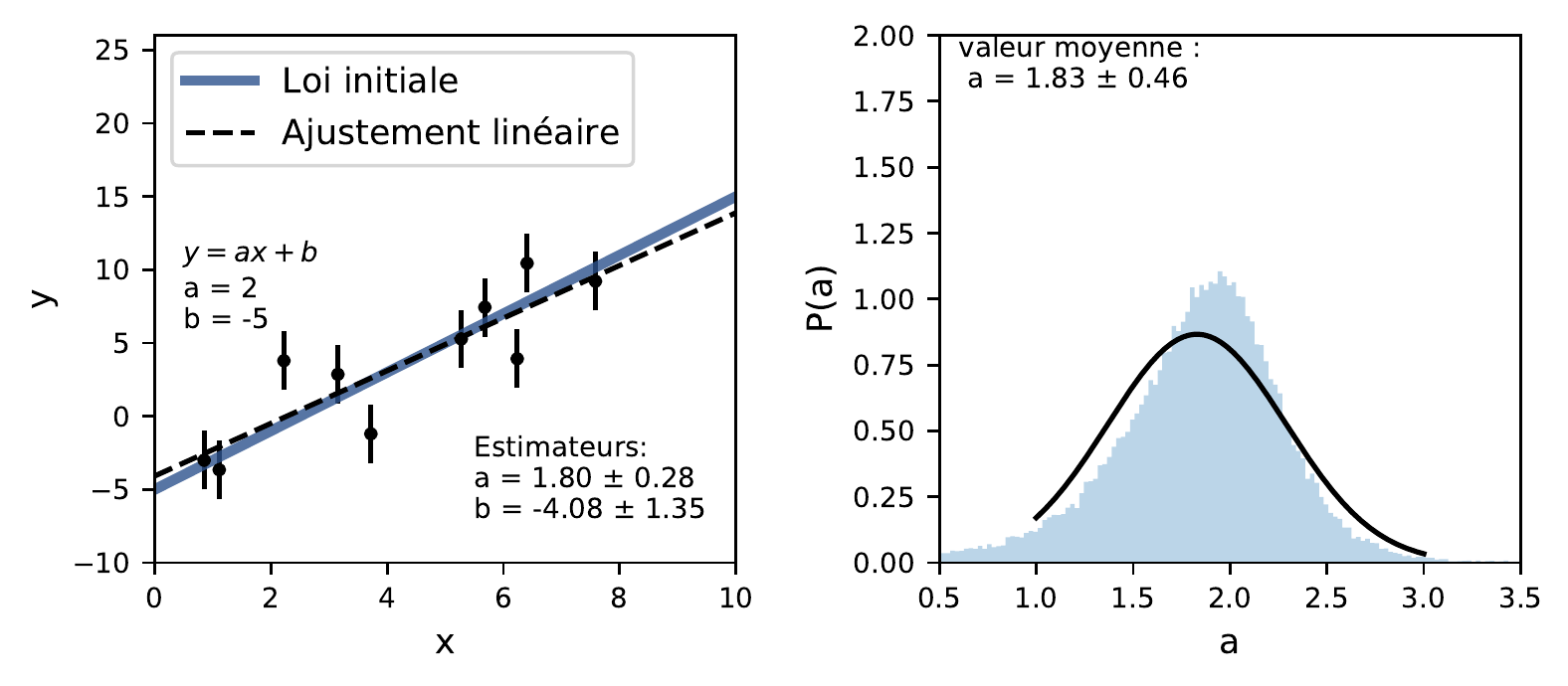}
\end{center}
\caption[Distribution statistique d'un coefficient déterminé par régression linéaire : influence d'une valeur aberrante]{Ajustement linéaire d'une relation affine. Les données ont été générées avec avec un bruit aléatoire constant. Un point au hasard n'est pas corrélé aux autres et joue le rôle d'une valeur aberrante. La distribution des valeur de $a$ ne suit plus une loi normale et n'est plus symétrique. La valeur moyenne de $a$ et l'écart type sont très éloignés des estimateurs. Il est à remarquer que la valeur la plus probable reste néanmoins égale à la valeur théorique du coefficient $a$.}
\label{LinearFitOutliers}
\end{figure}

La figure \ref{LinearFitOutliers} représente une telle situation. L'ensemble des données est généré similairement à celles présentées en figure \ref{LinearFit1}. Un point est choisi au hasard pour jouer le rôle d'une valeur aberrante. Ce dernier se voit attribuer une valeur aléatoire centrée sur la moyenne des valeurs $y$ et d'écart type constant mais élevé.

L'ajustement linéaire réalisé sur un jeu de donnée tiré au hasard est moins bon, mais est toujours compatible avec avec la loi initiale. Cependant, la statistique complète ne suit plus une loi normale et les intervalles de confiance ne sont plus valide.

\subsubsection{Élimination des valeurs aberrantes}

Une stratégie classique consiste à éliminer les données aberrantes. Ce processus d'élimination est fréquent, mais introduit des difficultés quand à la validité de la mesure en elle-même puisque ce processus est souvent à l'appréciation de la personne responsable de la mesure ou du traitement de données. Il existe des procédures d'élimination des données aberrantes, par exemple : 
\begin{itemize}
\item critère de Chauvenet;
\item critère de Peirce;
\item test de Grubb's.
\end{itemize}

Ces méthodes fournissent un procédé objectif et quantitatif pour l'élimination des données aberrantes, néanmoins cela ne les rends pas forcément scientifiquement et méthodologiquement correcte, en particulier dans les échantillons à faible population.

\section{Limites de la méthode}

Lors d'études de données, certains usages courants ne satisfont pas les conditions d'application strict de la régression linéaire :
\begin{itemize}
\item données aberrantes; 
\item loi non conforme au modèle : présence d'une non linéarité dans les données et modélisation par une loi affine;
\item incertitude suivant X et Y;
\item corrélation entre incertitudes et données
\end{itemize}

Les sous-parties suivantes présentent des exemples associés à la statistique complète du coefficient $a$ comme illustration de la déviation au modèle gaussien centré sur la valeur théorique de $a$. Ces déviations, bien que généralement faible, font qu'il est plus difficile de conclure sur la valeur finale et sur les incertitudes associées aux coefficients.

\subsection{Loi initiale non linéaire}

Le choix du modèle influe grandement sur les résultats de ce type de méthode. En figure \ref{FigureLimiteMethodeRegression}(a), la loi initiale n'est pas parfaitement linéaire, un terme quadratique de faible amplitude est ajouté à la loi initiale. Ceci a pour conséquence d'éloigner les valeurs des coefficients calculés par la méthode des moindres carrés. En figure \ref{FigureLimiteMethodeRegression}(a), la recherche d'une valeur $a=2$ conduit à une statistique d'allure gausienne centrée sur $a = \num{1.80(16)}$.

%\begin{figure}[h]
%\begin{center}
%\includegraphics[width = 1\textwidth]{LinearFitNonlinear.pdf}
%\end{center}
%\caption{Linear fit with non linear law}
%\label{LinearFitNonlinear}
%\end{figure}

\subsection{Présence d'incertitudes suivant $x$ et $y$}

Le modèle de la régression linéaire demande une incertitude suivant $x$ négligeable devant l'incertitude suivant $y$. La figure \ref{FigureLimiteMethodeRegression}(b) présente un exemple de régression réalisée avec une incertitude de même ordre de grandeur suivant les deux axes. De même que précédemment, la recherche d'une valeur $a=2$ conduit à une statistique d'allure gausienne centrée sur $a=\num{1.80(21)}$.

%\begin{figure}[h]
%\begin{center}
%\includegraphics[width = 1\textwidth]{LinearFitUncertaintyXY.pdf}
%\end{center}
%\caption{Linear fit with x and y uncertainty}
%\label{LinearFitUncertaintyXY}
%\end{figure}

\subsection{Incertitudes corrélées et changement de variable}

Les incertitudes ne doivent pas être corrélées aux données $x$ et $y$ pour pouvoir utiliser le modèle de la régression linéaire. Cette situation est fréquemment rencontrée expérimentalement : n'importe quelle mesure au multimètre possède une incertitude dépendant de la valeur lue.

Les changements de variables introduisent aussi une corrélation de ce type. Pour simplifier les analyses, il est courant de pratiquer un changement de variable pour se ramener à une relation linéaire. Cette pratique est aussi liée au fait que souvent, les logiciels utilisés pour réaliser les analyses de données ont peu d'options en terme de régression linéaire et souvent, seul un modèle affine est proposé.

Prenons le cas de la période $T$ d'un pendule de longueur $l$, à partir de laquelle il est extrait l'accélération de la pesanteur $g$.
\begin{equation}
T = 2\pi\sqrt{\dfrac{l}{g}}
\end{equation}

Le changement de variable classiquement introduit est $\theta = T^2 =  \dfrac{4\pi^2}{g} l$. Ainsi l'étude de $\theta = f(l)$ conduit à une relation affine dont le coefficient directeur permet une évaluation de $g$.

La difficulté est que l'incertitude associée à la variable $\theta$ est corrélée à $T$, en effet : 
\begin{equation}
\sigma(\theta) = 2T\sigma(T)
\end{equation}

Ainsi, même si l'incertitude suivant $T$ est constante, ce n'est pas le cas de $\theta$. La plupart des changement de variable conduisent à cette situation.

\begin{figure}[h]
\thisfloatpagestyle{empty}
\vspace*{-1cm}
\begin{center}
\subfloat[Loi initiale non linéaire.]{\includegraphics[width = 1\textwidth]{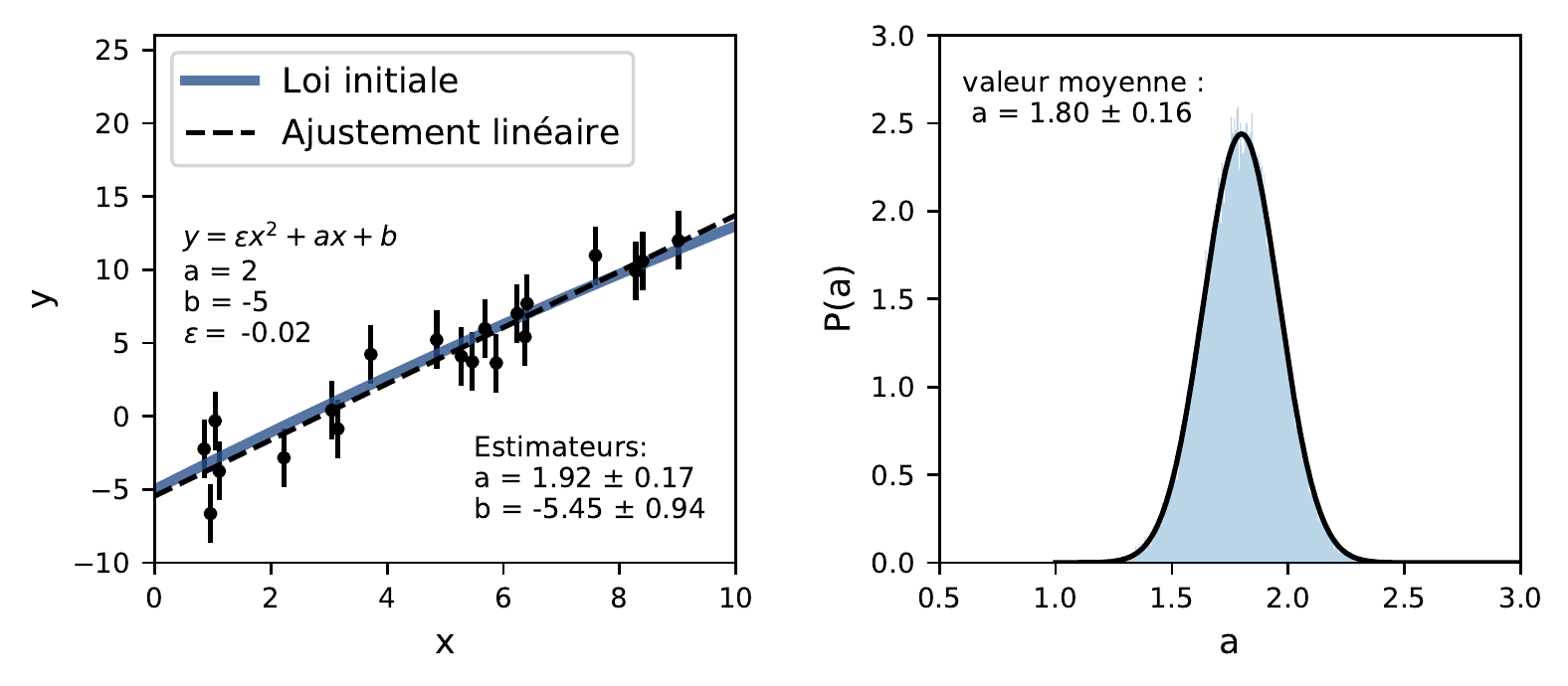}}\\
\subfloat[Incertitudes suivant $x$ et $y$.]{\includegraphics[width = 1\textwidth]{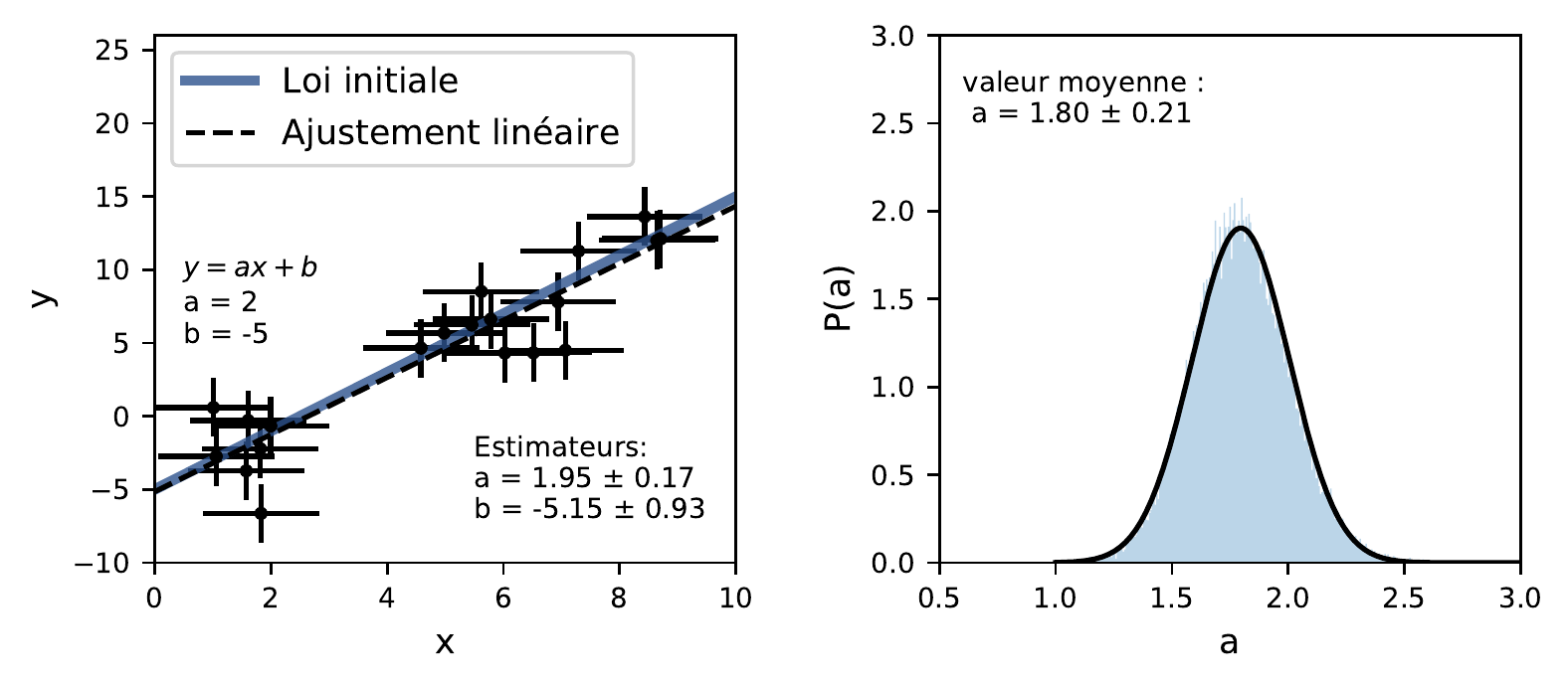}}\\
\subfloat[Incertitudes corrélée aux données : cas d'un changement de variables.]{\includegraphics[width = 1\textwidth]{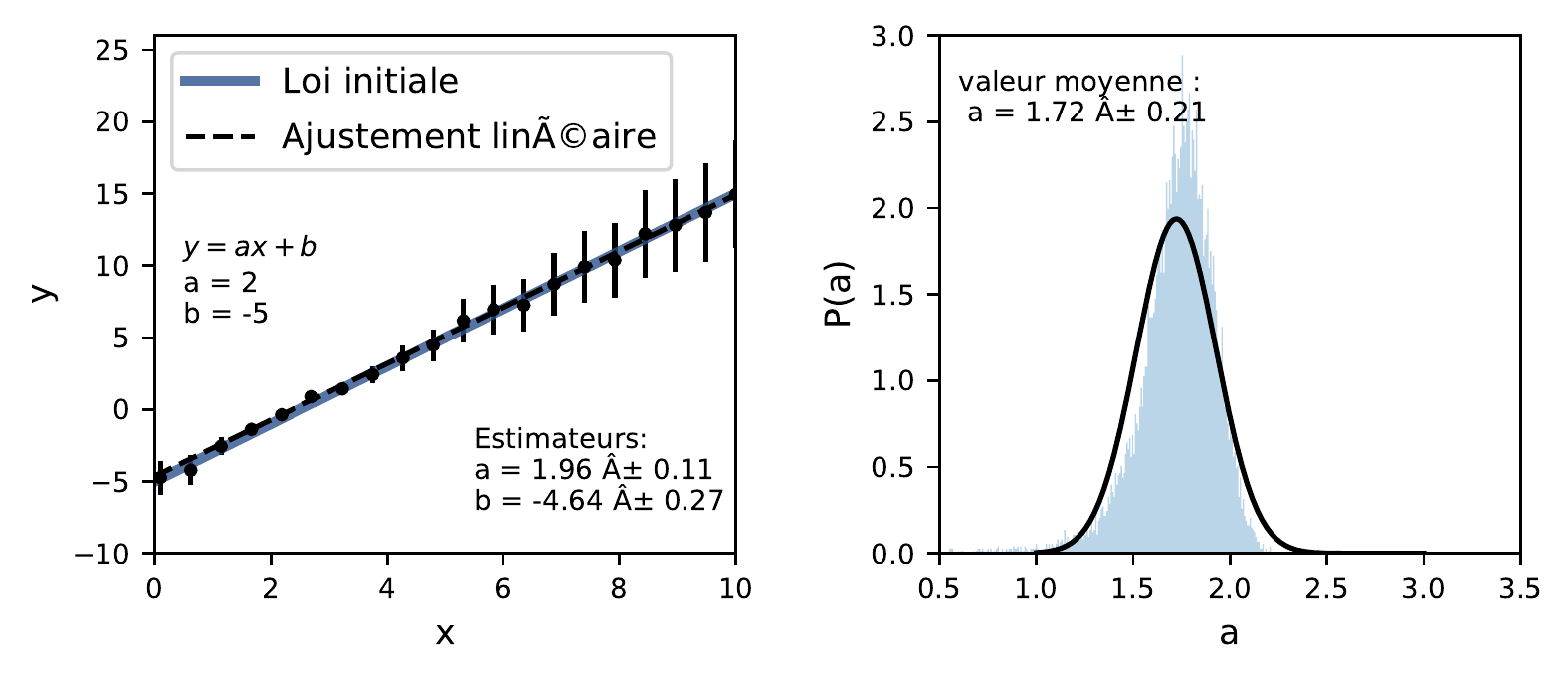}}
\end{center}
\caption[Distribution statistique d'un coefficient déterminé par régression linéaire : influence du non respect des conditions d'application]{Illustration de régression hors conditions strictes d'application. Les données sont générées aléatoirement avec un écart type non aléatoire. A gauche : exemple de régression linéaire et estimateurs associés. A droite : distribution complète du coefficient $a$.}
\label{FigureLimiteMethodeRegression}
\end{figure}

 La figure \ref{FigureLimiteMethodeRegression}(c) présente un exemple de régression réalisée avec une incertitude corrélée à la valeur $y$. La recherche d'une valeur $a=2$ conduit à une statistique centrée sur $a=\num{1.72(21)}$ dont l'allure n'est plus gausienne.

%LinearFitCorrelatedUncertainty
%
%\begin{figure}[h]
%\begin{center}
%\includegraphics[width = 1\textwidth]{LinearFitCorrelatedUncertainty.pdf}
%\end{center}
%\caption{Linear Fit Correlated Uncertainty}
%\label{LinearFitCorrelatedUncertainty}
%\end{figure}

%
%\section{Changements de variable}
%Une méthode classique pour démontrer une relation non linéaire entre deux grandeurs est un changement de variable. Prenons le cas de la période d'un pendule : 
%\begin{equation}
%T = 2\pi\sqrt{\dfrac{l}{g}}
%\end{equation}
%
%Le changement de variable consiste donc à étudier une loi de type :
%\begin{equation}
%Y = y^2 = \text{constante}\times x
%\end{equation}
%
%Avec l'incertitude suivant $y$ constante et indépendante des $y_i$.
%
%[Détailler le calcul : expression de l'incertitude; changer valeurs du calcul pour faire apparaitre g; le coefficient b est maintenu pour simuler un éventuel biais]
%
%\begin{figure}[h]
%\begin{center}
%\includegraphics[width = 1\textwidth]{ChangementDeVariable.pdf}
%\end{center}
%\caption{Ajustement linéaire avec changement de variable du type $Y = y^2$. L'incertitude n'est plus constante suivant $y_i$ mais est corrélée à la valeur $y_i$. Les conditions d'application de la régression linéaire ne sont plus respectées. Cela conduit à une évaluation de la valeur de $a$ erronée.}
%\label{ChangementDeVariable}
%\end{figure}
%

\clearpage

\section{Conditions d'applications de la méthode des moindres carrés}

Dans les parties précédentes, il a été mis en évidence le fait que les conditions d'application de la méthode des moindres carrés sont strictes et un non respect de ces dernières conduit rapidement à une estimation erronée des paramètres du modèle ainsi que leurs incertitudes.

La méthode des moindres carrés repose sur les hypothèses suivantes : 
\begin{itemize}
\item La loi physique sous-jacente aux données est connue;
\item L'incertitude suit une loi Normale uniquement suivant $y_i$ et non corrélée aux $y_i$;
\item Absence de points aberrants
\end{itemize}

Dans les faits, les conditions d'applications ne permettent souvent pas l'utilisation de cette méthode, mais dans un contexte scolaire, elles sont largement utilisées, faute de mieux. La grande difficulté étant qu'en dehors des conditions d'application strictes de cette méthode, il n'existe pas de consensus net permettant de réaliser l'ajustement des données par une loi connue de manière simple.

\section{Méthode de régression lorsque l'incertitude sur les données est inconnue}

\subsection{Estimateur de l'incertitude $\sigma_\epsilon$}

Un estimateur sans biais de l'écart type de l'incertitude est : 
\begin{equation}
\sigma_\epsilon = \dfrac{\sum \epsilon_i^2}{N-2}
\end{equation}

Avec $\epsilon_i$ les résidus définis dans la relation \ref{residu}.

\subsection{Intervalle de confiance pour $a$ et $b$}

Lorsque l'incertitude est estimée à partir des résidus, les facteurs d'élargissement à prendre en compte pour les intervalles de confiance ne sont plus ceux de la loi normale, mais ceux d'une loi de Student à $N-2$ degrés de liberté, les écarts types $\sigma_a$ et $\sigma_b$ sont identiques à ceux définis par les relations \ref{sigmaA} et \ref{sigmaB}.

\section{Conclusions}

Le traitement et la manipulation de données occupent une place fondamentale dans une démarche scientifique. Plus largement, l'utilisation d'outils statistiques est de plus en plus répandue car fortement démocratisés avec l'utilisation de tableurs dans des champs professionnels divers et variés. Ces outils grand public reposent tous sur le modèle de la régression linéaire dont les conditions d'utilisation sont assez strictes. De ce fait, il est important d'avoir des notions relatives à l'utilisation de cet outils et avoir conscience de leurs limites.

Concernant une utilisation pédagogique, il est important de ne pas négliger l'évaluation des incertitudes associées aux coefficients issues de la régression linéaire. Ces derniers sont simples à obtenir et si les conditions d'applications sont vérifiés, ils informent correctement sur l'incertitude des résultats.

%%Inférence Bayesienne
\chapter{Inférence bayésienne}
\label{InferenceBayesienne}

L'inférence est un procédé permettant d'\textit{induire} les caractéristiques générales d'une statistique à partir d'un échantillon. Ce procédé permet de calculer des estimateurs tout en évaluant le niveau de confiance de ces derniers.

L'ensemble des techniques présentés dans les chapitres \ref{MesureEtIncertitudes} et \ref{ModelesEtRegressions} sont des techniques d'inférences statistiques.

L'objectif de ce chapitre est d'introduire le concept d'inférence bayésienne. Cette technique d'inférence nécessite de renverser certains \textit{a priori} relatifs à la mesure et au traitement de données. Ce changement de paradigme sera expliqué dans la première partie de ce chapitre. 

La suite de ce chapitre sera consacrée à des études de cas permettant de mettre en lumière les techniques d'inférence bayésienne.

Références : 
\begin{itemize}
\item Frequentism and Bayesianism: A Python-drivenPrimer J.VanderPla
\item Data Analysis Recipes: Using Markov Chain Monte Carlo D.W. Hogg, D. Foreman-Mackey
\item emcee Documentation D. Foreman-Mackey
\item It is Time to Stop Teaching Frequentism to Non-statisticians W. M. Briggs
\item Bayesian Reasoning in Data Analysis G. D'Agostini
\item A Gentle Introduction to Bayesian Analysis: Applications to Developmental Research R. van de Shoot, D. Kaplan, J. Denissen, J.B. Asendorpf, F.J. Neyer, M.A.G. van Aken
\item Infolrxiation Theory and Statistical Mechanics E. T. Jaynes
\item Note de cours - Statistique Bayésiennes, J. Rousseau, ParisTech
\item Bayes Theorem, G. Sanderson
\item Bayésianisme versus fréquentisme en inférence statistique, J. Sprenger
\item BAYESIAN INDUCTIVE INFERENCE AND MAXIMUM ENTROPY S.F. Gull
\item Bayesian reasoning in data analysis, a critical introduction G.D. Agostini
\end{itemize}

\section{Le théorème de Bayes}

\subsection{Introduction}

\begin{center}
\includegraphics[width = 0.4\textwidth]{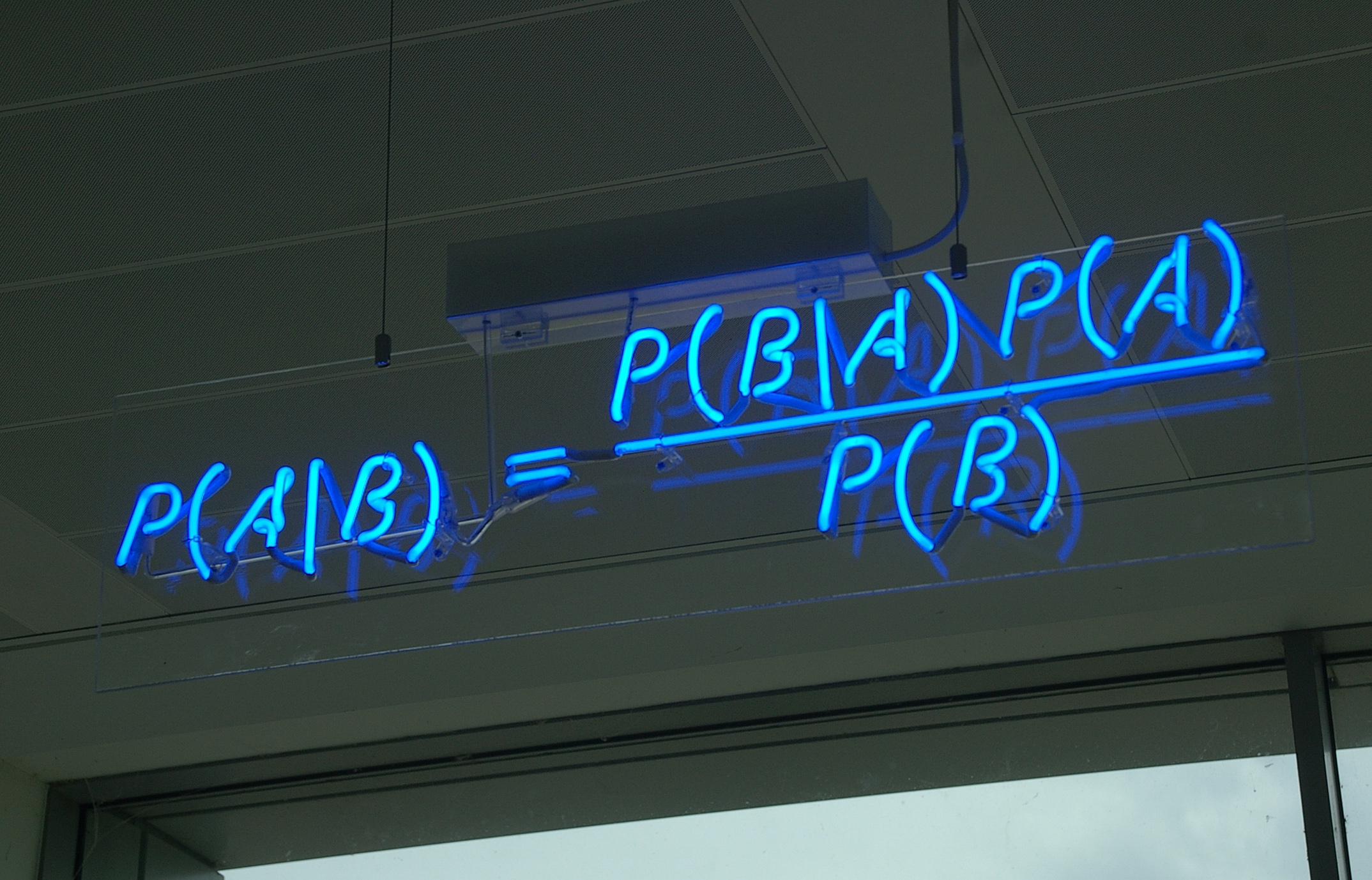}
\end{center}

Le théorème de Bayes permet l'estimation de probabilités. Il est utilisé dans des domaines variés comme l'intelligence artificielle en lien avec des algorithme de prise de décisions, dans les secteurs financiers pour les calculs de risques ou encore en sciences pour induire des informations concernant des hypothèses moyennant la connaissance de certaines données issues de mesures.

C'est un résultat de base en statistique permettant de manipuler des probabilités conditionnels.

\subsection{Quand utiliser ce théorème}

\subsubsection{Un premier cas}
Voyons en détail un exemple d'application de ce théorème.

\begin{quote}
Sur le campus d'une université, vous observez un groupe d'étudiants appeler un de leur camarade pour qu'il les rejoigne "Steve, par ici !". Steve, la démarche maladroite, chemise froissée et mal rangée, cheveux en batailles, les rejoint timidement. Vous l'observez dans sa course. Des lunettes rondes et repositionnées à la hâte barre son visage. Il porte une sacoche mal fermée entre ses bras.\\

Est-il plus probable qu'il s'agisse d'un doctorant en mathématiques fondamentales ou d'un étudiant d'école de commerce ?
\end{quote}

La première réponse qu'il vient à l'esprit est qu'il s'agit \textit{vraisemblablement} d'un étudiant en mathématiques. Cela semble raisonnable dans la mesure où la description colle à la représentation qu'il est possible de se faire de ce type d'étudiant. En étant honnête, la réponse va beaucoup dépendre des représentations et conception que l'on a de ces deux disciplines. Néanmoins, une étude similaire a été conduite par D. Kahneman et A. Tversky concernant le discernement et la prise de décision, l'énoncé diffère, mais l'esprit est identique. Les résultats de cette étude indique que les personnes interrogées pensent qu'il est hautement plus probable que Steve soit un étudiant en mathématique.

\begin{center}
\includegraphics[width = 0.7\textwidth]{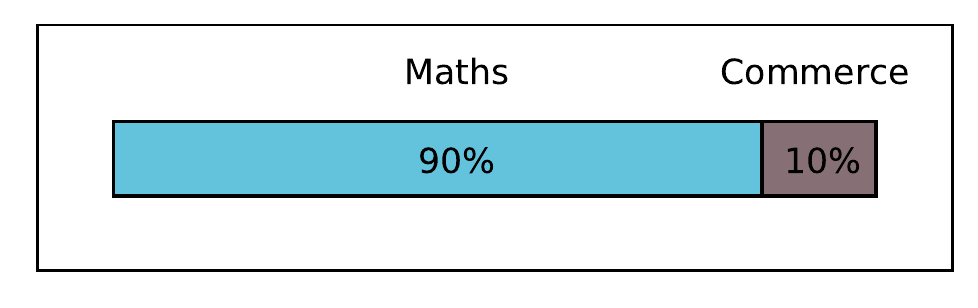}
\end{center}

Ce point de vu est largement biaisé par les représentations, les préjugés concernant les étudiants en mathématiques et en école de commerce. Et pour répondre correctement à cette questions, il faut inclure des informations concernant le ratio \textit{doctorant en mathématiques / étudiant en école de commerce}. C'est ce que permet le théorème de Bayes.

\subsubsection{La mécanique du théorème de Bayes}

Les données présentés dans ce paragraphe ne sont pas exactes mais cela n'a pas d'importance sur le raisonnement. Prenons un ratio de 1 doctorant en mathématique pour 20 étudiants en école de commerce. Ce ratio est représenté sous forme d'air coloré dans la figure suivante : 

\begin{center}
\includegraphics[width = 1\textwidth]{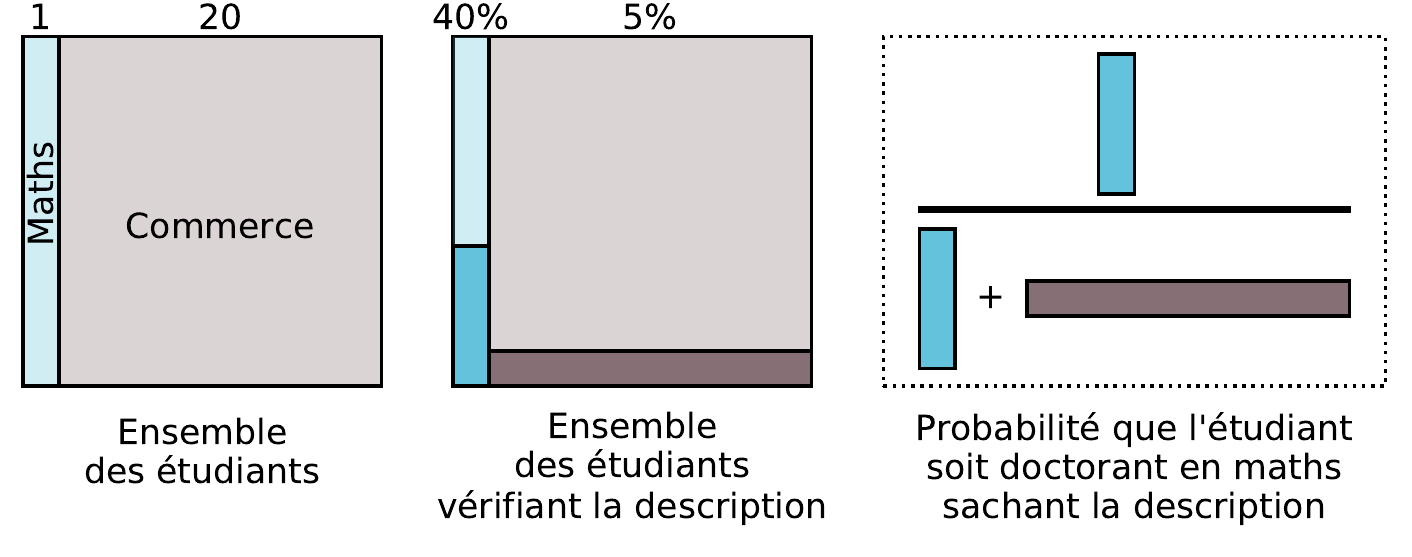}
\end{center}

Si nous devions mettre quelques chiffres, il n'est pas complètement absurde de dire que peut-être 40\% des doctorants en mathématiques fondamentales peuvent répondre à la description précédente alors que peut-être 5\% des étudiants en école vérifieraient cette description.

La probabilité de rencontrer un doctorant en maths vérifiant la description est donc : 
\begin{equation}
p = \dfrac{\text{doctorants en maths vérifiant la description}}{\text{Total des étudiants vérifiant la description}} = \dfrac{\num{0.4}\times 1}{\num{0.4}\times 1 +\num{0.05}\times 20} =\num{0.28}
\end{equation}

Ainsi, il est quand même moins probable de rencontrer un doctorant en mathématiques fondamentales qu'un étudiant en école de commerce ayant la description précédente.

\subsection{Vers un formalisme adapté à la mesure}

Les opérations que permettent ce formalisme sont les suivantes :
\begin{enumerate}
\item Nous cherchons à vérifier une hypothèse \textbf{H}, pour notre exemple, il s'agirait de \{ H : Steve est doctorant en mathématiques \}. 
\item Des mesures sont réalisés en lien avec cette hypothèse, ici il s'agit d'un ensemble d'observations décrivant Steve. Il s'agit d'un ensemble de données notées \textbf{D} pour datas.
\item Nous voulons connaître la probabilité que l'hypothèse \textbf{H} soit vrai \textit{sachant que} nous connaissons des données \textbf{D}.
\end{enumerate}

Cette dernière probabilité est une probabilité conditionnelle noté : $p(\textbf{H}|\textbf{D})$.

En reprenant l'analyse précédente, pour calculer cette probabilité, nous avons utilisé les termes suivant : 
\begin{itemize}
\item $p(\textbf{H})$ : la probabilité que l'hypothèse soit vraie. Il s'agit de la proportion de doctorant en mathématiques.
\item $p(\textbf{D}|\textbf{H})$ : la probabilité d'obtenir les données si l'hypothèse est vérifiée. Il s'agit de la proportion de doctorant en mathématiques vérifiant la description.
\item $p(\textbf{D})$ : la probabilité d'avoir les données. Il s'agit de la proportion d'individus vérifiant la description parmi les doctorants en mathématiques \textit{et} étudiant en école de commerce\footnote{Précédemment, cette probabilité a été exprimé sous la forme de la somme de deux termes : $p(\textbf{D}) = p(\textbf{D}|\textbf{H}) + p(\textbf{D}|\neg\textbf{H})$.}.
\end{itemize}

\begin{center}
\includegraphics[width = 0.7\textwidth]{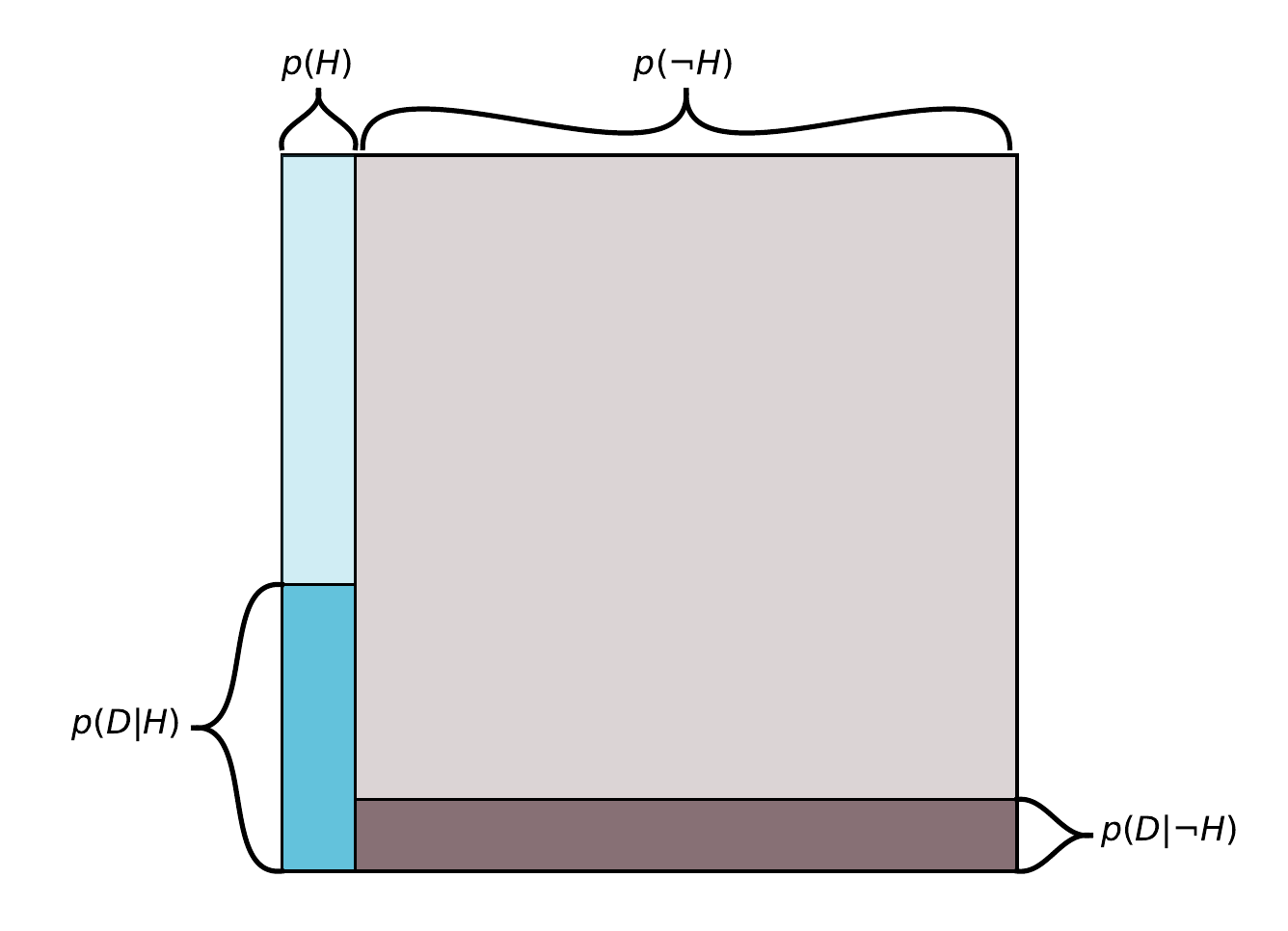}
\end{center}

Le théorème de Bayes s'écrit ainsi de la façon suivante :

\begin{equation}
p(\textbf{H}|\textbf{D}) = \dfrac{p(\textbf{D}|\textbf{H})\cdot p(\textbf{H})}{p(\textbf{D})}
\end{equation}

\subsubsection{Définitions}

\begin{itemize}
\item \textbf{H} est l'hypothèse pour laquelle un niveau de crédibilité est inférée. Dans le cadre d'une mesure, il s'agit d'un modèle associé à des paramètres $\theta_i$ qui permettent de définir des \textit{intervalles de crédibilité}.
\item $p(\textbf{H}|\textbf{D})$ est la \textbf{plausibilité} ou probabilité \textbf{postérieur} d'une hypothèse \textbf{H}. Elle est calculée en tenant compte du processus de mesure \textbf{D}. Il s'agit d'un degrés de confiance a postériori  prenant en compte toutes les informations à dispositions : données et connaissances \textit{a priori}. Il permet de quantifier la \textit{crédibilité} d'une hypothèse. 
\item $p(\textbf{D}|\textbf{H})$ est la fonction de \textbf{vraisemblance} des données \textbf{D} vis à vis de l'hypothèse \textbf{H}.  Cette fonction décrit la plausibilité des données \textbf{D} vis à vis d'une hypothèse \textbf{H} dans le cadre d'une réalisation aléatoire de ces données \textbf{D}.
\item $p(\textbf{H})$ est la probabilité \textit{a priori} qui précède toute mesures. Il peut s'agir d'une conviction estimée par l'opérateur ou des données antérieurs à la mesure et intégrées par ce biais dans le processus d'inférence.
\item $p(\textbf{D})$ agit comme une normalisation ou une fonction de partition en physique statistique. Nous verrons par la suite qu'il ne sera pas forcément utile de la prendre en compte et est souvent difficile à obtenir\footnote{analytiquement et numériquement}.
\end{itemize}

Le schéma de fonctionnement de ce formalisme est le suivant : 

\begin{equation*}
\text{Distribution \textit{a priori}} + \text{Données} \longrightarrow \text{Distribution \textit{a posteriori}}
\end{equation*}

\section{Mesures et inférence bayésienne}

Il est à noter que le formalisme présenté permet de calculer la \textit{probabilité} ou le \textit{niveau de crédibilité} d'une hypothèse moyennant la connaissance de \textit{données}. Ces probabilité se présentent sous la forme de distribution pour les paramètres de l'hypothèse ou du modèle considéré. Habituellement, dans un positionnement statistique classique, c'est l'inverse qui est recherché : quelle probabilité ont les données d'être vrai sachant {l'hypothèse \emph{qui est supposée vrai}}. De là et à partir de ces intervalles de confiance, un seuil de rejet est défini pour affirmer ou infirmer l'hypothèse choisie\footnote{Voir hypothèse nulle et facteur ou valeur p}.

Dans le formalisme bayésien, les données sont considérés comme vraies ou certaines\footnote{Après tout, ce sont des faits expérimentaux.}. Au moyen de la relation de Bayes, une distribution, la \textit{vraisemblance} d'une hypothèse, est inférée. À partir de cette vraisemblance, des \textit{intervalles de crédibilités} sont définis pour les paramètres de l'hypothèse.

%
%\subsection{Démarche pour l'analyse de données}
%
%[A faire]

\subsection{Premier cas d'analyse de données en formalisme bayésien : détermination d'un paramètre}

Cette exemple est inspiré de la documentation en ligne du module Python \textit{emcee} dédié à l'analyse de données par statistique bayésienne. Il s'agit d'une série de mesure de l'activité d'un échantillon radioactif. Cette activité $A_0$ est constante dans le temps. Nous réalisons une série de $n$ mesures : $D = \{ A_i, e_i\}$

\subsubsection{Génération des données}

Le bloque de code suivant permet de générer les données :

\begin{quote}
\begin{pyverbatim}
""" Création d'une série de données pour l'activité
radioactive d'un échantillon """
import numpy as np
from scipy import stats
# Les données sont aléatoires mais fixées avec ce paramètre
np.random.seed(123)  # Pour la répétabilité

A0 = 1000  # Valeur de référence de l'activité radioactive
N = 50 # Nombre de mesures
A = stats.poisson(A0).rvs(N)  # Réalisation des N mesures
ei = np.sqrt(A)  # écart type estimé pour une loi de Poisson
\end{pyverbatim}
\end{quote}

Les données sont générées avec une loi de Poisson et sont représentées à la figure \ref{ActiviteRadioactiveDonnees}. Cette dernière est pertinente pour décrire un comptage d'événements se produisant dans un intervalle de temps donné avec une fréquence moyenne connue et indépendant du temps. Ici, la dispersion statistique est supposée être uniquement due aux processus de mesure.

\begin{figure}[h]
\begin{center}
\includegraphics[width = 0.7\textwidth]{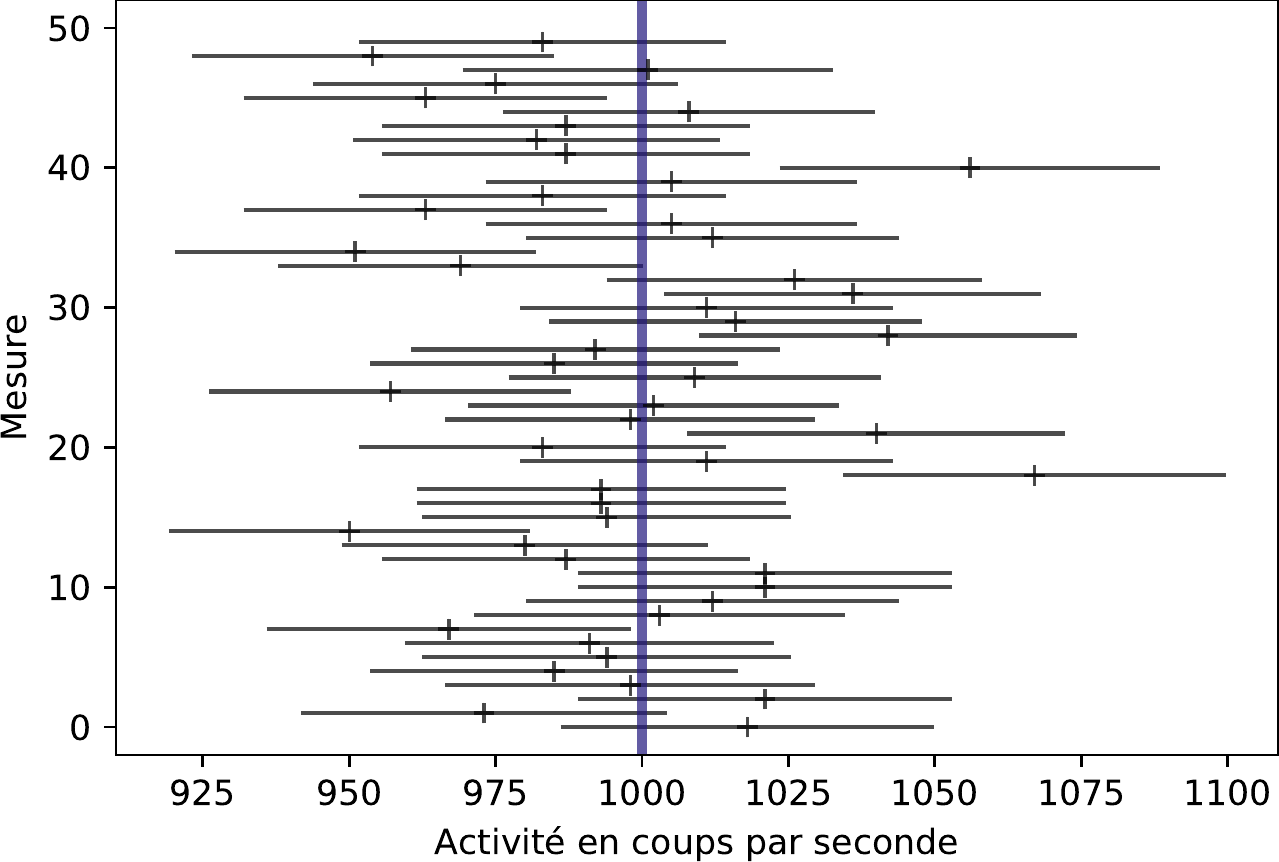}
\end{center}
\caption[Données brutes de l'activité d'un échantillon radioactif]{Données brutes de l'activité d'un échantillon radioactif $D = \{ A_i, e_i\}$. La barre centrale représente la valeur de référence. Une dispersion autour de la valeur de référence est observée.}
\label{ActiviteRadioactiveDonnees}
\end{figure}

\subsubsection{Approche classique}

Une fois les données acquises, le travail d'analyse consiste à répondre à la question suivante : \textit{Quelle est la meilleur estimation de l'activité réelle de l'échantillon.} L'approche statistique classique, détaillée dans les premiers chapitres, indique que le meilleur estimateur est la valeur moyenne empirique :

\begin{center}
\begin{tabular}{l}
 
$A_0 = \SI{1000}{\per\s}$ \\ 
 
$A_\text{estimée} = \dfrac{1}{n} \sum A_i =  \SI{997(4)}{\per\s}$ \\ 

\end{tabular} \\
\end{center}

Ainsi, la valeur estimée est compatible avec la valeur de référence pour 50 mesures.

\subsubsection{Approche bayésienne}

Maintenant, nous voulons calculer la probabilité de l'activité connaissant les mesures : $p(A|D)$ en utilisant le théorème de Bayes.

\begin{equation}
p(A|D) = \dfrac{p(D|A)\cdot p(A)}{p(D)}
\label{BayesActiviteRadioacive}
\end{equation}

\begin{itemize}
\item $p(A) \propto 1$ la probabilité \textit{a priori} est choisie uniforme et proportionnel à 1, il s'agit un d'un \textit{prior} non informatif.
\item $p(D|A) \propto \mathcal{L}\left(D|A\right)$ est la fonction de vraisemblance.
\end{itemize}

Cette fonction de vraisemblance est construite de façon à chiffrer la probabilité qu'une donnée $D_i$ soit vraie pour une activité $A$ réelle. Supposons une erreur de forme gausienne :
\begin{equation}
p(D_i | A) = \dfrac{1}{\sqrt{2\pi e_i^2}} e^{\left[-\dfrac{(A_i-A)^2}{2e_i^2}\right]}
\end{equation}

La fonction de vraisemblance est construite de la manière suivante : 
\begin{equation}
\mathcal{L}\left(D|A\right) = \prod_{i=1}^N p(D_i|A)
\end{equation}

En effet, les mesures étant indépendantes, la probabilité totale d'avoir ce jeu de données est simplement le produit des probabilités de chacune des données.

En combinant ces résultats avec l'équation \ref{BayesActiviteRadioacive} :

\begin{equation}
p(A|D) \propto \mathcal{L}\left(D|A\right)
\end{equation}

La distribution \textit{postérieur} est simplement proportionnelle à la fonction de vraisemblance $\mathcal{L}$. Il n'est pas utile de calculer tous les termes du théorème de Bayes. Le terme $p(D)$ est une normalisation qui est indépendante des paramètres de l'hypothèse, il n'apporte rien à l'analyse de données.

\begin{quote}
\begin{pyverbatim}
# Definition logarithmique des probabilités
def log_prior(theta):
    return 1  # prior non informatif

def log_likelihood(theta, Ai, ei):
    return -0.5 * np.sum(np.log(2 * np.pi * ei ** 2)
                         + (Ai - theta) ** 2 / ei ** 2)

def log_posterior(theta, Ai, ei):
    return log_prior(theta) + log_likelihood(theta, Ai, ei)
\end{pyverbatim}
\end{quote}

Les probabilités utiles sont définis par leur logarithmique car il est plus simple de réaliser une somme qu'un produit numériquement.

\begin{quote}
\begin{pyverbatim}
# Calcul la probabilité a posteriori p
A = np.linspace(975, 1020, num = 500) #Paramètre du modéle
logp=[] #initialisation du logp

for Avar in A:
    logp.append(log_posterior(Avar, Ai, ei))

# Normalisation à l'unité de p    
from scipy.integrate import simps 
p = np.exp(logp)
p /=simps(p,A)

# Tracé de la probabilité postérieure
fig, ax = plt.subplots()
plt.plot(A,p)
plt.xlabel("Activité en coups par seconde")
plt.ylabel("p(A|D)")
\end{pyverbatim}
\end{quote}

\begin{figure}[h]
\begin{center}
\includegraphics[width = 0.7\textwidth]{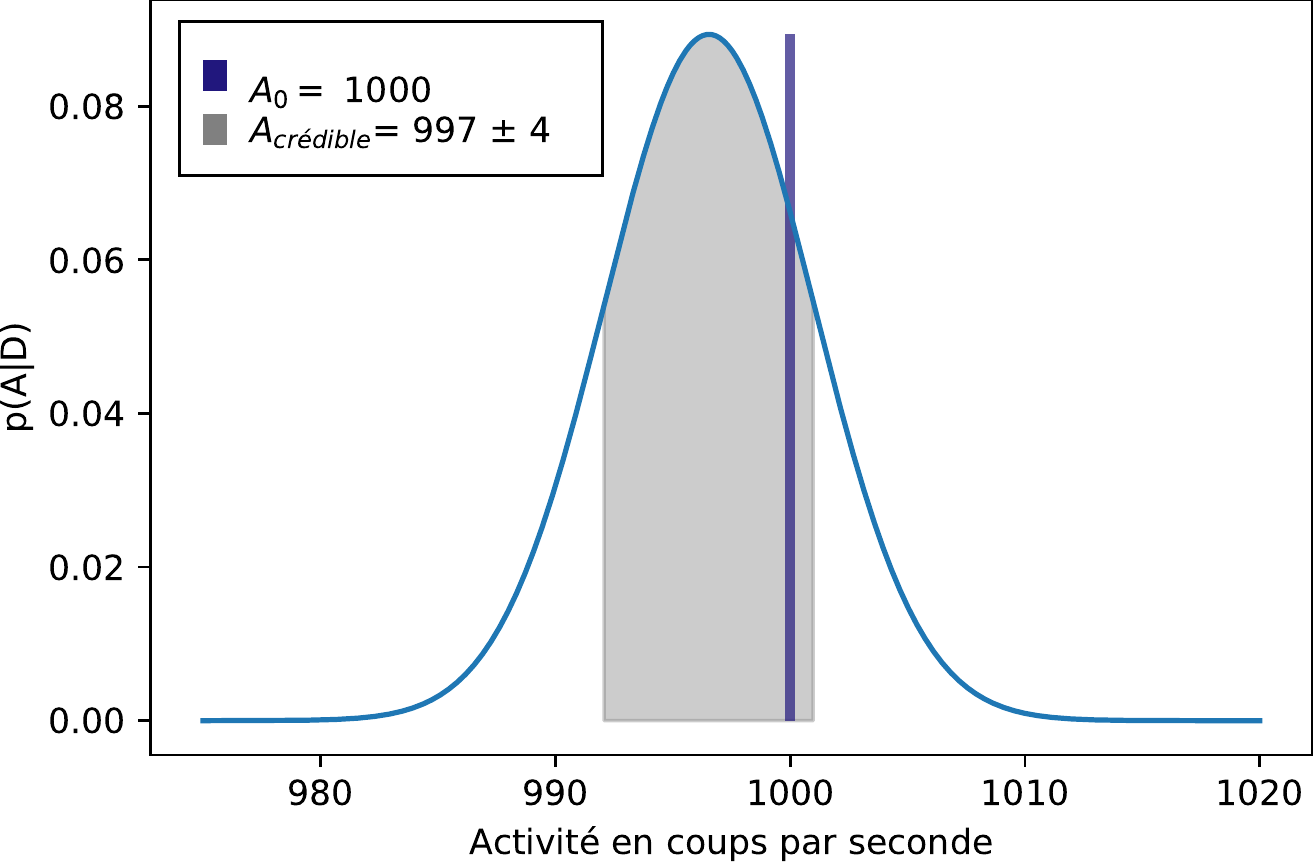}
\end{center}
\caption[Inférence bayésienne : activité d'un échantillon radioactif]{Activité de l'échantillon inférée par méthode bayésienne à partir d'un échantillon $D = \{ A_i, e_i\}$. La zone centrale grise représente l'intervalle de crédibilité.}
\label{ActiviteRadioactiveDonnees}
\end{figure}

Nous obtenons un ensemble de distribution de probabilité concernant le paramètre $A$ du modèle. L'hypothèse la plus probable et son intervalle de crédibilité sont identiques à la valeur moyenne et à l'intervalle de confiance calculés avec des technique statistique classique.

\begin{center}
\begin{tabular}{l}
 
$A_0 = \SI{1000}{\per\s}$ \\ 
 
$A_\text{crédible} =\SI{997(4)}{\per\s}$ \\ 

\end{tabular} \\
\end{center}

L'inférence bayésienne est plus complexe à mettre en œuvre, et fournit un résultat rigoureusement identique à l'inférence classique. L'enjeu réside dans le positionnement de ce dernier et surtout dans l'universalité de la méthode. La partie suivante propose une extension de l'étude réalisé ci-dessus et le code algorithmique mis en place sera à peine modifié pour répondre à un problème bien plus complexe.

Concernant la procédure numérique, la probabilité postérieure est facile à déterminer par calcul direct. Ce n'est plus le cas quand le modèle possède beaucoup de variable. Dans la partie suivante, nous utiliserons le module \textit{emcee} utilisant des techniques d'échantillonnage avec méthode de Monte-Carlo par chaîne de Markov. Ces techniques sont particulièrement adaptés lorsque le modèle se complexifie.

\subsection{Deuxième cas : détermination de deux paramètres et leur incertitude}
\label{DeuxPara}

Cet exemple est similaire au cas précédant. Nous allons étudier l'activité radioactive d'un échantillon. A la différence de précédemment, cet activité $A_0$ fluctue aléatoirement dans le temps. Nous réalisons une série de $n$ mesures : $D = \{ A_i, e_i\}$.

Nous cherchons donc à déterminer deux paramètre : $\theta = [\mu_A, \sigma_A]$, avec $\mu_A$ la valeur moyenne de l'activité et $\sigma_A$ l'écart type associée aux fluctuations intrinsèques de la source. L'activité suit donc le modèle suivant : 
\begin{equation}
A_0 \propto \dfrac{1}{\sqrt{2\pi\sigma_A^2}} e^{\left[-\dfrac{(A-\mu)^2}{2\sigma_A^2}\right]}
\end{equation}
\subsubsection{Génération des données}

Le bloque de code suivant permet de générer les données :
\begin{quote}
\begin{pyverbatim}
# Création d'une série de données pour l'accélération de la pesanteur
import numpy as np
from scipy import stats

# Pour la répétabilité:
# les données sont aléatoires mais fixées avec ce paramètre
np.random.seed(42) 

N = 50 # Nombre de mesures
# Valeur de référence de l'activité radioactive
# de l'échantillon en coups par seconde et écart type du modèle
mu_A, sigma_A = 1000, 30  

A0 = stats.norm(mu_A, sigma_A).rvs(N) #Activité de référence fluctuante
Ai = stats.poisson(A0).rvs(N)  # Réalisation des N mesures
ei = np.sqrt(Ai)  # écart type estimé pour une loi de Poisson
\end{pyverbatim}
\end{quote}

\subsubsection{Approche classique d'inférence}

Une approche classique consiste à définir une fonction de deux paramètres maximisant la probabilité que les données vérifient le modèle. Cette technique est identique à celle utilisée pour la régression linéaire qui minimise l'erreur entre une courbe et une série de point, et donc consiste à trouver des paramètres maximisant la probabilité que les données vérifie le modèle.

La fonction de vraisemblance utilisé est l'association des distributions statistique de la source et les incertitudes de mesure. Les deux processus ne sont pas corrélés, nous allons utiliser un modèle gaussien dont l'écart type est : $\sigma_A^2 + e_i$ :

\begin{equation}
\mathcal{L}(D|\theta) = \prod_{i=1}^N \dfrac{1}{\sqrt{2\pi(\sigma_A^2 + e_i^2)}} e^{\left[-\dfrac{(A_i-\mu_A)^2}{2(\sigma_A^2 +e_i^2)}\right]}
\label{Likelyhood2parameters}
\end{equation}

Ici, $\theta$ représente le modèle, donc l'hypothèse à vérifier.

Il n'est pas possible d'utiliser des modèles analytiques pour trouver une solution. Des procédures numériques existent et permettent de déterminer les valeurs optimales de $\mu_A$ et $\sigma_A$ avec leur incertitude associées.

Le principe consiste à générer un grand nombre de sous-ensembles de données à partir des données initiales\footnote{algorithme de bootstrap ou jacknife}. Chacun de ces sous-ensembles est alors traité comme une réalisation des mesures pour lequel la valeur moyenne et l'écart type est calculé. Il est ainsi possible d'obtenir un ensemble de valeurs moyennes et d'écarts types et ainsi d'obtenir l'incertitude sur ces deux grandeurs.\\

Résultats avec techniques bootstrap et maximisation d'une fonction de vraisemblance :
\begin{center}
\begin{tabular}{lll}
Valeurs théoriques& $\mu = \SI{1000}{\per\s}$ & $\sigma = \SI{10}{\per\s}$ \\ 
Valeurs inférées & $\mu = \SI{1000(4)}{\per\s}$ & $\sigma = \SI{14(6)}{\per\s}$ \\ 
\end{tabular} 
\end{center}

Les intervalles données ci-dessus sont les intervalles de confiance à 68\%. Ces derniers supposent que la statistique d'incertitude est gaussienne suivant $\mu$ et $\sigma$. Cela donne une zone de confiance rectangulaire centrée sur $\mu = \num{1000}$ et $\sigma =  \num{14}$ de côtés les intervalles de confiance. Nous allons voir que l'approche bayésienne permet d'aller plus loin.

%
%\begin{pyverbatim}
%'''
%Traitement classique d'un modèle à deux paramètres via Boostrap
%'''
%
%# Definition de la fonction vraisemblance
%def log_likelihood(theta, Ai, ei):
%    return -0.5 * np.sum(np.log(2 * np.pi * (theta[1] ** 2 + ei ** 2))
%                         + (Ai - theta[0]) ** 2 / (theta[1] ** 2 + ei ** 2))
%
%# maximiser la fonction vraisemblance <-> minimiser son opposée
%def neg_log_likelihood(theta, F, e):
%    return -log_likelihood(theta, F, e)
%
%#Modules
%import numpy as np
%from scipy import optimize     
%from astroML.resample import bootstrap
%
%# initialisation des paramètres pour la fonction optimize
%theta_init = [900, 5]
% 
%# Fonction pour calculer les paramètres d'optimisation de la
%# fonction vraisemblance
%def fit_SousEchantillon(SousEchantillon):
%    # Calcul des paramètres de maximisation de la fonction
%    # de vraisemblance pour chaque échantillonage du boostrap
%    return np.array([optimize.fmin(neg_log_likelihood, theta_guess,
%                                   args=(Ai, np.sqrt(Ai)), disp=0)
%                     for Ai in SousEchantillon])
%
%# Re-échantillonage via un algorithme boostrap
%# Et calcul des paramètres d'optimisation pour chaque sous-échantillon
%# du bootstrap
%EchantillonsBootstrap = bootstrap(Ai, 500, fit_SousEchantillon)   
%
%mu_Bootstrap = EchantillonsBoostrap[:, 0]
%sig_Bootstrap = abs(EchantillonsBoostrap[:, 1])
%
%print("mu    ={0:.0f}+/-{1:.0f}".format(mu_Boostrap.mean(),mu_Bootstrap.std()))
%print("sigma ={0:.0f}+/-{1:.0f}".format(sig_Boostrap.mean(),sig_Bootstrap.std()))
%
%\end{pyverbatim}

\subsubsection{Approche par inférence bayésienne}
De même que précédemment, nous définissons le logarithme du prior, de la fonction de vraisemblance et de la probabilité postérieure. La fonction de vraisemblance est identique (équation \ref{Likelyhood2parameters}) à celle définie pour une approche classique.

\begin{quote}
\begin{pyverbatim}
# Definition logarithmique des termes
def log_prior(theta):
        # theta[0] représente mu_A, et theta[1] est sigma_A
        # sigma_A DOIT être positif.
    if theta[1] <= 0:
        return -np.inf
    else:
        return 0

def log_likelihood(theta, Ai, ei):
    return -0.5*np.sum(np.log(2*np.pi*(theta[1]**2 + ei**2))
                         + (Ai - theta[0])**2 / (theta[1]**2 + ei**2))

def log_posterior(theta, Ai, ei):
    return log_prior(theta) + log_likelihood(theta, Ai, ei)
\end{pyverbatim}
\end{quote}

Le calcul du postérieur est encore envisageable avec une procédure naïve pour deux paramètres à inférer. La technique consister à calculer les valeurs de \textit{log(p)} pour chaque couple $(\mu, \sigma)$ dans une fenêtre donnée.

\begin{quote}
\begin{pyverbatim}
# Calcul du postérieur

# Nombre de points
Nb_mu = 200
Nb_sig = 200

# Création des tableaux pour mu et sigma
mu_line = np.linspace(990,1010,Nb_mu)
sig_line = np.linspace(0,30,Nb_sig)

# Création du maillage
mu_mesh, sig_mesh = np.meshgrid(mu_line, sig_line)

# Initialisation du log du postérieur
logP = np.zeros([Nb_mu,Nb_sig])

# Calcul du postérieur avec double boucle
for i in tqdm(range(Nb_mu)):
    var_mu = mu_line[i]
    for j in range(Nb_sig):
        var_sig = sig_line[j]
        logP[j,i] += log_posterior(var_mu, var_sig, Ai, ei)
        
# Normalisation à 1
logP -= np.max(logP)
p = np.exp(logP)

# Valeur la plus probable
result = np.where(p == np.amax(p))
\end{pyverbatim}
\end{quote}

Et le tracé de la distribution postérieur comme présenté en figure \ref{ActiviteRadioactiveDonnees} :

\begin{quote}
\begin{pyverbatim}
# Affichage graphique
fig, ax = plt.subplots()
ax.pcolor(mu_mesh, sig_mesh, p)
levels = [0.68, 0.95]
CS = ax.contour(mu_mesh, sig_mesh,p, levels, colors='k', linewidths = 0.7)
ax.plot([mu_A], [sigma_A], 'o', color='red', ms=4);
ax.text(mu_line[result[1]][0],sig_line[result[0]][0], '+' )
\end{pyverbatim}
\end{quote}

\begin{figure}[h]
\begin{center}
\includegraphics[width = 0.7\textwidth]{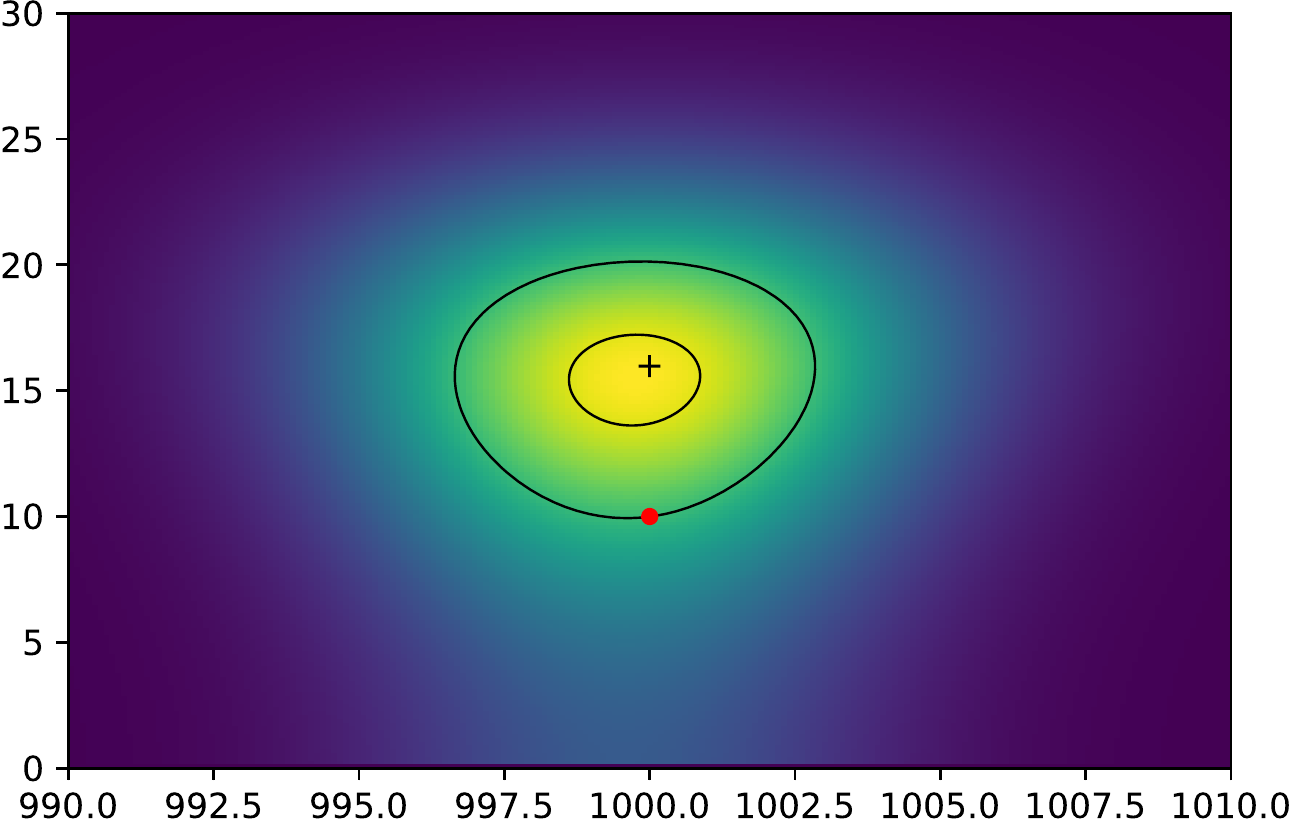}
\end{center}
\caption[Inférence bayésienne : détermination de deux paramètres]{Modèle aléatoire de l'activité de l'échantillon inféré par méthode bayésienne à partir d'un ensemble de mesures $D = \{ A_i, e_i\}$. La croix représente la valeur la plus probable, le point rouge représente la valeur théorique.}
\label{ActiviteRadioactiveDonnees}
\end{figure}

Les courbes représentent les zones de crédibilité à 68\% et 95\%. La forme générale de la distribution n'est plus gaussienne et les zones de crédibilités sont plus restreintes. Le résultat final est identique à celui obtenu par méthode classique :

Résultats avec techniques d'inférence bayésienne :
\begin{center}
\begin{tabular}{lll}
Valeurs théoriques& $\mu = \SI{1000}{\per\s}$ & $\sigma = \SI{10}{\per\s}$ \\ 
Valeurs inférées & $\mu = \SI{1000(4)}{\per\s}$ & $\sigma = \SI{14(6)}{\per\s}$ \\ 
\end{tabular} 
\end{center}

%
%
%Le théorème de Bayes est une relation entre probabilités conditionnelles :
%
%\begin{equation}
%P(A|B) = \dfrac{P(B|A)\cdot P(A)}{P(B)}
%\end{equation}

\section{Prise en compte de données \textit{a priori}}

Cette partie est consacrée à l'influence de la probabilité \textit{a priori}. L'exemple proposé est l'analyse de données d'une mesure de résistance. Les données sont simulées numériquement, mais les incertitudes associées sont celles fournies par la documentation du multimètre numérique agilent 34401.

\subsection{Positionnement du problème}
Une résistance à 5\% de valeur nominale $R_\text{nom} = \SI{500}{\ohm}$ est déterminée au moyen d'une méthode volt-ampèremétrique. La valeur réelle de la résistance est $R\text{vrai} = \SI{512}{\ohm}$. Les caractéristiques de la mesure sont données dans le tableau ci-contre :

\begin{center}
\begin{tabular}{rc}
 
$\sigma_U$ & \SI{0.002}{\milli\V}   \\ 

$\sigma_I$ &\SI{0.01}{\milli\A}    \\ 

$\sigma_R$& \SI{5}{\ohm}\\

\end{tabular} 
\end{center} % = R\sqrt{(\frac{\sigma_U}{U})^2+(\frac{\sigma_I}{I})^2 }$

Les valeurs $U$ et $I$ sont tirées aléatoirement avec une loi normale d'écart type donnée par les caractéristiques de la mesure.

\subsection{Fonction de vraisemblance}
La fonction de vraisemblance reflète la loi normale choisie pour représenter l'incertitude sur les mesures. Pour une mesure, la probabilité $p(R_i = \frac{U_i}{I_i} | R_0)$ est donnée par la relation suivante :
\begin{equation}
p(R_i = \tfrac{U_i}{I_i} | R_0, \sigma_R) = \dfrac{1}{2\pi\sigma_R}\exp\left[-\frac{1}{2}\left(\frac{R_i - R_0}{\sigma_R}\right)^2\right]
\end{equation}

Pour une ensemble de $n$ mesures $\lbrace R_i \rbrace$, la fonction de vraisemblance est le produit des probabilités $p(R_i = \tfrac{U_i}{I_i} | R_0, \sigma_R)$ : 
\begin{equation}
p(\lbrace R_i \rbrace | R_0, \sigma_R) = \prod_{i=0}^n  \dfrac{1}{2\pi\sigma_R}\exp\left[-\frac{1}{2}\left(\frac{R_i - R_0}{\sigma_R}\right)^2\right]
\end{equation}

\subsection{Choix du prior}

L'objectif est d'étudier l'influence de la probabilité \textit{a priori} $p(R_0)$ sur la probabilité postérieure.

Nous allons utiliser deux priors différents se basant sur les connaissances suivantes de la valeur de résistance : 
\begin{itemize}
\item données constructeurs : prior quasi non informatif, une fonction uniforme sur l'intervalle défini par la tolérance de la résistance est utilisée
\item mesure précédente : prior suivant une loi normale $R_\text{prior} = \SI{490(2)}{ohm}$.
\end{itemize}

Ces deux priors correspondent à des situations concrètes de mesure, dans le premier cas seuls les informations constructeurs servent de base, dans le second cas une mesure rapide a été réalisée avec un ohmmètre donnant un résultat cohérent mais éloigné de la valeur réelle.

\subsection{Tracé de la probabilité \textit{a prosteriori}}
Comme dans les parties précédentes, la probabilité \textit{a posteriori} est définie par : 
\begin{equation}
p(R_0 | \lbrace R_i \rbrace, \sigma_R) \propto p(R_0) \times p(\lbrace R_i \rbrace | R_0, \sigma_R)
\end{equation}

\begin{figure}[h]
\begin{center}
\includegraphics[width = 1\textwidth]{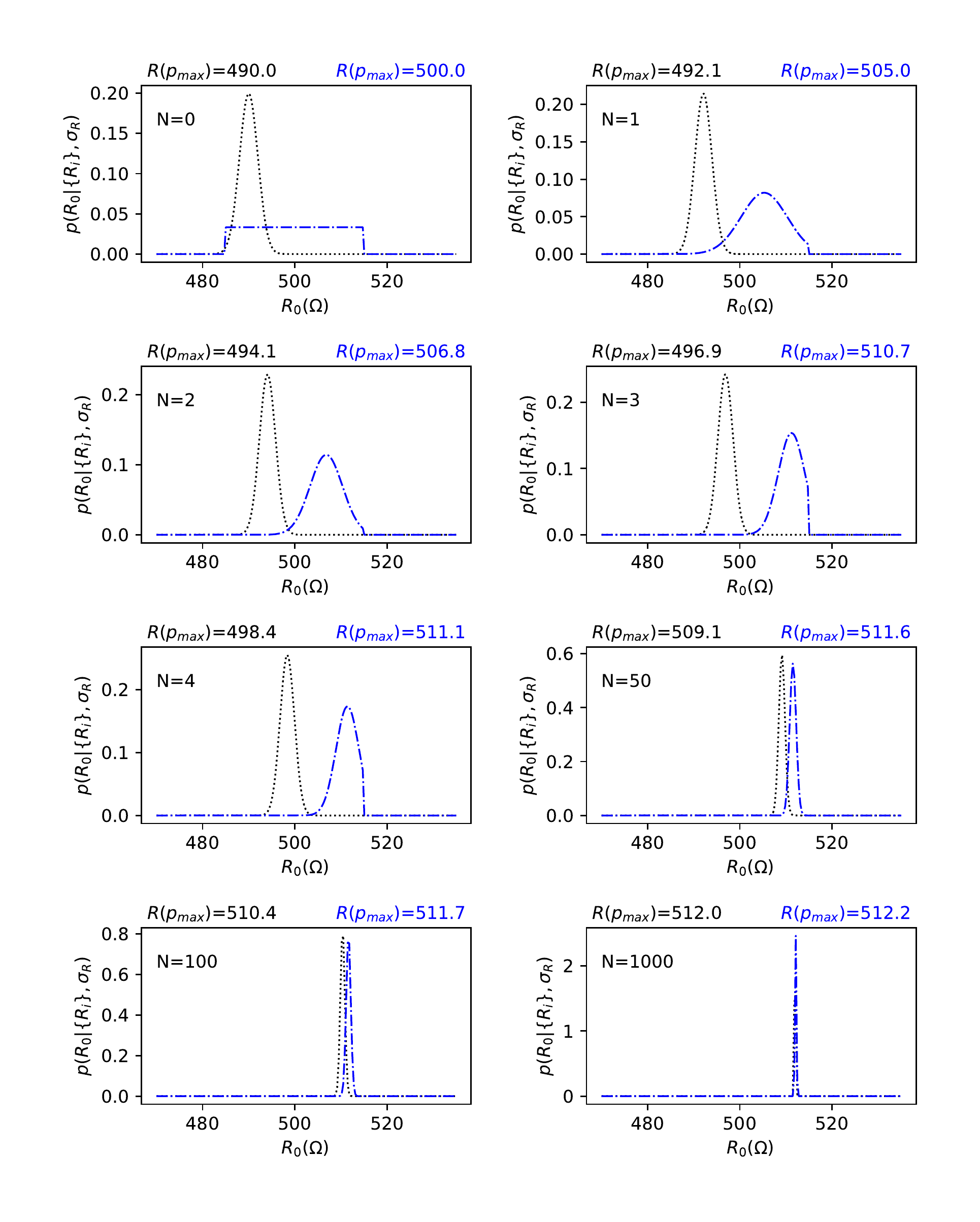}
\end{center}
\caption[Influence du nombre de mesures et du prior sur le résultat final]{Influence du nombre de mesures et du prior sur le résultat final de la probabilité \textit{a posteriori}.}
\label{PriorInfluence}
\end{figure}

Une normalisation est ensuite réalisée pour obtenir une distribution. Les résultats sont tracées en figure \ref{PriorInfluence}.

La probabilité \textit{a posteriori} est tracée avec un nombre différent de mesure. Le cas $N=0$ correspond au tracé du prior, une distribution normale et uniforme est bien retrouvé et correspondent aux informations précédant toutes mesures. Avec l'augmentation du nombre de mesures, les probabilités obtenues s'affinent de plus en plus et tendent toutes les deux vers une seule et même valeur la plus probable : $R_0 (p_\text{max}) = \SI{512}{\ohm}$ qui est bien la valeur vraie de la résistance.

\subsection{Code Python}

\begin{pyverbatim}
import numpy as np
import matplotlib.pyplot as plt
from scipy import stats

# Pour la répétabilité
np.random.seed(10) 

#Nombre de mesures : 
N =1

#Valeur réelle de la résistance en Ohm
R_vrai = 512   

#Etude d'une résistance nominal de 500 Ohm à 5%
R_nom = 500
Rtol = 0.03

#Mesure : méthode voltampéremétrique 
#I = 1mA fixée et connue à 0.1mA
#U mesuré à 1mV
#Incertitude gaussienne, incertitude = agilent multimeter 34401a

I0 = 1  #mA
sigmaI0 = 0.01 #mA  résistance de shunt = 0.1 Ohm

U0 = 512  #mV
sigmaU0 = 0.002 #mV

I_mes = stats.norm.rvs(size = N, loc = I0, scale = sigmaI0)
U_mes = stats.norm.rvs(size = N, loc = U0, scale = sigmaU0)
R_mes = U_mes / I_mes
sigma_R = 5


# Définition des probabilités pour inférence bayésienne
def log_prior_uniforme(theta):
    #g_i needs to be between 0 and 1
    if (theta < R_nom*(1+Rtol)) and (theta > R_nom*(1-Rtol)):
        return 0
    else:
        return -np.inf  # recall log(0) = -inf


def log_likelihood(theta):
    return - np.sum(((theta-R_mes)**2)/(2*sigma_R**2))


def log_posterior(theta):
    return log_prior_uniforme(theta) + log_likelihood(theta)


# Calcul de la probabilité postérieure
R_line = np.linspace(470, 535, 200)
logpost = [] #initialisation du logarithme du postérieur
for R in R_line:
    logpost.append(log_posterior(R))
post = np.exp(logpost)
post /=np.trapz(post, R_line)

# Tracé graphique
plt.plot(R_line, post)
\end{pyverbatim}

\subsection{Conclusions}

Le prior a un effet important quand le nombre de mesures est faible, mais il tend à s'effacer rapidement avec l'augmentation de données empiriques. Les conclusions sont identiques et indépendantes du choix initiales pour le prior. 

A l'inverse, il est très important de réaliser un choix judicieux et éclairé du prior quand le nombre de mesure est faible. Ce dernier permet d'intégrer toute les connaissances liées à la grandeur mesurée, et permet d'une certaine manière une mise à jour de cette grandeur sous l'éclairage de nouvelles mesures.

Cette notion de probabilité \textit{a priori} est une critique récurrente associée à la démarche d'inférence bayésienne. Cependant, il est souvent possible de formaliser des choix rationnels et objectifs pour le prior et reflétant les connaissances préalables à la mesure. L'avantage est qu'il permet justement de prendre en compte un certain nombre de connaissances liées à la mesure car il est rare de réaliser une mesure sans aucune intuition ou contrainte sur le résultat.

%%Application Inférence Bayesienne

\chapter{Applications de l'inférence bayésiennes à l'analyse de données}
\label{ApplicationsInferencesBayesiennes}
Ce chapitre portera sur quelques exemples d'application de l'inférence bayésiennes. Le chapitre \ref{InferenceBayesienne} a permis d'introduire cette méthodes et de montrer sa cohérence avec les techniques d'inférence classique. Ce chapitre porte sur des situations où les techniques classiques sont inadaptées.

\section{Analyse statistique d'une panne de machine}

\subsection{Présentation du problème}

Cette partie est un exemple détaillé par E.T Jaynes dans \textit{Confidence Intervals vs Bayesian Intervals}.

 Prenons une installation industrielle dont le fonctionnement nécessite des pièces d'usures. A partir d'un instant $\theta$, l'usure devient telle que la machine devient sujette à des pannes. La probabilité qu'une panne se déclenche suit une loi exponentielle. 
 
 Il est trop couteux pour l'industriel de réaliser des vérifications régulières des pièces ainsi que les changer trop régulièrement.

\begin{quote}
A device will operate without failure for a time $\theta$ because of a protective chemical inhibitor injected into it; but at time $\theta$ the supply of the chemical is exhausted, and failures then commence, following the exponential failure law. It is not feasible to observe the depletion of this inhibitor directly; one can observe only the resulting failures. From data on actual failure times, estimate the time $\theta$ of guaranteed safe operation...
\end{quote}

La probabilité qu'une panne se produite à un instant $t$ est donnée par la distribution suivante :
\begin{equation}
p(t|\theta) = \begin{cases}
\exp\left(\theta - t \right) &\text{, } t>\theta\\
0 &\text{, } t<\theta\\
\end{cases}
\end{equation}

\begin{figure}[h]
\begin{center}
\includegraphics[width = 0.7\textwidth]{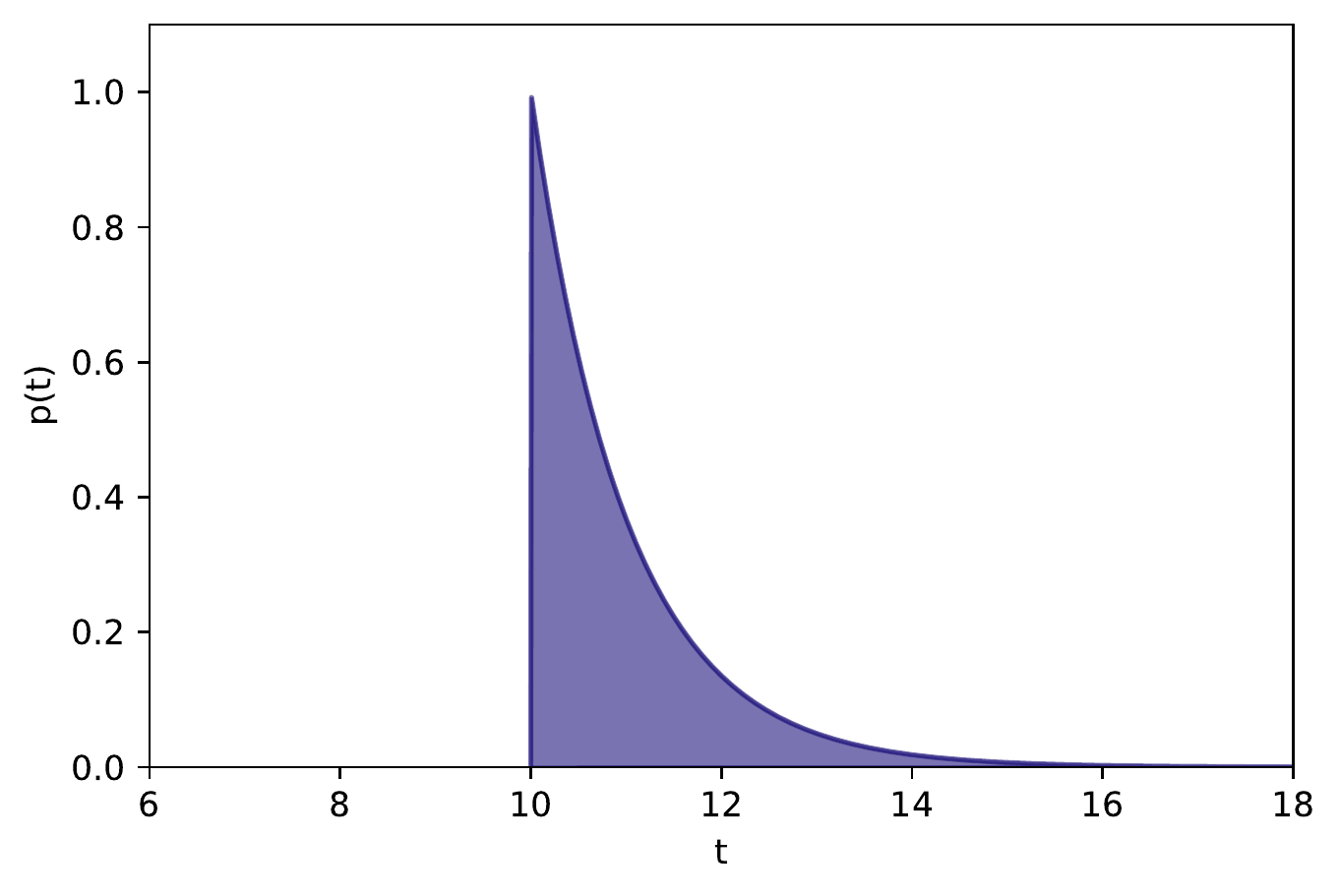}
\end{center}
\caption[Distribution exponentielle : modélisation d'un risque aléatoire de panne]{Distribution exponentielle : modélisation d'un risque aléatoire de panne. Ici $\theta = 10$.}
\label{TruncatedExp}
\end{figure}

Les données recueilli par l'industriel sont : $D = \lbrace 10, 12, 15 \rbrace$ en semaines. L'objectif est de trouver une estimation de $\theta$ connaissant $D = \lbrace t_i \rbrace$.

Étant donnée la forme générale de la distribution, il est évident que $\theta$ doit être plus petit que la plus petite valeur observée : $\theta \leq \text{min}(D)$.

\subsection{Approche classique}

Connaissant la forme de la distribution, il est montré qu'un estimateur de $\theta$ est\footnote{On montre que $E(t) = \int_0^\infty t p(t) dt = \theta +1$} :
\begin{equation}
\hat{\theta} = \dfrac{1}{n}\sum_{i=1}^n t_i -1
\end{equation}

Il est possible d'approcher l'\textit{intervalle de confiance} par une loi normale d'écart type $\sigma = 1$, ce n'est pas rigoureusement l'intervalle de confiance réel pour cette loi, mais cela ne modifie pas les conclusions. l'intervalle de confiance est donc : 
\begin{equation}
\hat{\theta} \pm \dfrac{1}{\sqrt{n}}
\end{equation}

Soit : $\theta_\text{estimé} = \num{12.3(6)} = \left[\num{11.7} ; \num{12.9}\right]$

L'intervalle de confiance est entièrement en dehors de la zone dans laquelle doit se trouver $\theta$. C'est principalement lié au faible nombre de données. Ceci étant, pour ce type de problème, le nombre de données ne sera jamais élevé puisqu'il faut attendre une panne arrivant à un délais supérieur à 10 semaines pour ajouter un point à la statistique.

\subsection{Approche bayésienne}

Commençons par écrire le théorème de Bayes :
\begin{equation}
p(\theta|D) = \dfrac{p(D|\theta)\cdot p(\theta)}{p(D)}
\end{equation}

Nous allons utiliser un prior non informatif : $p(\theta) = 1$ et une fonction de vraisemblance de la forme : 
\begin{equation}
p(D|\theta) = \prod_{i=1}^n p(t|\theta)
\end{equation}

Le produit d'exponentielles tronquées donne : 
\begin{equation}
p(D|\theta)\begin{cases}
n\exp\left[n\left(\theta - \text{min}(D) \right)\right] &\text{, } \theta<\text{min}(D)\\
0 &\text{, } \theta>\text{min}(D)\\
\end{cases}
\end{equation}

Dans la mesure où l'exponentielle est une fonction croissante, le plus petit intervalle $\left[\theta_\text{min}; \theta_\text{max}\right]$ contenant 65\% des valeurs est donnée par : 
\begin{equation}
\int_{\theta_\text{min}}^{\theta_\text{max}} n\exp\left[n\left(\theta - \text{min}(D) \right)\right] d\theta = \num{0.65}
\end{equation}
 avec $\theta_\text{max} = \text{min}(D)$ qui vérifie l'approche de bon sens.

Et donc : 
 \begin{equation}
 \theta_\text{min} = \theta_\text{max} + \dfrac{\log\left(1-\num{0.65}\right)}{n}
 \end{equation}

Ce qui donne l'\textit{intervalle de crédibilité} suivant : 
$\theta_\text{estimé} = \left[\num{9.65} ; \num{10}\right]$

Sans surprise, nous constatons que l'approche bayésienne vérifie le bon sens et permet de définir un intervalle pendant lequel planifier l'intervention sur l'installation industrielle avant que cette dernière ne tombe en panne.

\section{Problème de la position du phare}

Un phare est positionné sur des récifs au large d'une côte. Sa position et sa distance à la côte sont inconnue. Il émet aléatoirement des faisceaux étroits de lumière dans des directions aléatoires. Une série de photo-détecteurs sont placés le long de la côte et permettent simplement de détecter qu'un faisceau a été émis : la direction dans laquelle il a été émis est inconnue.

\begin{figure}[h]
\begin{center}
\includegraphics[width = 1\textwidth]{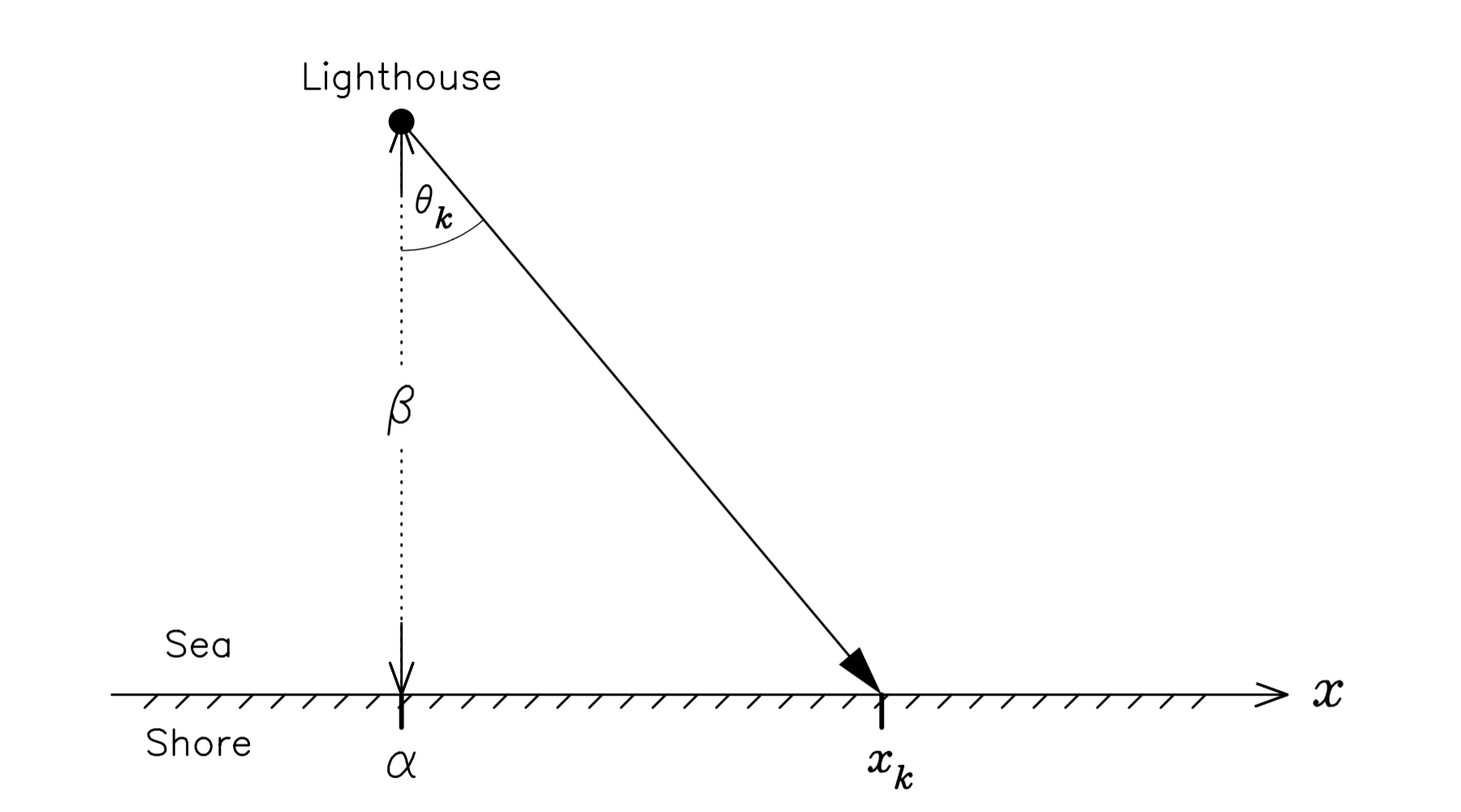}
\end{center}
\caption[Illustration de la géométrie du problème de la détermination de la position d'un phare]{Illustration de la géométrie du problème de la détermination de la position d'un phare.}
\label{LightHousePb}
\end{figure}

Les données sont un ensemble $D = \lbrace x_k \rbrace$ de position pour laquelle un flash a été enregistré. Où est situé le phare ?

\subsection{Analyse du problème}
Il est raisonnable de penser que les flashs sont émis uniformément suivant un angle $\theta_k$ autour du phare, en direction de la côte :
\begin{equation}
p(\theta_k | \alpha, \beta) = \dfrac{1}{\pi}
\label{p(theta)}
\end{equation}

Les notations utilisées sont celles définies dans la figure \ref{LightHousePb}. Les flashs ayant lieu vers le demi espace supérieur ne sont reçus par aucun capteur et sont perdus, tout se passe comme si le phare n'émettait que vers la côte.

En reliant $\theta_k$ à $x_k$ :
\begin{equation}
\beta \tan \theta_k = x_k - \alpha
\end{equation}

Ainsi, en réécrivant l'équation \ref{p(theta)} :
\begin{equation}
p(x_k|\alpha, \beta) = \dfrac{1}{\pi}\dfrac{\beta}{\beta^2 + (x_k-\alpha)^2}
\end{equation}

\begin{figure}[h]
\begin{center}
\includegraphics[width = 1\textwidth]{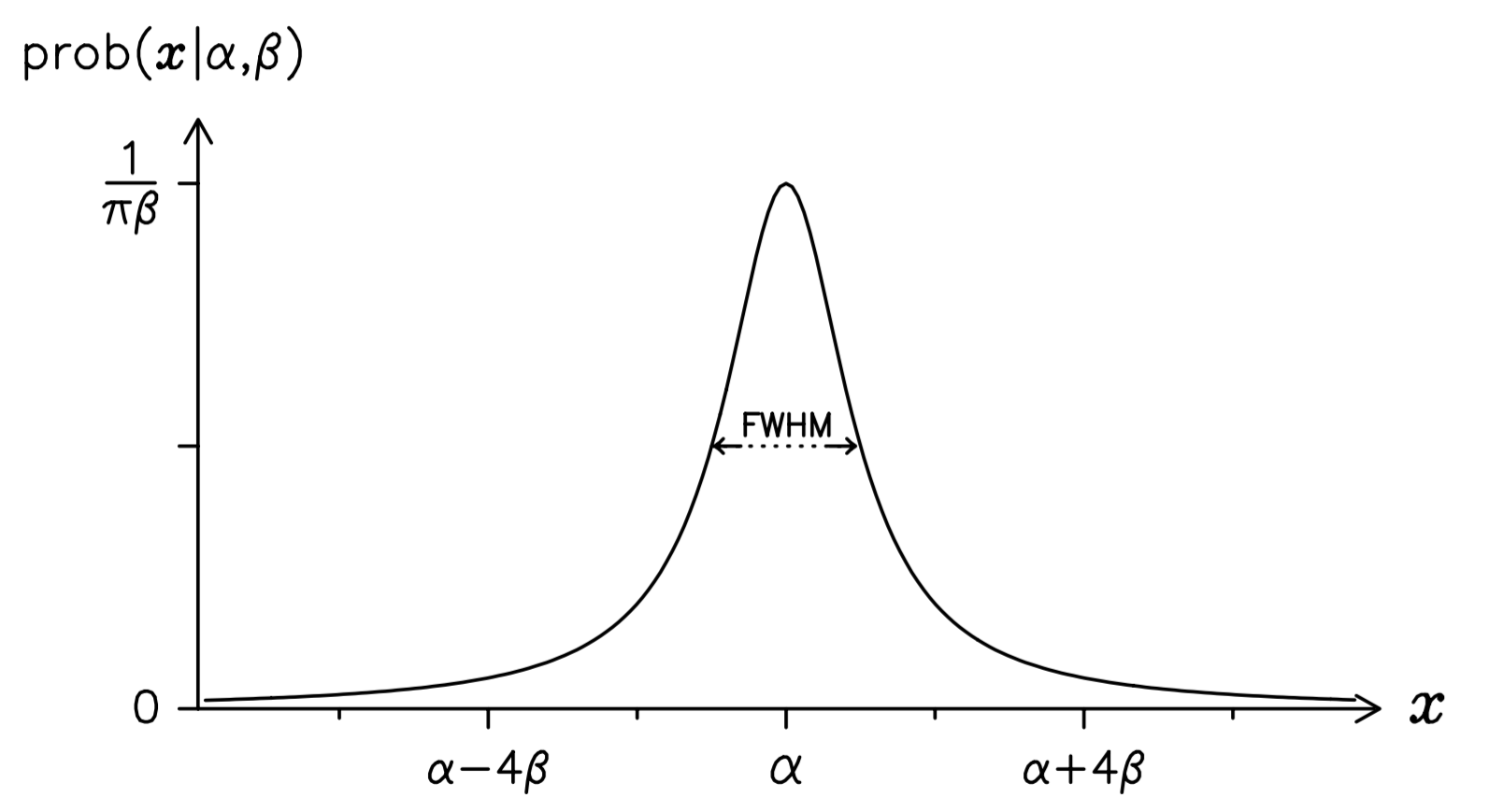}
\end{center}
\caption[Distribution de Cauchy]{Distribution de Cauchy.}
\label{CauchyDistribution}
\end{figure}

Dans ce problème, la probabilité de mesurer un flash à la position $x_k$ connaissant les coordonnées $(\alpha, \beta)$ du phare est décrite pas une distribution de Cauchy\footnote{C'est une fonction de Lorentz.}. Cette distribution possède un maximum en $x_k = \alpha$ et une largeur à demi hauteur de $2\beta$ comme montré sur la figure \ref{CauchyDistribution}.

L'objectif est d'estimer la position $(\alpha, \beta)$ du phare. Pour des raisons de clarté, nous allons considérer $\beta$ comme connue et fixée dans un premier temps et chercher à déterminer $\alpha$ uniquement. L'application du théorème de Bayes donne la relation suivante : 
\begin{equation}
p(\alpha |\lbrace x_k \rbrace, \beta) \propto p(\lbrace x_k \rbrace|\alpha, \beta) \cdots p(\alpha, \beta)
\end{equation}

\subsubsection{Le prior $p(\alpha, \beta)$}
Nous allons choisir un prior faiblement informatif : le phare est situé dans une plage $\lbrace \alpha_\text{min} ; \alpha_\text{max}\rbrace \times\lbrace \beta_\text{min} ; \beta_\text{max}\rbrace$

Les valeurs minimal et maximal sont complètement arbitraires et peuvent être aussi larges que souhaité si nous n'avons aucune idée de la position du phare ou au contraire plutôt réduites si une zone plus précise dans laquelle peut se situer le phare est définie.

\begin{equation}
p(\alpha, \beta) = \begin{cases}
1, &\text{si~} \alpha_\text{min} < \alpha < \alpha_\text{max} \text{~et~} \beta_\text{min} < \beta < \beta_\text{max}\\
0, &\text{sinon.}
\end{cases}
\end{equation}

\subsubsection{Fonction de vraisemblance $p(\lbrace x_k \rbrace|\alpha, \beta)$}

Les données mesurées proviennent d'événements indépendants, la probabilité d'obtenir un ensemble $\lbrace x_k \rbrace$ est simplement le produit des probabilités $x_k$ :
\begin{equation}
p(\lbrace x_k \rbrace|\alpha, \beta) = \prod_\text{k=1}^n p(x_k|\alpha, \beta)
\end{equation}

Ce qui donne :
\begin{equation}
\log(p(\lbrace x_k \rbrace|\alpha, \beta)) = n\log\beta - \sum_\text{k=1}^n \log\left(\beta^2 + (x_k-\alpha)^2\right)
\end{equation}

\subsection{La probabilité a postériori et recherche de la valeur de $\alpha$ la plus crédible}

Nous allons considérer que le prior est suffisamment large pour pour ne pas s'en soucier dans l'écriture du postérieur. Numériquement, ce prior permet de délimiter les zones de tracé ou de recherche des solutions.

\begin{equation}
\log(p(\alpha|\lbrace x_k \rbrace, \beta)) = \text{constante} - \sum_\text{k=1}^n \log\left(\beta^2 + (x_k-\alpha)^2\right)
\label{PharePosterieurAlpha}
\end{equation}

\begin{figure}[h]
\begin{center}
\includegraphics[width = 1\textwidth]{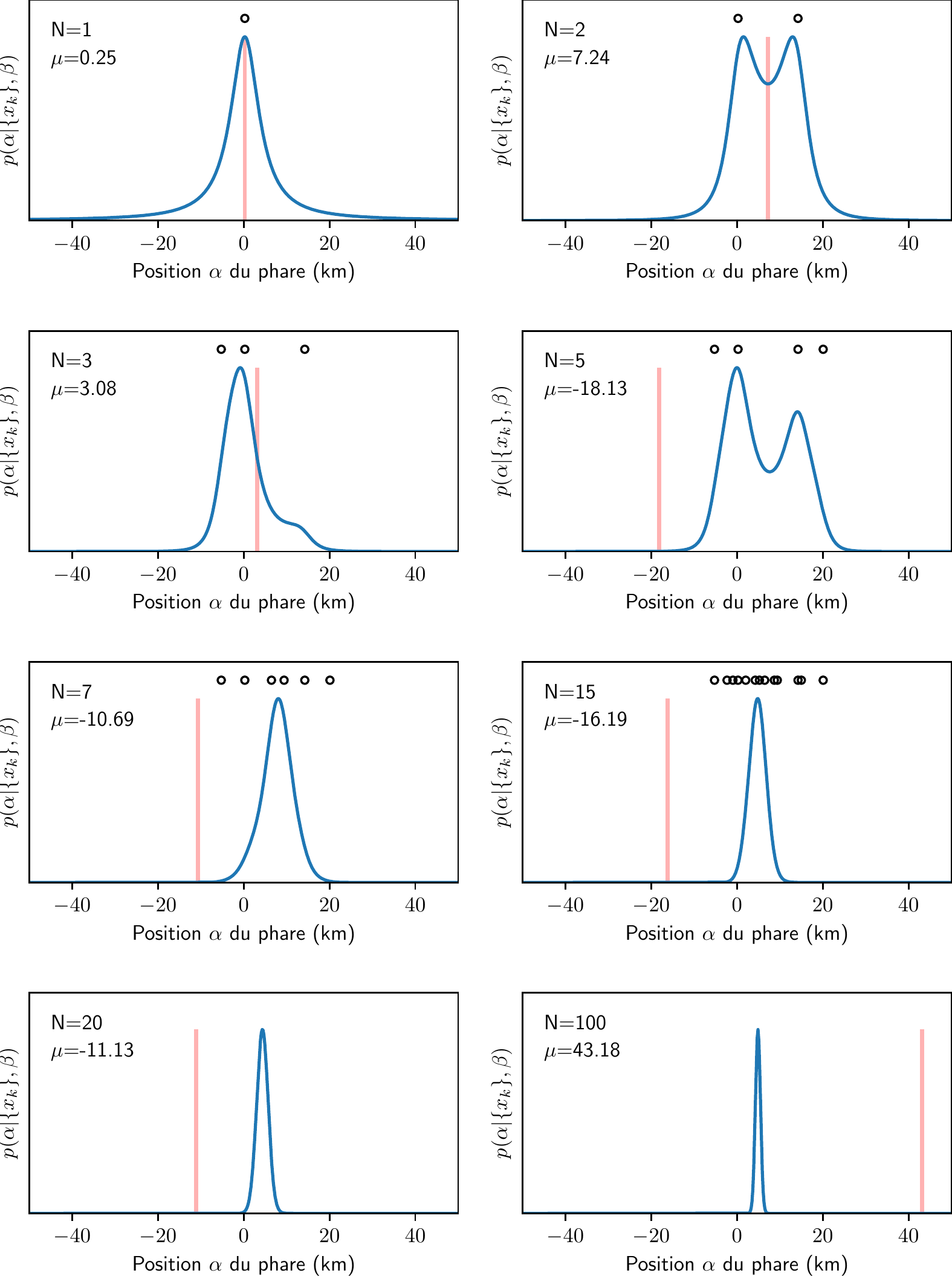}
\end{center}
\caption[Probabilité postérieure de la position $\alpha$ du phare]{Probabilité postérieure de la position $\alpha$ du phare le long de la côte en fonction du nombre de données. Le nombre de données et la moyenne de ces dernières sont indiqués en haut à gauche de chaque graphique. La valeur moyenne est représentée par un trait verticale.}
\label{PharePosterieur}
\end{figure}

La meilleur estimation pour la valeur de $\alpha$ est donnée pour le maximum de la relation donnée en \ref{PharePosterieurAlpha}. Analytiquement, cette relation est extrêmement difficile à inverser pour exprimer $\alpha$ en fonction des données du problème. Numériquement, c'est un problème simple à résoudre : il suffit de tracer cette probabilité en fonction de $\alpha$ et de rechercher la valeur maximum graphiquement. Ce tracé est réalisé en figure \ref{PharePosterieur} pour des valeurs $(\alpha, \beta) = (\num{5}, \num{4})$.

La figure \ref{PharePosterieur} montre l'évolution du postérieur en fonction du nombre de données. Le maximum de probabilité vaut $\alpha_\text{max} = \SI{4.70}{\kilo\m}$ pour $N = 100$. La distribution que suit $\alpha$ en fonction du nombre de point $N$ tend vers une forme piquée sur la valeur vraie de $\alpha$. Il est à remarquer que pour de faibles valeurs de $N$, cette probabilité peut devenir multimodale : des maxima principaux et secondaires apparaissent. Il très difficile de pouvoir utiliser les outils classiques que sont la valeur moyenne et l'écart-type pour décrire cette distribution ou en tirer des informations.

L'analyse bayésienne présente ainsi l'avantage de prendre en compte toute la complexité du problème et d'en donner une réponse adéquate.

En ce qui concerne la valeur moyenne, il peut sembler surprenant de constater qu'elle ne semble pas tendre vers une valeur centrale, comme le laisserait supposer une application du théorème de la limite centrale. De plus, cette valeur est très éloignée de la zone la plus probable pour $\alpha$, ce qui en fait un mauvais estimateur pour ce problème. 

Ici, la distribution statistique des positions sur la côte est pilotée par la distribution de Cauchy. Cette dernière ne possède ni valeur moyenne, ni écart type, ce qui explique le comportement erratique de la valeur moyenne et le fait que le théorème de la limite centrale ne s'applique pas.

\subsubsection{Recherche de $\alpha$ et $\beta$}
\begin{figure}[h]
\begin{center}
\includegraphics[width = 1\textwidth]{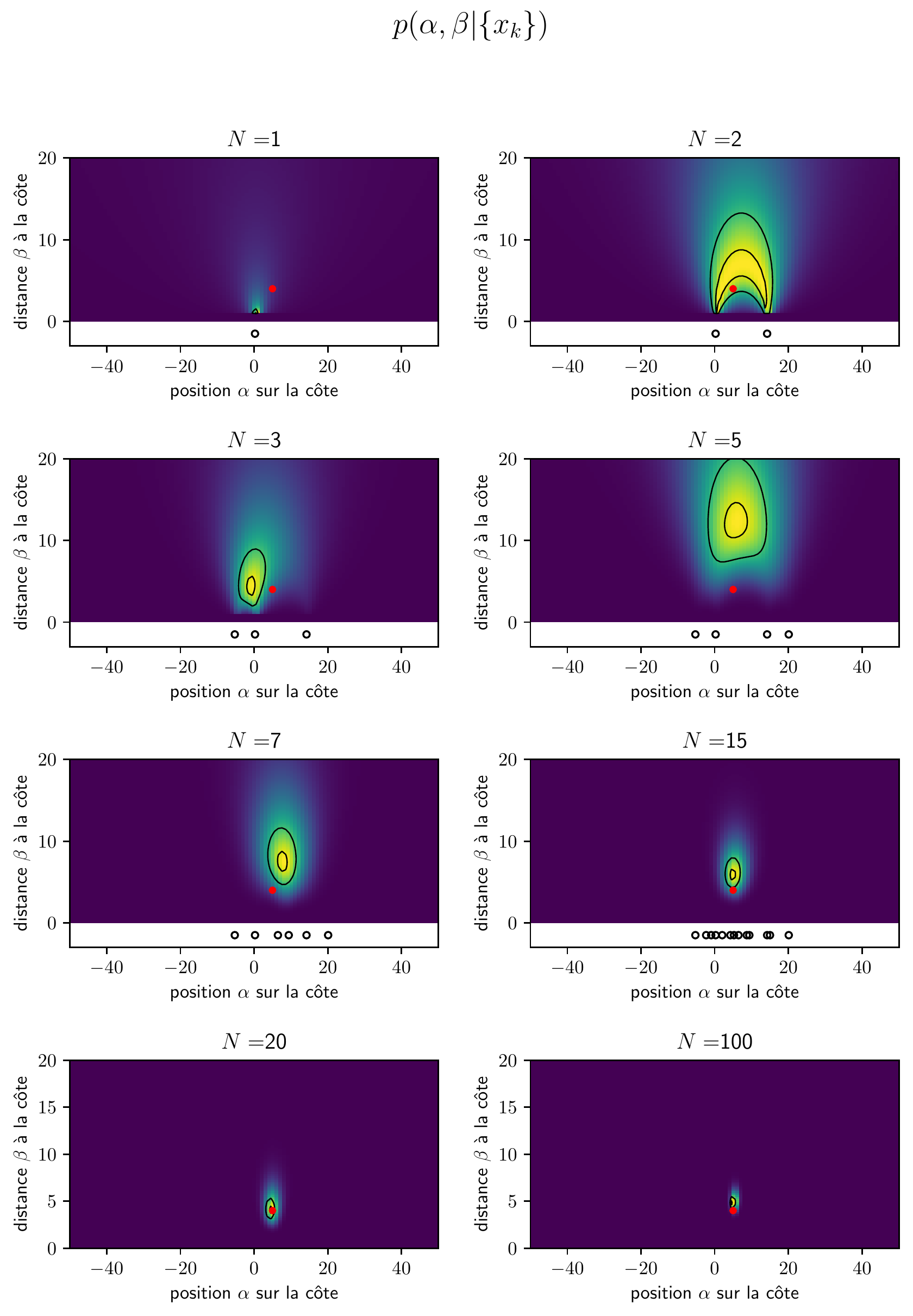}
\end{center}
\caption[Probabilité postérieure des coordonnées $\alpha$ et $\beta$ du phare]{Probabilité postérieure des coordonnées $\alpha$ et $\beta$ du phare. Le nombre de données est indiqué en haut de chaque graphique. La position exacte du phare est représentée par un point rouge.}
\label{DeuxParaPharePosterieur}
\end{figure}

La démarche précédente s'étend sans difficultés aux paramètres $(\alpha, \beta)$. Il s'agit d'un problème à deux dimensions dont la démarche de résolution est similaire à celui présenté en \ref{DeuxParaPharePosterieur}.

Cette technique permet de déterminer les positions les plus probables en fonction des deux dimensions de la position du phare. Avec suffisamment de points, la position du phare est déterminée avec une bonne précision.

\subsection{Conclusion}

L'analyse de données ne peut se limiter à fournir une valeur moyenne et un écart type ou une série de paramètres issues d'une régression. Il faut garder à l'esprit qu'un tel résultat sous-entend toujours une loi normale, or ce n'est pas toujours le cas. Prenons simplement l'exemple d'une mesure de masse, est-il raisonnable d'écrire $m = \SI{7(8)}{\g}$ pour un  intervalle de confiance à 95\% ? Il est évident qu'une masse négative n'a pas de sens.

 En mettant de côté la \textit{bizarrerie} de la distribution de Cauchy\footnote{Qui est utilisée pour décrire les raies spectroscopiques.} de ne pas posséder de valeur moyenne ou d'écart type, il existe des situations où la probabilité postérieure ne saurait se résumer à une simple loi normale. Dans beaucoup de situations, cette probabilité est asymétrique ou multimodale, ce qui rend non pertinent la description par une simple valeur moyenne.

\FloatBarrier

\section{Module EMCEE pour Python}

EMCEE\footnote{module Python sous licence MIT} est une implantation Python d'un méthode de Monte-Carlo par chaîne de Markov. Cette méthode permet d'échantillonner des distributions statistiques et d'en déterminer les caractéristiques.

Références : 
\begin{itemize}
\item Ensemble samplers with affine invariance, J. Goodman and J. Weare
\item EMCE documentation
\item lecture Astrostat C. Miller. www.astro.umd.edu/~miller/teaching/astrostat/lecture10.pdf
\item Emcee : the MCMC Hammer, D. Foreman-Mackey, D. W. Hogg, D. Lang, J. Goodman
\end{itemize}

L'intérêt de cet algorithme est sa capacité à rapidement converger vers une distribution satisfaisant la probabilité postérieure et d'en dessiner une image. Jusqu'à présent, il a été possible de tracer numériquement cette probabilité dans la mesure où les problèmes rencontrés ont peu de paramètres (maximum 2 jusqu'à présent). La difficulté de cette méthode directe est qu'elle est trop couteuse en temps de calcul pour des modèles avec beaucoup de paramètres, et certains nécessitent autant de paramètres que de données.

\subsection{Retour sur le problème de la position du phare}

Reprenons l'analyse précédente en traitant numériquement le problème avec le module EMCEE. L'analyse du problème est strictement identique, seul la recherche des paramètres $\alpha$ et $\beta$ optimaux et le tracé diffèrent.

\begin{pyverbatim}
#algorithme emce : détermination des paramètres a et b
ndim = 2  # number of parameters in the model
nwalkers = 200  # number of MCMC walkers
nburn = 3000  # "burn-in" period to let chains stabilize
nsteps = 6000  # number of MCMC steps to take

# Estimation pour l'initialisation des "walkers"
starting_guesses = np.random.rand(nwalkers, ndim)    
starting_guesses[:,0] = 50 *starting_guesses[:,0] -25
starting_guesses[:,1] *= 25
starting_guesses[:,1] += 0

sampler = emcee.EnsembleSampler(nwalkers, ndim, log_posterior, args=[c_pos])
sampler.run_mcmc(starting_guesses, nsteps)

sample = sampler.chain 
sample = sampler.chain[:, nburn:, :].reshape(-1, 2)

# Tracé graphique
from astroML.plotting import plot_mcmc
fig, ax = plt.subplots(figsize=(10, 10))
plot_mcmc(sample.T, fig=fig, labels=[r'position a sur la côte',...
 r'distance b à la côte'], colors='k')
plt.plot(sample[:, 0], sample[:, 1], ',k', alpha=0.1)
plt.ylim(-2, 50)
plt.xlim(-50,50)
plt.plot(a, b, 'o', color='red', ms=10);
\end{pyverbatim}

L'algorithme EMCEE fonctionne de la manière suivante : des \textit{walkers} sont initialisés à des positions estimées et proche de la solution recherchée. Ici, les \textit{walkers} pour a et b sont initialisés aléatoirement dans la zone $a = [-50,50]$ et $b = [0, 50]$. Ensuite, un nombre d'étapes est défini pour laisser le temps aux \textit{walkers} d'explorer la distribution et de perdre la mémoire de leur position initiale, c'est la phase de \textit{burn-in}. Les données acquises durant cette phase sont simplement éliminées. Enfin, après cette première phase, un certain nombre d'étapes sont laissées aux \textit{walkers} pour explorer pleinement la distribution.

Le nombre d'étape de chacune des deux phases dépend de la complexité du problème.

\begin{figure}[h]
\begin{center}
\includegraphics[width = 1\textwidth]{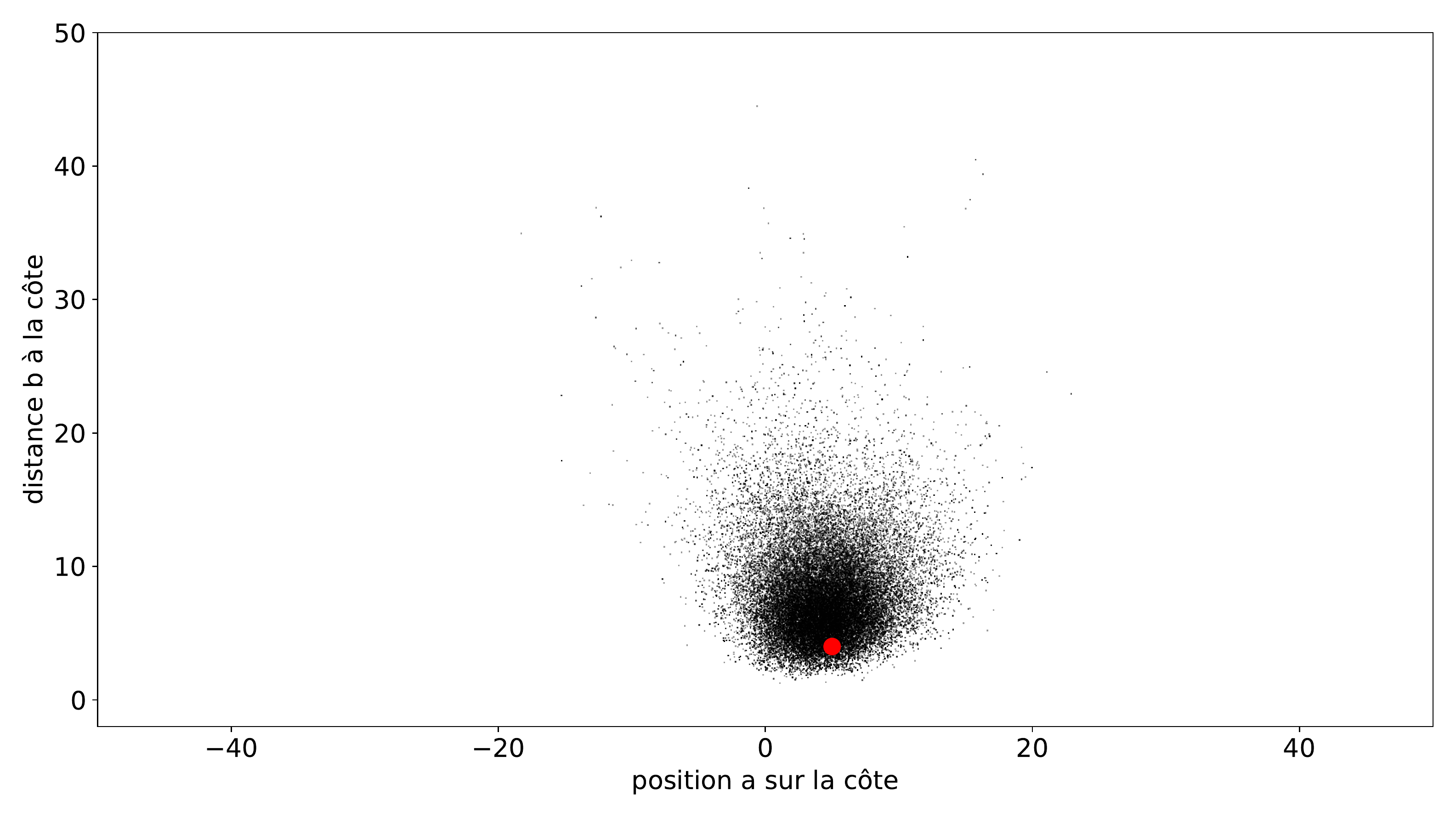} %ExempleEMCE.pdf
\end{center}
\caption[Probabilité postérieure déterminée au moyen de l'algorithme EMCEE]{Probabilité postérieure des coordonnées $\alpha$ et $\beta$ du phare calculé au moyen de l'algorithme EMCE.}
\label{ExempleEMCE}
\end{figure}

Ensuite la distribution est tracée comme montré en figure \ref{ExempleEMCE}. Les zones sombres sont des zones fortement explorées par les \textit{walkers} et correspondent aux zones où la probabilité postérieure est grande.

\section{Régression et élimination des données aberrantes}

Références : 
\begin{itemize}
\item Notice en ligne EMCE
\item Data analysis recipes: Fitting a model to data, D.W. Hogg, J. Bovy, D. Lang
\end{itemize}

Avec le module EMCEE et les techniques d'inférences bayésiennes, il est possible de définir des modèles de régression suffisamment complexes pour détecter et éliminer les valeurs aberrantes d'un ensemble de données. En effet, l'enjeu est important dans la mesure où la régression linéaire classique est très sensible aux valeurs aberrantes. Il existe des techniques de correction pour rendre ces régressions moins sensible \footnote{Fonction objectif d'Huber, par exemple.} aux valeurs aberrantes, mais ces techniques ont des limites et à défaut d'éliminer complètement les valeurs aberrantes, elles réduisent le poids de ces dernières.

\subsection{Positionnement du problème}

L'objectif est de trouver les meilleurs coefficients a et b d'un modèle affine $y = ax + b$ pour une série de données obtenues par les relations suivantes :

\begin{equation}
\begin{cases}
\lbrace x_i \rbrace =& \text{distribution uniforme sur l'intervalle} \left[0, 100\right]\\
\lbrace\sigma_i \rbrace =& \text{distribution uniforme sur l'intervalle} \left[2, 22\right]\\
\lbrace y_\text{err} \rbrace = &\mathcal{N}(0, \lbrace \sigma_i \rbrace )\\
\lbrace y_i \rbrace =& a\cdot \lbrace x_i \rbrace  + b + \lbrace y_\text{err} \rbrace\\

\end{cases}
\end{equation}
Les données aberrantes sont ensuite inclues dans l'ensemble des données.  Le code Python indiqué ci-dessous permet la génération de données décrite précédemment.

\begin{pyverbatim}

import numpy as np
import matplotlib.pyplot as plt
from scipy import stats
from tqdm import tqdm

# Pour la répétabilité 
np.random.seed(10) 

# Paramètre du modèle
a_true = 2
b_true = -5

# Generation de données à partir du model
N = 20
x = 0.5 + np.sort(99 * np.random.rand(N))
sigma = 2 + 20 * np.random.rand(N)
y = a_true * x + b_true
yerr = stats.norm.rvs(size = N, loc = 0, scale = abs(sigma))
y+= yerr

# Ajout de valeurs aberrantes
NbOutliers = 3
Nrand = np.random.randint(0,N-1,size = NbOutliers)
y[Nrand] = [174.5, 115.9, 95.9 ]

\end{pyverbatim}

\begin{figure}[h]
\begin{center}
\includegraphics[width = 1\textwidth]{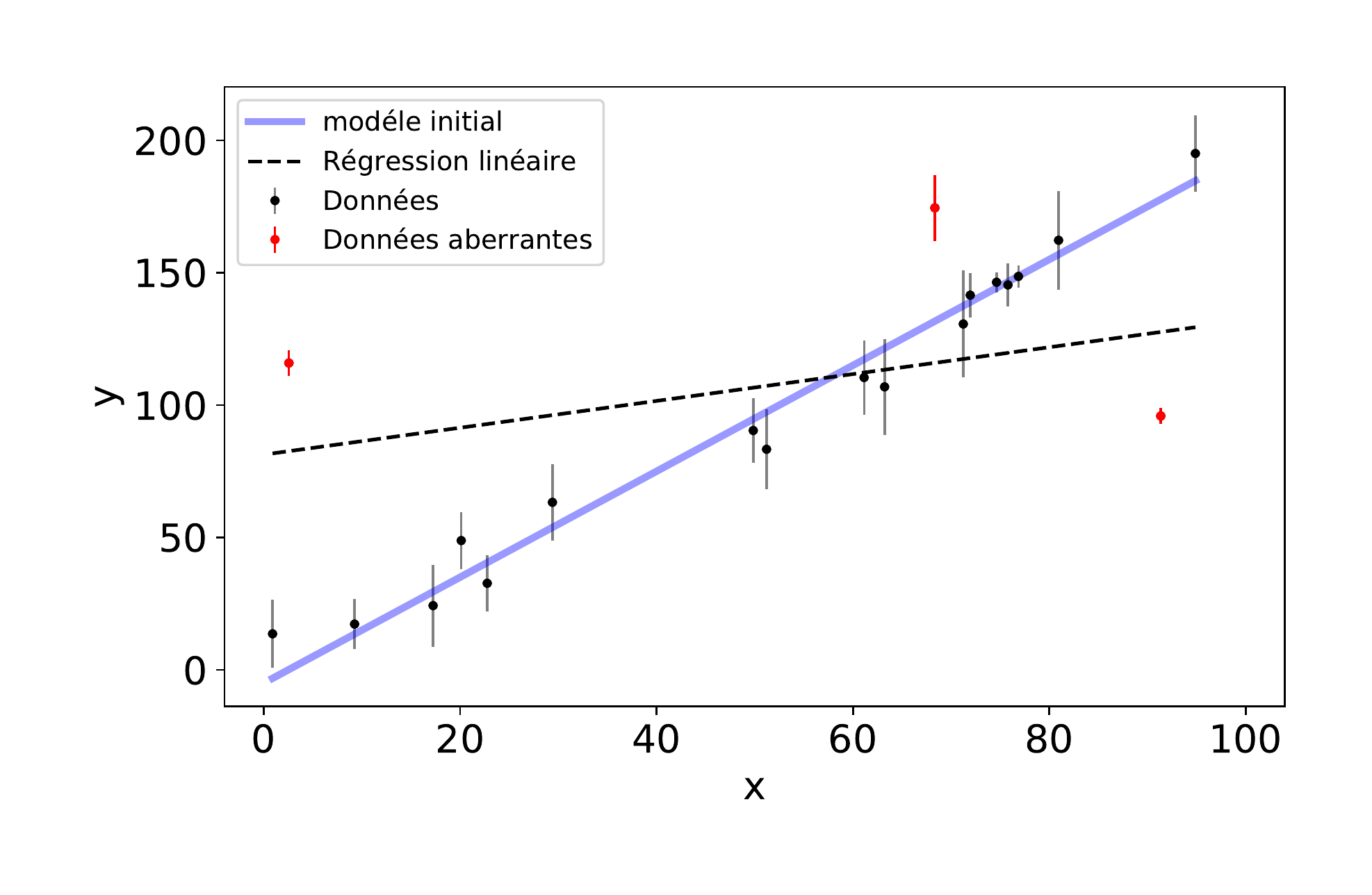} %ExempleEMCE.pdf
\end{center}
\caption[Données issues d'un tirage aléatoire : influence des données aberrantes sur la régression linéaire]{Données issues d'un tirage aléatoire suivant une loi normale d'écart type $\sigma \approx 10$ et centrée sur chaque point $y_i = ax_i + b$, avec $a=2$ et $b = -5$. Des données aberrantes décorrélées de la relation affine initiale sont inclues dans l'ensemble des données et représentées en rouge.}
\label{Outliers_Datas}
\end{figure}

Un modèle de régression linéaire tel que décrit dans le chapitre \ref{ModelesEtRegressions} est ajoutée à la représentation graphique des données en figure \ref{Outliers_Datas}. Ce dernier est lourdement influencé par les quelques données aberrantes et ne permet pas de décrire la relation linéaire qui se dessine intuitivement avec les données valides.

\subsection{Construction du modèle}

\subsubsection{Caractéristique des données valides}
Le modèle initiale est similaire à celui décrit par la relation \ref{Chi2AvecSigma} : Il faut chercher un extremum d'une fonction chiffrant l'écart du modèle aux données.

Nous allons supposer que l'incertitude associée à chaque point suit une loi normale. La probabilité d'avoir les données $D = \lbrace x_i, y_i, \sigma_i \rbrace$ connaissant le modèle $H = \left(a, b\right)$ est : 

\begin{equation}
p\left(\lbrace x_i, y_i, \sigma_i \rbrace | \left(a, b\right) \right) = p(D|H) = \prod_{i=0}^n  \dfrac{1}{2\pi\sigma_i}\exp \left[-\frac{1}{2}\left(\frac{ a x_i +b - y_i}{\sigma_i}\right)^2\right]
\end{equation}

Chaque terme du produit est simplement une probabilité donnée par une loi normale d'écart type $\sigma_i$ et de valeur centrale $a x_i +b$ donnée par le modèle. En utilisant cette probabilité comme fonction de vraisemblance et en prenant un prior uniforme, l'utilisation du théorème de Bayes conduit à minimiser la même grandeur que celle définie par la relation \ref{Chi2AvecSigma}, et cela conduit au modèle de la régression linéaire.

\subsubsection{Caractéristique des données aberrantes}

Une donnée aberrante est une donnée qui n'est pas corrélée\footnote{Ce choix est discutable pour une mesure réelle, mais par soucis de simplification, c'est le modèle qui est retenu pour cet exemple.} au phénomène dont on mesure une grandeur mais qui suit une statistique indépendante du modèle :

\begin{equation}
\dfrac{1}{2\pi\sigma_A}\exp \left[-\frac{1}{2}\left(\frac{ Y_A - y_i}{\sigma_A}\right)^2\right]
\end{equation}

\subsubsection{Association des deux caractéristiques}

Pour combiner les deux aspects, il est possible de construire une probabilité reposant sur des paramètres $\lbrace g_i \rbrace$ chiffrant le caractère aberrant ou non d'une données. $g_i$ est un paramètre supplémentaire associé à chaque point et variant de 0 à 1 :
\begin{itemize}
\item $g_i$ proche de 1, le point est une donnée valide;
\item $g_i$ proche de 0, le point est une donnée aberrante.
\end{itemize}

L'association des données aberrantes et valides se fait donc au travers de la probabilité suivante : 
\begin{equation}
\begin{cases}
f(g_i) = \begin{cases}
1,& \text{si~} g_i > g_0\\
0,& \text{si~} g_i < g_0\\
\end{cases}\\
p\left( x_i, y_i, \sigma_i | \left(a, b\right), g_i \right) = \dfrac{f(g_i)}{2\pi\sigma_i}\exp \left[-\frac{1}{2}\left(\frac{ a x_i +b - y_i}{\sigma_i}\right)^2\right] + \dfrac{1-f(g_i)}{2\pi\sigma_A}\exp \left[-\frac{1}{2}\left(\frac{ Y_A - y_i}{\sigma_A}\right)^2\right]
\end{cases}
\end{equation}

Suivant la valeur de $g_i$, la probabilité suivie est soit celle d'une donnée valide, soit celle d'une donnée aberrante.

\subsubsection{Fonction de vraisemblance, prior et probabilité postérieure}

Un prior uniforme\footnote{Concernant le paramètre $a$, le prior n'est pas non informatif. En effet, il faudrait choisir une probabilité $a$ tel que la probabilité que la droite fasse un angle $\alpha$ avec l'horizontale soit uniforme.} est utilisé, par soucis de simplicité :
\begin{equation}
p(a, b) \propto 1
\end{equation}

La fonction de vraisemblance est simplement le produit des probabilités $p$ :
\begin{equation}
p\left(\lbrace x_i, y_i, \sigma_i \rbrace| \left(a, b\right), \lbrace g_i\rbrace \right) =  \prod_{i=0}^n p\left( x_i, y_i, \sigma_i | \left(a, b\right), g_i \right)
\end{equation}

La probabilité postérieure est donc : 

\begin{equation}
p\left(\left(a, b\right), \lbrace g_i\rbrace | \lbrace x_i, y_i, \sigma_i \rbrace \right) = p(H,\lbrace g_i\rbrace|D) \propto  \prod_{i=0}^n p\left( x_i, y_i, \sigma_i | \left(a, b\right), g_i \right)
\end{equation}

\subsubsection{Code python}

Les probabilités précédentes sont définies numériquement au moyen de logarithme pour en facilité le calcule numérique.

\begin{pyverbatim}
# Probabilité inférence bayésienne
# theta est un tableau contenant tous les paramètres :
# theta[0]  : b
# theta[1]  : a
# theta[2:] : gi
def log_prior(theta):
    #g_i needs to be between 0 and 1
    if (all(theta[2:] > 0) and all(theta[2:] < 1)):
        return 0
    else:
        return -np.inf  # recall log(0) = -inf

def log_likelihood(theta, x, y, e, sigma_B):
    dy = y - theta[0] - theta[1] * x
    dyA= np.mean(y) - y
    g = np.clip(theta[2:], 0, 1)  # g<0 or g>1 conduit à des NaN 
    for i in range(len(g)): # fonction f stocké dans g
        if g[i]<0.5 : g[i]=0
        else : g[i]=1
    logL1 = np.log(g) - 0.5*np.log(2*np.pi*e**2) - 0.5*(dy/e)**2
    logL2 = np.log(1 - g) - 0.5*np.log(2*np.pi*sigma_B**2) - 0.5*(dyA/sigma_B)**2
    return np.sum(np.logaddexp(logL1, logL2))

def log_posterior(theta, x, y, e, sigma_B):
    return log_prior(theta) + log_likelihood(theta, x, y, e, sigma_B)


\end{pyverbatim}

\subsection{Paramètres de nuisance}

Les paramètres $ \lbrace g_i\rbrace$ sont des paramètres de nuisance. Un paramètre de nuisance est nécessaire au modèle retenu, mais n'apporte aucune information concernant le résultat final. L'opération visant à les faire disparaitre est appelée \textit{marginalisation des paramètres de nuisance} et consiste à intégrer la probabilité sur l'ensemble des valeurs de ces paramètres :

\begin{equation}
p\left(\left(a, b\right)|\lbrace x_i, y_i, \sigma_i \rbrace \right)=p(H|D)= \int_{\lbrace g_i \rbrace} p\left(\left(a, b\right), \lbrace g_i\rbrace | \lbrace x_i, y_i, \sigma_i \rbrace \right) dg_i
\end{equation}

\subsection{algorithme EMCEE}

\begin{pyverbatim}
# Paramètre du modèle et du module EMCEE
ndim = 2 + len(x)  # Nombre de paramètre du modèle
nwalkers = 50  # Nombre de "walkers"
nburn = 10000  # periode de "burn-in"
nsteps = 15000 # Nombre d'étapes de l'algorithme

# Estimation pour l'initialisation des "walkers"
np.random.seed(0)
starting_guesses = np.zeros((nwalkers, ndim))
starting_guesses[:, 0] = np.random.normal(-5, 10, (nwalkers))
starting_guesses[:, 1] = np.random.normal(2, 5, (nwalkers))
starting_guesses[:, 2:] = np.random.normal(0.5, 0.25, (nwalkers, ndim - 2))

# Fonctions principales EMCEE
import emcee
sampler = emcee.EnsembleSampler(nwalkers,ndim,log_posterior,args=[x,y,sigma,100])
sampler.run_mcmc(starting_guesses, nsteps, progress = True)

#Mise en forme de l'échantillon "sample"
#Sample est une image de la distribution postérieur
sample = sampler.chain  # shape = (nwalkers, nsteps, ndim)
sample = sampler.chain[:, nburn:, :].reshape(-1, ndim)
\end{pyverbatim}

L'opération de marginalisation est cachée par le principe de l'algorithme. En effet, une des propriétés des chaines de Monte-Carlo est d'\textit{imager} la distribution postérieur, et l'opération de marginalisation se fait simplement en ignorant les paramètres $\lbrace g_i\rbrace$.

\textit{sample} est un tableau dont l'une des dimensions contient les paramètres du système : les deux premiers sont les coefficients a et b, tous les suivants sont les $ \lbrace g_i\rbrace$. 

\begin{pyverbatim}
# Marginalisation des paramètres de nuisance
a_sample = sample[:,1]
b_sample = sample[:,0]

# Tracé graphique
plt.plot(b_sample, a_sample, ',k', alpha=0.1)
plt.xlabel('Ordonnée à l'origine b')
plt.ylabel('Pente a');
\end{pyverbatim}

\begin{figure}[h]
\begin{center}
\includegraphics[width = 1\textwidth]{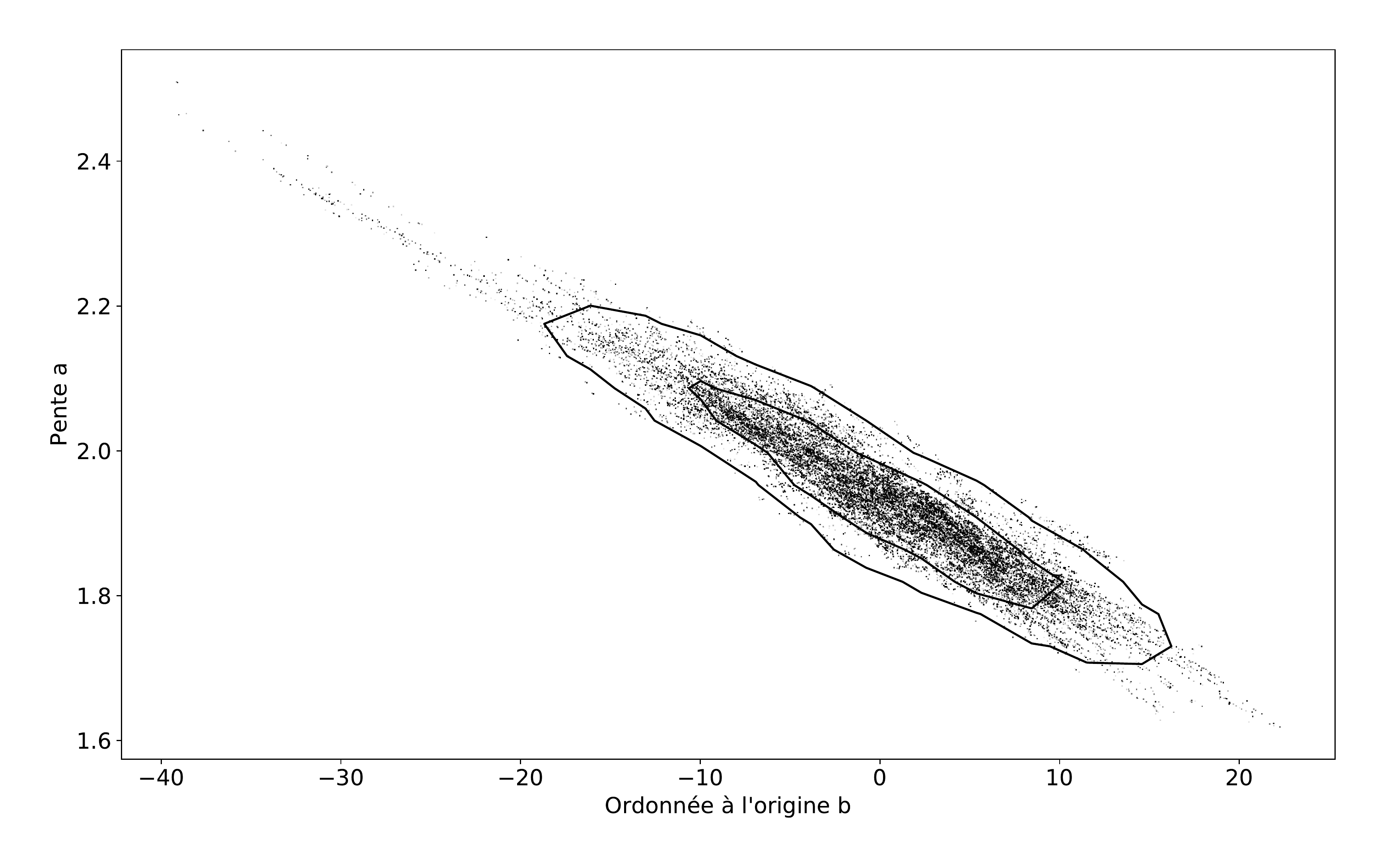} %ExempleEMCE.pdf
\end{center}
\caption[Inférence des coefficients a et b ajustement d'un modèle affine]{Inférence de la zone la plus crédible pour les pentes $a$ et ordonnées à l'origine $b$ du modèle.}
\label{SlopeIntercept}
\end{figure}

Les données a et b issues de l'échantillonnage sont représentées en figure \ref{SlopeIntercept}. Une zone centrée sur $a=2$ et $b=0$ ressort des données et correspond aux zones de crédibilité les plus élevées.

\subsection{Élimination des données aberrantes}

Le modèle sélectionne les valeurs de $g_i$ minimisant l'erreur commise entre la droite moyenne et les données. Lorsque les $g_i$ dépassent une valeur seuil $g_0$ arbitrairement fixée à \num{0.5}, ces derniers ne participent plus à l'évaluation des coefficients $a$ et $b$. 

\begin{figure}[h]
\begin{center}
\includegraphics[width = 1\textwidth]{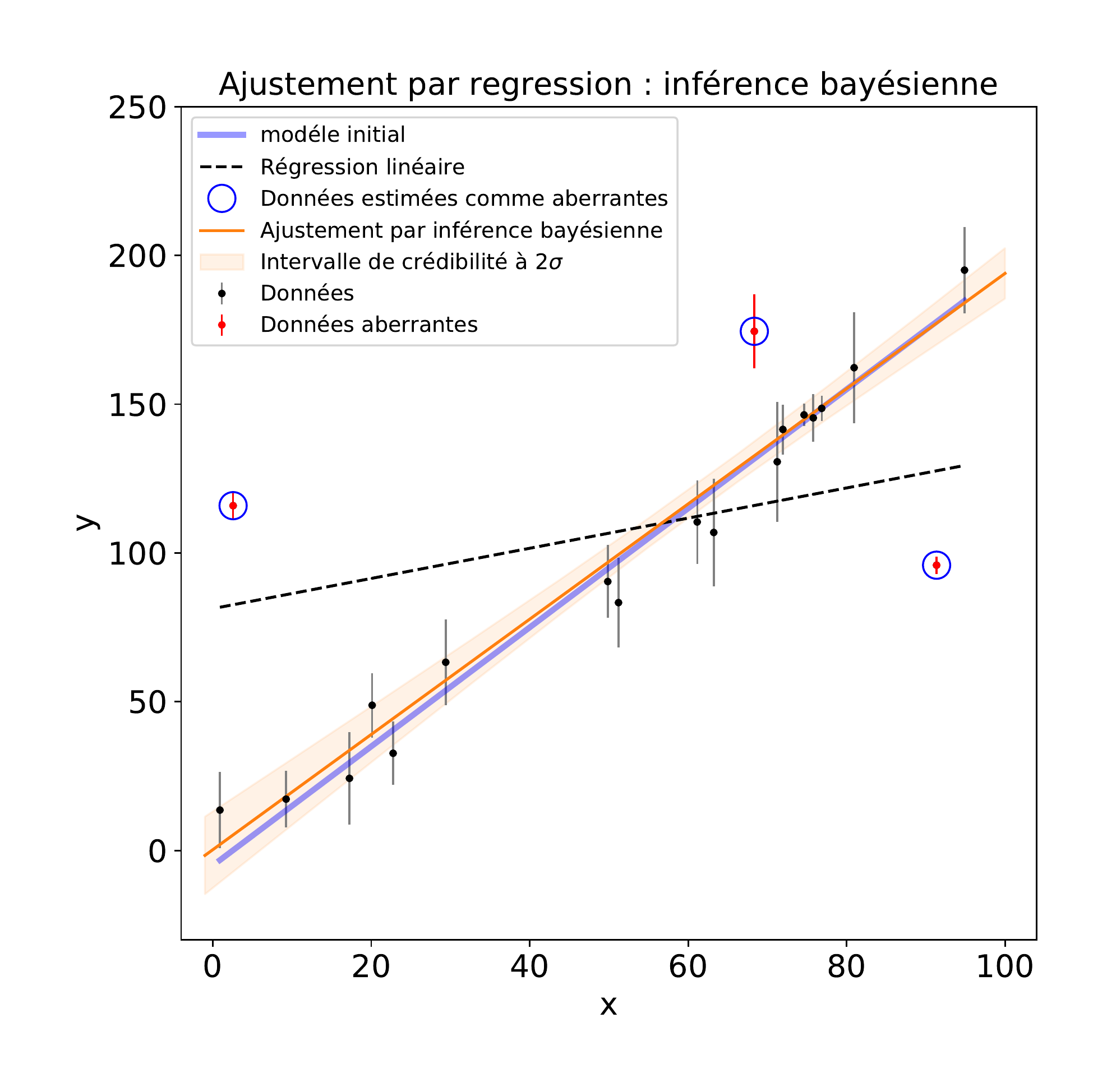} %ExempleEMCE.pdf
\end{center}
\caption[Ajustement d'un modèle affine par inférence bayésienne]{Ajustement d'un modèle affine par inférence bayésienne. Des valeurs aberrantes sont ajoutées à un ensemble de données bruitées dont le modèle sous-jacent est une fonction affine. La méthode permet d'éliminer les valeurs aberrantes pour ne prendre en compte que les données valides contrairement à une régression linéaire simple sensible aux valeurs aberrantes.}
\label{AjustementParInferenceBayesienne}
\end{figure}

Cela signifie que les données aberrante possèdent un paramètre $g_i$ inférieur à \num{0.5}. Il est donc possible de mettre en évidence les valeurs aberrantes. Le graphique \ref{AjustementParInferenceBayesienne} reprend l'ensemble de l'analyse réalisée. Les données entourée en bleu sont celles détectées comme valeurs aberrantes par la méthode. Une faisceau de courbes représentant l'intervalle de crédibilité à 95\% est ajouté. La droite servant de modèle initiale est bien inclue dans ce faisceau.

\begin{pyverbatim}

# Fonctions de tracé graphique
def plot_MCMC_model(ax, xdata, ydata, trace):
    """Plot the linear model and 2sigma contours"""

    alpha, beta = trace[:2]
    xfit = np.linspace(-1, 100, 10)
    yfit = alpha[:, None] + beta[:, None] * xfit
    mu = yfit.mean(0)
    sig = 2 * yfit.std(0)

    ax.plot(xfit, mu, "C1", label = "Ajustement par inférence bayésienne")
    ax.fill_between(xfit, mu - sig, mu + sig, color = 'C1', alpha=0.1,...
     label = r"Intervalle de crédibilité à $2\sigma$")#, color='lightgray')

    ax.set_xlabel('x')
    ax.set_ylabel('y')

# Retrait des données issues de la phase de "burn-in"
emcee_trace = sampler.chain[:, nburn:, :].reshape(-1, ndim).T

fig, ax = plt.subplots()
plot_MCMC_model(ax, x, y, emcee_trace)

\end{pyverbatim}

L'intérêt de cette méthode est de pouvoir normaliser et rendre transparent le processus d'élimination des valeurs aberrantes. La richesse du formalisme d'analyse bayésienne rend cette opération possible. Bien que plus complexe à mettre en œuvre, une fois implémenté ces outils peuvent s'utiliser dans une grande variété de situations.

\clearpage
\section{Quelques mots pour conclure}
Références : 
\begin{itemize}
\item Nature, volume 506, issue 7487, 13 February 2014, "Statistical errors", R. Nuzzo
\item Bayesian Reasonning in Data Analysis, G. D'Agostini
\end{itemize}

Je reviendrai simplement sur la différence de point de vu entre les traitements classiques et par inférences bayésiennes. Cette différence réside dans l'interprétation donnée dans l'analyse statistique d'un problème. D'un côté, nous nous intéressons à la probabilité\footnote{Cette probabilité se résume souvent au fameux facteur $p$.} d'obtenir un jeu de données particulier moyennant la connaissance d'un modèle. D'un autre côté, nous déterminons le degrés de probabilité du modèle connaissant un jeu de données. 

\begin{quote}
A P value measures whether an observed result can be attributed to chance. But it cannot answer a researcher’s real question: what are the odds that a hypothesis is correct? Those odds depend on how strong the result was and, most importantly, on how plausibile the hypothesis is in the first place.
\end{quote}

Ces deux approches sont différentes et ne répondent fondamentalement pas aux mêmes questions. 

La difficulté réside dans le fait que la plupart du temps l'analyse de données conduit aux mêmes résultats chiffrés quelque soit la méthode employée. De ce fait, l'aspect utilitaire l'emporte : pourquoi faire compliqué ?

L'approche bayésienne est complexe à mettre en œuvre auprès d'élèves ou étudiants et nécessite beaucoup de prérequis. Le cadre des travaux pratiques en temps limité ne permet souvent pas de réaliser l'analyse de données ou même de s'y arrêter quelques minutes alors qu'elle fait partie intégrante du travail expérimental. 

Cependant, après ce modeste\footnote{Au vu de la littérature universitaire existant sur le sujet} tour des capacités d'analyse de l'approche bayésienne et des limites de certaines pratiques, j'ose espérer que l'analyse de données occupera une place plus importante, et qu'a défaut de rentrer dans les détails, le cadre d'application des méthodes d'analyses statistiques sera défini et le sens à donner à ces analyses sera introduit.

%----------------------------------------------------------------------------------------
%	BIBLIOGRAPHY
%----------------------------------------------------------------------------------------

%\bibliography{MesureBiblio} % Use the bibliography.bib file for the bibliography
%%\bibliographystyle{plainnat} % Use the plainnat style of referencing
%\bibliographystyle{plain}

%---------------------------------------------------------------------------------------

\end{document}